\documentclass[10pt,journal,cspaper,compsoc]{IEEEtran}
\usepackage{cqiqe19}

\ifCLASSOPTIONcompsoc
  \usepackage[nocompress]{cite}
\else
  \usepackage{cite}
\fi

\makeatletter
\@ifundefined{lhs2tex.lhs2tex.sty.read}%
  {\@namedef{lhs2tex.lhs2tex.sty.read}{}%
   \newcommand\SkipToFmtEnd{}%
   \newcommand\EndFmtInput{}%
   \long\def\SkipToFmtEnd#1\EndFmtInput{}%
  }\SkipToFmtEnd

\newcommand\ReadOnlyOnce[1]{\@ifundefined{#1}{\@namedef{#1}{}}\SkipToFmtEnd}
\usepackage{amstext}
\usepackage{amssymb}
\usepackage{stmaryrd}
\DeclareFontFamily{OT1}{cmtex}{}
\DeclareFontShape{OT1}{cmtex}{m}{n}
  {<5><6><7><8>cmtex8
   <9>cmtex9
   <10><10.95><12><14.4><17.28><20.74><24.88>cmtex10}{}
\DeclareFontShape{OT1}{cmtex}{m}{it}
  {<-> ssub * cmtt/m/it}{}

\DeclareFontShape{OT1}{cmtt}{bx}{n}
  {<5><6><7><8>cmtt8
   <9>cmbtt9
   <10><10.95><12><14.4><17.28><20.74><24.88>cmbtt10}{}
\DeclareFontShape{OT1}{cmtex}{bx}{n}
  {<-> ssub * cmtt/bx/n}{}

\newcommand{\Conid}[1]{\mathit{#1}}
\newcommand{\Varid}[1]{\mathit{#1}}
\newcommand{\anonymous}{\kern0.06em \vbox{\hrule\@width.5em}}

\renewcommand{\leq}{\leqslant}

\usepackage{polytable}

\@ifundefined{mathindent}%
  {\newdimen\mathindent\mathindent\leftmargini}%
  {}%

\def\resethooks{%
  \global\let\SaveRestoreHook\empty
  \global\let\ColumnHook\empty}
\newcommand*{\savecolumns}[1][default]%
  {\g@addto@macro\SaveRestoreHook{\savecolumns[#1]}}
\newcommand*{\restorecolumns}[1][default]%
  {\g@addto@macro\SaveRestoreHook{\restorecolumns[#1]}}
\newcommand*{\aligncolumn}[2]%
  {\g@addto@macro\ColumnHook{\column{#1}{#2}}}

\resethooks

\newcommand{\onelinecommentchars}{\quad-{}- }
\newcommand{\commentbeginchars}{\enskip\{-}
\newcommand{\commentendchars}{-\}\enskip}

\newcommand{\visiblecomments}{%
  \let\onelinecomment=\onelinecommentchars
  \let\commentbegin=\commentbeginchars
  \let\commentend=\commentendchars}

\newcommand{\invisiblecomments}{%
  \let\onelinecomment=\empty
  \let\commentbegin=\empty
  \let\commentend=\empty}

\visiblecomments

\newlength{\blanklineskip}
\newcommand{\hsindent}[1]{\quad}
\let\hspre\empty
\let\hspost\empty

\EndFmtInput
\makeatother
\ReadOnlyOnce{polycode.fmt}%
\makeatletter

\newcommand{\hsnewpar}[1]%
  {{\parskip=0pt\parindent=0pt\par\vskip #1\noindent}}

\newcommand{\hscodestyle}{}

\newcommand{\sethscode}[1]%
  {\expandafter\let\expandafter\hscode\csname #1\endcsname
   \expandafter\let\expandafter\endhscode\csname end#1\endcsname}

  {\par\noindent
   \advance\leftskip\mathindent
   \hscodestyle
   \let\\=\@normalcr
   \let\hspre\(\let\hspost\)%
   \pboxed}%
  {\endpboxed\)%
   \par\noindent
   \ignorespacesafterend}

  {\hsnewpar\abovedisplayskip
   \advance\leftskip\mathindent
   \hscodestyle
   \let\hspre\(\let\hspost\)%
   \pboxed}%
  {\endpboxed%
   \hsnewpar\belowdisplayskip
   \ignorespacesafterend}

  {\hsnewpar\abovedisplayskip
   \advance\leftskip\mathindent
   \hscodestyle
   \let\\=\@normalcr
   \(\pboxed}%
  {\endpboxed\)%
   \hsnewpar\belowdisplayskip
   \ignorespacesafterend}

\newcommand{\plainhs}{\sethscode{plainhscode}}

\plainhs

  {\hsnewpar\abovedisplayskip
   \advance\leftskip\mathindent
   \hscodestyle
   \let\\=\@normalcr
   \(\parray}%
  {\endparray\)%
   \hsnewpar\belowdisplayskip
   \ignorespacesafterend}

  {\parray}{\endparray}

  {\(\parray}{\endparray\)}

\def\codeframewidth{\arrayrulewidth}
\RequirePackage{calc}

  {\parskip=\abovedisplayskip\par\noindent
   \hscodestyle
   \arrayrulewidth=\codeframewidth
   \tabular{@{}|p{\linewidth-2\arraycolsep-2\arrayrulewidth-2pt}|@{}}%
   \hline\framedhslinecorrect\\{-1.5ex}%
   \let\endoflinesave=\\
   \let\\=\@normalcr
   \(\pboxed}%
  {\endpboxed\)%
   \framedhslinecorrect\endoflinesave{.5ex}\hline
   \endtabular
   \parskip=\belowdisplayskip\par\noindent
   \ignorespacesafterend}

\newcommand{\framedhslinecorrect}[2]%
  {#1[#2]}

  {\(\def\column##1##2{}%
   \let\>\undefined\let\<\undefined\let\\\undefined
   \newcommand\>[1][]{}\newcommand\<[1][]{}\newcommand\\[1][]{}%
   \def\fromto##1##2##3{##3}%
   }{\) }%

  {\let\orighscode=\hscode
   \let\origendhscode=\endhscode
   \def\endhscode{\def\hscode{\endgroup\def\@currenvir{hscode}\\}\begingroup}
   \orighscode\def\hscode{\endgroup\def\@currenvir{hscode}}}%
  {\origendhscode
   \global\let\hscode=\orighscode
   \global\let\endhscode=\origendhscode}%

\makeatother
\EndFmtInput
\def\fun#1{\mathsf{#1}}
\def\quanta#1{\mathopen{\langle\!|}#1\mathclose{|\!\rangle}}
\def\rcb#1#2#3#4{\def\nothing{}\def\range{#3}\mathopen{\langle}#1 \ #2 \ \ifx\range\nothing::\else: \ #3 :\fi \ #4\mathclose{\rangle}}
\def\ket#1{\vert#1\rangle}
\def\kcomp{\mathbin{\bullet}}
\def\meither#1#2{\left[\begin{array}{c|c}{#1}&{#2}\end{array}\right]} 
\def\matrix#1#2#3#4{\begin{bmatrix}#1&#2\\#3&#4\end{bmatrix}}

\def\arrows{\ar@<1.8ex>[r]\ar@<-1.8ex>[r]}
\def\fstcbox#1#2#3#4#5{%
\xymatrix{%
\nobox{\raisebox{1em}{$#1$}\hskip-1.9ex\raisebox{-1em}{$#2$}}%
		\arrows
&
	\wbox{$#3$}%
		\arrows
&
\nobox{\raisebox{1em}{\hskip 1ex$#4$}\hskip -1.5ex\raisebox{-1em}{$#5$}}%
}}

\usepackage{tikz}
\usepackage{qcircuit}
\usepackage{hyperref}

\tikzstyle{block} = [rectangle, draw, text centered]
\tikzstyle{arrow} = [draw, -latex']

\newcount\td \td=0 
\long \def\ToDo#1{\vskip 1ex\noindent\fbox{
\global\advance\td by 1 \footnotesize
\textbf{To-do (\number\td):}\\~
\begin{minipage}{.35\textwidth}
#1
\end{minipage}
}\vskip 1ex}
\begin{document}
\title{
	Compiling quantamorphisms \\ for the IBM Q Experience
}
\author{ Ana~Neri, Rui~Soares~Barbosa, and~Jos\'e~N.~Oliveira%
\IEEEcompsocitemizethanks{
	\IEEEcompsocthanksitem A. Neri and J. N. Oliveira are with HASLab/INESC TEC, Universidade do Minho, Braga, Portugal.\protect\\
E-mail: \url{ana.neri@quantalab.uminho.pt}, \url{jno@di.uminho.pt}
	\IEEEcompsocthanksitem R. S. Barbosa is with INL -- International Iberian Nanotechnology Laboratory, Braga, Portugal. This work was done in part while RSB was with the Department of Computer Science, University of Oxford, United Kingdom.\protect\\
E-mail: \url{rui.soaresbarbosa@inl.int}}%
}

\IEEEcompsoctitleabstractindextext{%
\begin{abstract}
Based on the connection between the categorical derivation of classical programs
from specifications and the category-theoretic approach to quantum physics,
this paper contributes to extending the laws of classical program algebra to quantum
programming.
This aims at building \emph{correct-by-construction} quantum circuits to
be deployed on quantum devices such as those available at the IBM Q Experience.

Quantum circuit reversibility is ensured by minimal complements, extended
recursively.  Measurements are postponed to the end of such recursive computations,
termed \emph{``quantamorphisms"}, thus maximising the quantum effect.

Quantamorphisms are classical catamorphisms which, extended to ensure quantum
reversibility, implement quantum cycles (vulg.\ for-loops) and
quantum folds on lists. By Kleisli correspondence, quantamorphisms can be
written as monadic functional programs with quantum parameters. This enables
the use of Haskell, a monadic functional programming language, to perform
the experimental work.

Such calculated quantum programs prepared in Haskell are pushed through Quipper
to the Qiskit interface to IBM Q quantum devices. The generated quantum circuits -- often quite
large -- exhibit the predicted behaviour. However, running them
on real quantum devices incurs into a significant amount of errors. As quantum
devices are constantly evolving, an increase in reliability is likely in
the near future, allowing for our programs to run more accurately.
\end{abstract}

}

\maketitle

\newsavebox{\toffolicore}
\sbox{\toffolicore}{
\begin{circuitikz}
\draw
(0,1) node (and) [xshift=1cm,and port, scale=.8]   {}
(3,0.5) node (xor) [xor port, scale=.8]            {}
(and.in 1) node (A1)     [anchor=east,xshift=-1cm]  {\ensuremath{\Varid{a}}}
(and.in 2) node (B1)     [anchor=east,xshift=-1cm]  {\ensuremath{\Varid{b}}}
(xor.in 2) node (C1)     [anchor=east,xshift=-3cm]  {\ensuremath{\Varid{c}}}
(xor.out) node (Z1)      [anchor=east,xshift= 1.6cm]  {\ensuremath{\Varid{z}}};
\draw
	(and.out) |- (xor.in 1)
	(and.in 1) -- (A1)
	(and.in 2) -- (B1)
	(xor.in 2) -- (C1)
	(xor.out) -- (Z1);
\end{circuitikz}
}


\section{Introduction}

As is well known, there are highly complex problems that cannot be efficiently
solved by classical computers. On the other hand, classical system design is under pressure to decrease
the size of circuits as much as possible. In this context, \emph{quantum
technologies} appear as prime candidates to support a new computing era ---
the \emph{quantum computing} age \cite{DowlingMilburn2003,QuantumManifesto2016}.

This prospect is attracting both industry and academia, the former primarily
interested in understanding the potential advantages of quantum computing
for their business and the latter interested in pushing quantum science and
technology even further. Companies at the vanguard of quantum technology
(namely IBM, Google, and Microsoft) are already exploring it,
leading to new consortia between industry and academia. An example of this
is the IBM Quantum Network (IBM Q for short), which involves a number of
companies (e.g.\ Mitsubishi Chemical) as well as academic institutions (e.g.\
the University of Oxford) and is aimed primarily at sharing know-how.

The question arises: how much of the \emph{classical} way of programming can
evolve and contribute to quantum programming?
The current paper
proposes one such evolution, termed \emph{quantamorphism}, that enables
the construction of quantum (recursive) programs in a structured way.
To motivate this concept, it is worth looking back to the past and (briefly)
reviewing how similar strategies arose in classical programming. Indeed,
as history goes, there are striking similarities between the evolution of
quantum computing and that of its classical forerunner.

Classical computing is rooted on mathematical abstractions that led in particular
to the \emph{Turing machine} \cite{Tu36} -- which is still regarded
as the canonical abstract notion of a programmable computer -- and to the
$\lambda$-calculus \cite{Ch36} -- a mathematical system that provided the basis
for functional programming.

A step from abstraction to reality was made possible by advances in physics,
such as the invention of triodes (1912) and then of transistors (1948),
leading to the integrated circuits that are the basis of the \emph{in silico} technology of today \cite{Ne09,Ro04,La13}.

Once such devices were first employed to store information in realistic situations, it became clear that further abstraction was required. This led to the explicit adoption of formal logic, a very important abstraction still in use today. As the aphorism says, \emph{``logic is the language of computing''}.

Analogously to classical computing, but several decades later, quantum
computing was also born out of mathematical abstractions, this time
with the description of the first universal quantum computer by Deutsch \cite{De97}.
And the parallel goes on: nowadays, quantum physicists and engineers are testing strategies to implement such abstract concepts, linking theory to reality once again.

Soon ideas for quantum programming arose \cite{SZ00}, not only at the flowchart level
\cite{Se04,YF11} but also in the functional programming style \cite{AG05}.
And so, in a similar fashion to what happened for classical computation, \emph{software} started finding its way into quantum computation's history.

The birth of software as an independent technology took place in the 1950s.
But it soon was faced with a crisis because an effective discipline of programming
was lacking. The term \emph{software engineering} appeared in the 1960s and
was the subject of a conference supported by NATO that took place in Garmisch,
Germany in 1968. People at this conference expressed concerns and called
for theoretical foundations. This resulted in the birth of the principles
of \emph{structured programming} that became popular in the 1970s. But, in
a sense, the 1968 crisis is not over yet:
the problem with software engineering is that quality control is based on \emph{testing} software artifacts \emph{after} they have been built, and not on ensuring quality in a stepwise manner, as advocated by academia since the 1970s.

Some believe that the problem is lack of mathematical abstraction once again \cite{Kr07}. Stepping back to the original computational abstractions of the 1930s, the $\lambda$-calculus was developed with the aim of creating a model of computation based exclusively on \emph{function} abstraction and application. This led to a mathematically robust style of programming known as functional programming (FP), which has become a reliable paradigm for software production.
The \emph{correct-by-construction} programming techniques proposed in this field have had a significant impact on software theory. Such techniques promise a significant reduction in development costs by avoiding dependence on testing and debugging.

Correct-by-construction design techniques advocate the calculation of programs
from problem specifications. This is the main aim of the so-called ``Mathematics
of Program Construction'' (MPC) discipline \cite{Ba04a}, a branch of mathematics
applied to program calculation based on logic and relational algebra.

In the functional setting, such a discipline has led to the so-called \emph{``Algebra of Programming”} (AoP) which is the subject of textbook \cite{BM97}. The branch of mathematics that supports the AoP abstractions is category theory \cite{Ma71}.

Despite its strong algebraic basis -- cf.\ Hilbert spaces, linear algebra, etc. \cite{NC11} -- quantum mechanics is still a counter-intuitive theory and one that will require further abstractions for programmers. In quantum mechanics, every observation implies the destruction of superposed states, spoiling the quantum advantage altogether. This renders current step-by-step \emph{debugging} strategies obsolete and nearly impossible: one needs to \emph{get it right} from the very beginning!

\section{Research questions}
Because testing and debugging cannot apply to quantum programming, at least
in current standards, the traditional life-cycle based on
\emph{edit-compile-run} is not an option. This further increases the need
for correct-by-construction methods, leading us into the main research questions
addressed by the current paper:
\begin{enumerate}
\item
Is it possible to extend the MPC culture, principles and constructions --
which have been so effective in disciplining the whole field of (classical)
recursive functional programming and data structuring -- to quantum
programming?
\item
Is it viable to apply such constructions to derive programs down to the level
of actually \emph{running} them on the experimental quantum devices of today?
\end{enumerate}

An important requirement to take into account when scaling classical paradigms
to the quantum level is \emph{reversibility}, because quantum programs are limited
to \emph{unitary transformations} \cite{YM08}, which are special
cases of reversible operations. Therefore, this research largely intersects with that
on \emph{reversible} computing.

A similar extension of the MPC paradigm to \emph{probabilistic programming}
has been shown to be viable in practice \cite{MuO15}, although in a very different
context: that of reasoning about program reliability in the presence of faulty
hardware. The laws of that approach require \emph{typed} linear algebra
rather than just the algebra of \emph{relations}\footnote{This has been referred
to by the acronym LAoP (``linear algebra of programming'') \cite{Ol12}.}
in order to reason about probabilistic functions (Markov chains). On the experimental
side, this requires programming over the \emph{distribution monad}.\footnote{As implemented
by \cite{EK06} in Haskell.}

The recursive programming construction studied in \cite{MuO15} is the
so-called (probabilistic) \emph{catamorphism} \cite{BM97}. The idea
in the current paper is to generalise from such catamorphisms to
unitary transformations over a vector space monad implementing finite-dimensional
Hilbert spaces \cite{Co11}. The corresponding extension of the catamorphism
concept, to be developed further in this paper, is termed ``quantamorphism"
-- a restricted form of \emph{recursive} quantum control of \emph{quantum}
data \cite{Ol18}.

A half-way concept between classical functions and unitary transformations
is that of a reversible function, also known as \emph{isomorphism} or \emph{bijection}.
The paper will contribute to the current investment in reversible computing
by extending a technique known as \emph{complementation} \cite{MHNHT07} to
recursive programs. The background of all this research is also enhanced
by studies in quantum functional programming (QFP) \cite{AG05,GLRSV13} and
extensive research in categorical quantum physics \cite{Co11,HH16,DBLP:conf/lics/AbramskyBKM19}.


\section{Algebra of Programming}
The standard algebra of programming \cite{BM97} is an evolution of the binary
relation algebra pioneered by Augustus de Morgan (1806--71).
Later, Peirce (1839--1914) invented \emph{quantifier} notation to explain de Morgan's
algebra of relations.\footnote{See e.g.\ \cite{Mad91} for a comprehensive
historical overview.} De Morgan's pioneering work was ill-fated: the language
invented to explain his calculus of relations became eventually more popular
than the calculus itself -- it is nowadays known as first-order logic (FOL).

Alfred Tarski (1901--83), who had a life-long struggle with quantifier notation,
revived relation algebra. Together with Steve Givant he wrote a book (published
posthumously) on \emph{set theory without variables} \cite{TG87}.

Meanwhile, category theory \cite{Ma71} was born, stressing the
description of mathematical concepts in terms of abstract \emph{arrows}
(morphisms) and \emph{diagrams}, unveiling a compositional, abstract language
of universal \emph{combinators} that is inherently \emph{generic} and \emph{pointfree}.

The category of sets and functions immediately provided a basis for pointfree
functional reasoning, but this was by and large ignored by John Backus (1924--2007)
in his FP algebra of programs \cite{Ba78}. In any case, Backus's landmark FP paper
was the first to show how relevant this reasoning style is to programming.
This happened four decades ago.

A bridge between the two pointfree schools -- the relational and the categorical --
was eventually established by Freyd and Scedrov \cite{FS90} in their proposal
of the concept of an \emph{allegory}, which instantiates to \emph{typed} relation algebra.
The pointfree algebra of programming (AoP) as it is understood today \cite{BM97} stems
directly from \cite{FS90}.

\section{Reversibility}
Standard program design relies on program \emph{refinement} techniques \cite{Mor90,OF13}.
A program (or specification, or model) is refined wherever it leads to a
more \emph{defined} version of it, in a double sense: more
deterministic and more responsive. The limit of a refinement process is
always a \emph{function}: a totally defined and fully deterministic computational process.

Quantum programming brings with it a new concern in programming, that of
\emph{reversibility}. This concern is relatively new in the traditional algebra
of programming. In fact, classical \emph{program} design, e.g.\ by source-to-source
\emph{transformation}, primarily seeks time and space \emph{efficiency}
but not \emph{reversibility}.

Reversible computations are \emph{functions} that are injective and surjective
-- that is, \emph{bijective}. Recall that a function \ensuremath{\Varid{f}\mathbin{:}\Conid{X}\to \Conid{Y}} is injective iff
\begin{eqnarray*}
	\ensuremath{\Varid{f}\;\Varid{x}\mathrel{=}\Varid{f}\;\Varid{x'}\Rightarrow \Varid{x}\mathrel{=}\Varid{x'}}
\end{eqnarray*}
holds for every \ensuremath{\Varid{x},\Varid{x'}\;{\in}\;\Conid{X}}, and \emph{surjective} iff, for all \ensuremath{\Varid{y}\;{\in}\;\Conid{Y}},
there exists some \ensuremath{\Varid{x}} such that \ensuremath{\Varid{y}\mathrel{=}\Varid{f}\;\Varid{x}}.

Refinement is normally expressed in terms of a  \emph{preorder} \ensuremath{\Varid{p}\leq \Varid{q}} meaning
that program \ensuremath{\Varid{q}} is more refined than program \ensuremath{\Varid{p}}, that is, \ensuremath{\Varid{q}} is
closer to an implementation of \ensuremath{\Varid{p}}.\footnote{This ensures
that a program always refines itself (reflexivity) and that a refinement
of a refinement is also a refinement (transitivity).} Refinement towards reversibility
calls for an \emph{injectivity} pre-{order} -- one that will enable us to order the functions in the picture
below in the way shown:
\begin{eqnarray*}
	\includegraphics[height=0.09\textheight]{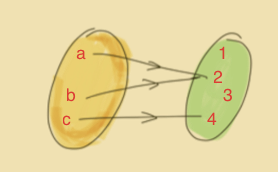}
 \makebox(15,65){\ensuremath{\leq }}
	\includegraphics[height=0.09\textheight]{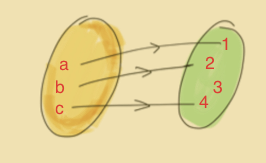}
\end{eqnarray*}
The intuition is that injective functions \emph{discriminate} more: in the
picture, \ensuremath{\Varid{a}} and \ensuremath{\Varid{b}} are mapped onto the same output (number \ensuremath{\mathrm{2}}) by the
less injective function on the left-hand side, while on the right-hand side
the fact that \ensuremath{\Varid{a}} and \ensuremath{\Varid{b}} are different is preserved at image level (\ensuremath{\mathrm{1}\not=\mathrm{2}}).

In general, \ensuremath{\Varid{g}} is said to be \emph{less injective} than \ensuremath{\Varid{f}} if
\begin{eqnarray*}
	\ensuremath{\Varid{g}\leq \Varid{f}} & \equiv & \ensuremath{\Varid{f}\;\Varid{x}\mathrel{=}\Varid{f}\;\Varid{x'}\Rightarrow \Varid{g}\;\Varid{x}\mathrel{=}\Varid{g}\;\Varid{x'}}
\end{eqnarray*}
holds.

A problem with the definitions just given, which are standard
in mathematics, is that they are declarative but not calculational. Moreover,
there are simpler ways of saying the same things. For instance, function \ensuremath{\Varid{g}} in
\begin{quote}
	\ensuremath{\Varid{f} \comp \Varid{g}\mathrel{=}\Varid{id}}
\end{quote}
is \emph{injective} because it has a left inverse \ensuremath{\Varid{f}} (which is \emph{surjective}).
An equivalent way of writing \ensuremath{\Varid{f} \comp \Varid{g}\mathrel{=}\Varid{id}} is
\begin{quote}
\ensuremath{\Varid{g}\; \subseteq \;\conv{\Varid{f}}}
\end{quote}
using the relation algebra \emph{converse} operator: \ensuremath{\Varid{b}\mathrel{=}\Varid{f}\;\Varid{a}} means the same as
\ensuremath{\Varid{a}\;\conv{\Varid{f}}\;\Varid{b}}.
This says that \ensuremath{\Varid{g}} injective because \ensuremath{\Varid{g}} is smaller than the \emph{converse} of a
function (\ensuremath{\conv{\Varid{f}}}), and function converses are always injective. In case \ensuremath{\Varid{g}\mathrel{=}\conv{\Varid{f}}} then \emph{both} \ensuremath{\Varid{f}} and \ensuremath{\Varid{g}} will
be injective and surjective, i.e.\ bijections.

Instead of relying directly on first-order logic, this style of argumentation
relies on \emph{relation algebra} \cite{BM97}, as detailed next.

\section{Functions and relations}
In the same way that we declare a function \ensuremath{\Varid{f}\mathbin{:}\Conid{A}\to \Conid{B}} by specifying its input type
\ensuremath{\Conid{A}} and output type \ensuremath{\Conid{B}}, and write \ensuremath{\Varid{f}\mathbin{:}\Conid{B}\leftarrow \Conid{A}} to mean exactly the same thing,
so we write \ensuremath{\Conid{R}\mathbin{:}\Conid{B}\leftarrow \Conid{A}} or \ensuremath{\Conid{R}\mathbin{:}\Conid{A}\to \Conid{B}}, or even \ensuremath{\rarrow{\Conid{A}}{\Conid{R}}{\Conid{B}}} or \ensuremath{\larrow{\Conid{A}}{\Conid{R}}{\Conid{B}}},
to declare the type of a relation \ensuremath{\Conid{R}}.  Moreover, we use infix notation \ensuremath{\Varid{b}\;\Conid{R}\;\Varid{a}} to denote \ensuremath{(\Varid{b},\Varid{a})\;{\in}\;\Conid{R}}, in the tradition of \ensuremath{\Varid{b}\leq \Varid{a}}, \ensuremath{\Varid{b}\;{\in}\;\Varid{s}}, and
so on.

Functions are special cases of relations. We use lowercase letters (e.g.\
\ensuremath{\Varid{f}}, \ensuremath{\Varid{g}}, ...) to denote functions and uppercase letters (e.g.\ \ensuremath{\Conid{R}}, \ensuremath{\Conid{S}},
...) to denote relations. The singularity of functions as relations is captured
by \ensuremath{\Varid{b}\;\Varid{f}\;\Varid{a}\Leftrightarrow\Varid{b}\mathrel{=}\Varid{f}\;(\Varid{a})}.

Given relations \ensuremath{\larrow{\Conid{C}}{\Conid{R}}{\Conid{B}}} and \ensuremath{\larrow{\Conid{A}}{\Conid{S}}{\Conid{C}}}, their composition \ensuremath{\larrow{\Conid{A}}{\Conid{R} \comp \Conid{S}}{\Conid{C}}} is defined by \ensuremath{\Varid{b}\;(\Conid{R} \comp \Conid{S})\;\Varid{a}} iff, for some \ensuremath{\Varid{c}\;{\in}\;\Conid{C}}, \ensuremath{\Varid{b}\;\Conid{R}\;\Varid{c}} and
\ensuremath{\Varid{c}\;\Conid{S}\;\Varid{a}} hold. In the case of functions, this specialises to the familiar composition of functions:
\ensuremath{\Varid{b}\;(\Varid{f} \comp \Varid{g})\;\Varid{a}} means \ensuremath{\Varid{b}\mathrel{=}\Varid{f}\;(\Varid{g}\;\Varid{a})}. The
unit of composition is the identity function \ensuremath{\Varid{id}\;\Varid{x}\mathrel{=}\Varid{x}}, that is, \ensuremath{\Conid{R} \comp \Varid{id}\mathrel{=}\Conid{R}\mathrel{=}\Varid{id} \comp \Conid{R}}.

Relations of the same type are ordered by entailment, i.e.\ inclusion. This is denoted by
\ensuremath{\Conid{R}\; \subseteq \;\Conid{S}} meaning that \ensuremath{\Varid{b}\;\Conid{R}\;\Varid{a}} logically implies \ensuremath{\Varid{b}\;\Conid{S}\;\Varid{a}} for all \ensuremath{\Varid{b},\Varid{a}}.

The converse \ensuremath{\larrow{\Conid{B}}{\conv{\Conid{R}}}{\Conid{A}}} of a relation \ensuremath{\rarrow{\Conid{A}}{\Conid{R}}{\Conid{B}}} is such that \ensuremath{\Varid{a}\;\conv{\Conid{R}}\;\Varid{b}} means the same as \ensuremath{\Varid{b}\;\Conid{R}\;\Varid{a}}.
In the case of a function \ensuremath{\Varid{f}\mathbin{:}\Conid{A}\to \Conid{B}}, its \emph{converse} is the relation \ensuremath{\conv{\Varid{f}}\mathbin{:}\Conid{A}\leftarrow \Conid{B}} such that \ensuremath{\Varid{a}\;\conv{\Varid{f}}\;\Varid{b}\Leftrightarrow\Varid{b}\mathrel{=}\Varid{f}\;\Varid{a}}.

\begin{figure}
\footnotesize\centering
\(
\xymatrix@R=1ex@C=-4.0ex{
&&&
	\mbox{binary relation}
		\ar@{-}[drrr]
		\ar@{-}[dr]
		\ar@{-}[dl]
		\ar@{-}[dlll]
\\
	\mbox{injective}
		\ar@{-}[dr]
&
&
	\mbox{entire}
		\ar@{-}[dl]
		\ar@{-}[dr]
&
&
	\mbox{simple}
		\ar@{-}[dr]
		\ar@{-}[dl]
&
&
	\mbox{surjective}
		\ar@{-}[dl]
\\
&
	\mbox{representation}
		\ar@{-}[dr]
&
&
	\mbox{\emph{function}}
		\ar@{-}[dr]
		\ar@{-}[dl]
&
&
	\mbox{abstraction}
		\ar@{-}[dl]
\\
&
&
	\mbox{injection}
		\ar@{-}[dr]
&
&
	\mbox{surjection}
		\ar@{-}[dl]
\\
&&&
	\mbox{\bf bijection 
	}
}
	\label{eq:fig:031230a}
\)
\caption{\footnotesize Binary relation taxonomy (relation `bestiary').\label{fig:190223a}}
\end{figure}
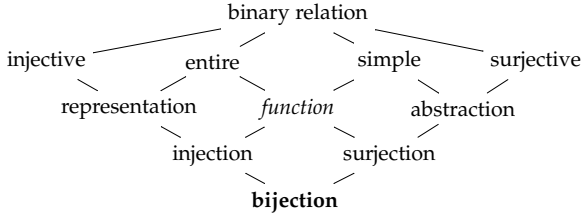

\paragraph{A taxonomy of relations}
A relation \ensuremath{\Conid{R}\mathbin{:}\Conid{A}\to \Conid{B}} is said to be \emph{injective} whenever
	\ensuremath{\conv{\Conid{R}} \comp \Conid{R}\; \subseteq \;\Varid{id}}
holds.\footnote{For functions (\ensuremath{\Conid{R}\mathbin{:=}\Varid{f}}), \ensuremath{\conv{\Varid{f}} \comp \Varid{f}\; \subseteq \;\Varid{id}} means precisely
what we had before, \ensuremath{\Varid{f}\;\Varid{x}\mathrel{=}\Varid{f}\;\Varid{x'}\Rightarrow \Varid{x}\mathrel{=}\Varid{x'}}.} Moreover:
\begin{eqnarray*}
\start
	\mbox{\ensuremath{\Conid{R}} \emph{injective}} \ensuremath{\Leftrightarrow} \ensuremath{\underbrace{\conv{\Conid{R}} \comp \Conid{R}}_{\ker{\Conid{R}}}\; \subseteq \;\Varid{id}}
\more
	\mbox{\ensuremath{\Conid{R}} \emph{simple}} \ensuremath{\Leftrightarrow} \mbox{\ensuremath{\conv{\Conid{R}}} \emph{injective}}
\more
 	\mbox{\ensuremath{\Conid{R}} \emph{entire}} \ensuremath{\Leftrightarrow} \ensuremath{\Varid{id}\; \subseteq \;\underbrace{\Conid{R} \comp \conv{\Conid{R}}}_{\img{\Conid{R}}}}
\more
 	\mbox{\ensuremath{\Conid{R}} \emph{surjective}} \ensuremath{\Leftrightarrow} \mbox{\ensuremath{\conv{\Conid{R}}} \emph{entire}}
\end{eqnarray*}
leading to the taxonomy of Figure \ref{fig:190223a}. Below we shall use the
definition of \emph{kernel} of a relation:
\begin{eqnarray*}
	\ensuremath{\ker{\Conid{R}}} \deff \ensuremath{\conv{\Conid{R}} \comp \Conid{R}}
\end{eqnarray*}
In the case of functions, \ensuremath{\Varid{a'}\;(\ker{\Varid{f}})\;\Varid{a}} means \ensuremath{\Varid{f}\;\Varid{a'}\mathrel{=}\Varid{f}\;\Varid{a}}, that is, \ensuremath{\Varid{a'}}
and \ensuremath{\Varid{a}} have the same image under \ensuremath{\Varid{f}}. For any \ensuremath{\Varid{f}}, \ensuremath{\ker{\Varid{f}}} is always an \emph{equivalence}
relation \cite{Ol17}. If \ensuremath{\Varid{f}} is injective, this equivalence is the identity.
Moreover,
\begin{eqnarray}
\mbox{
	\ensuremath{\Varid{f}} is bijective
} & \equiv &
\mbox{
	\ensuremath{\ker{\Varid{f}}\mathrel{=}\Varid{id}\mathrel{\wedge}\img{\Varid{f}}\mathrel{=}\Varid{id}}
}
\end{eqnarray}
that is, \ensuremath{\conv{\Varid{f}} \comp \Varid{f}\mathrel{=}\Varid{id}} and \ensuremath{\Varid{f} \comp \conv{\Varid{f}}\mathrel{=}\Varid{id}}.

\paragraph{Relations as matrices}
Following \cite{Sc10}, it is helpful to depict relations using (Boolean) \emph{matrices},
for instance Boolean \emph{negation}
\begin{eqnarray}
\ensuremath{\larrow{\B}{\neg }{\B}} =
\begin{array}{c|cc}
	& 0	& 1	\\\hline
0	& 0	& 1	\\
1	& 1	& 0
\end{array}
	\label{eq:not}
\end{eqnarray}
(a {\em bijection}, also known as the \ensuremath{\Conid{X}}-gate), \emph{exclusive-or}
\begin{eqnarray}
\ensuremath{\larrow{\B\times\B}{\xsor }{\B}} =
\begin{array}{c|cccc}
	& 0	& 0	& 1	& 1 \\
	& 0	& 1	& 0	& 1 \\\hline
0	& 1	& 0	& 0	& 1 \\
1	& 0	& 1	& 1	& 0 \\
\end{array}
\end{eqnarray}
(surjective but not injective), and so on. Clearly:
\begin{itemize}
\item	
\emph{Function} matrices have \emph{exactly one} \ensuremath{\mathrm{1}} in every column.
\item	
\emph{Bijections} are square matrices with exactly one \ensuremath{\mathrm{1}} in every \emph{column} and in every \emph{row}.
\end{itemize}

\section{Injectivity preorder}

The injectivity preorder \cite{Ol14a} is defined by
\begin{eqnarray*}
	\ensuremath{\Conid{R}\leq \Conid{S}} & \equiv & \ensuremath{\ker{\Conid{S}}\; \subseteq \;\ker{\Conid{R}}}
\end{eqnarray*}
As an example, take two list functions: \ensuremath{\Varid{elems}} computing the set of all
items of a list and \ensuremath{\Varid{bagify}} keeping the bag of such elements. The former
loses more information (order and multiplicity) than the latter (which 
forgets the order only). Thus \ensuremath{\Varid{elems}\leq \Varid{bagify}}.

Below, we shall explore this preorder as a \emph{refinement
ordering} guiding us towards more and more \emph{injective} computations,
heading to \emph{reversibility}.

This injectivity preorder is rich in properties.
For instance, it is upper-bounded\footnote{See e.g.\ \cite{Ol14a} for more details.}
\begin{eqnarray}
	\ensuremath{{\Conid{R}}\kr{\Conid{S}}} \leq X & \equiv &
	R \leq X \land S \leq X
	\label{eq:050112a}
\end{eqnarray}
by relation \emph{pairing}, which is defined in the expected way
\begin{eqnarray}
	\ensuremath{(\Varid{b},\Varid{c})\;({\Conid{R}}\kr{\Conid{S}})\;\Varid{a}} &\ensuremath{\Leftrightarrow}& \ensuremath{\Varid{b}\;\Conid{R}\;\Varid{a}\mathrel{\wedge}\Varid{c}\;\Conid{S}\;\Varid{a}}
\end{eqnarray}
specialising, in the case of functions, to
\begin{eqnarray}
	\ensuremath{({\Varid{f}}\kr{\Varid{g}})\;\Varid{a}\mathrel{=}(\Varid{f}\;\Varid{a},\Varid{g}\;\Varid{a})}.
\end{eqnarray}
Cancellation over \eqref{eq:050112a} means that
\emph{pairing} always \emph{increases injectivity}:
\begin{eqnarray}
	\ensuremath{\Conid{R}\leq {\Conid{R}}\kr{\Conid{S}}} \mbox{ ~ and ~ } \ensuremath{\Conid{S}\leq {\Conid{R}}\kr{\Conid{S}}}.
	\label{eq:180120a}
\end{eqnarray}
Facts \eqref{eq:180120a} are jointly equivalent to
\ensuremath{\ker{({\Conid{R}}\kr{\Conid{S}})}\; \subseteq \;{(\ker{\Conid{R}})}\mathbin\cap{(\ker{\Conid{S}})}},
which in fact is an equality:
\begin{eqnarray}
	\ensuremath{\ker{({\Conid{R}}\kr{\Conid{S}})}\mathrel{=}{(\ker{\Conid{R}})}\mathbin\cap{(\ker{\Conid{S}})}} .
	\label{eq:030218a}
\end{eqnarray}
This is a corollary of the following more general law:\footnote{Details in  \cite{BM97}.}
\begin{eqnarray}
	\ensuremath{\conv{({\Conid{R}}\kr{\Conid{S}})} \comp ({\Conid{Q}}\kr{\Conid{P}})\mathrel{=}{(\conv{\Conid{R}} \comp \Conid{Q})}\mathbin\cap{(\conv{\Conid{S}} \comp \Conid{P})}} .
\end{eqnarray}

Injectivity \emph{shunting laws} also arise by standard relational algebra
calculation \cite{Ol14a}, for instance:
\begin{eqnarray*}
	\xarrayin{ \ensuremath{\Conid{R} \comp \Varid{g}\leq \Conid{S}} & \equiv & \ensuremath{\Conid{R}\leq \Conid{S} \comp \conv{\Varid{g}}} }
\end{eqnarray*}

Restricted to \emph{functions}, the preorder \ensuremath{(\leq )} is \emph{universally} bounded by
\begin{quote}
	\ensuremath{\mathop{!}\leq \Varid{f}\leq \Varid{id}}
\end{quote}
where \ensuremath{\larrow{\Conid{A}}{\mathop{!}}{\mathrm{1}}} is the unique function of its type, where symbol
\ensuremath{\mathrm{1}} denotes the singleton type. Moreover,
\begin{itemize}
\item	
A function is \emph{injective} iff
\(
	\ensuremath{\Varid{id}\leq \Varid{f}}
\) holds, that is, \ensuremath{\ker{\Varid{f}}\mathrel{=}\Varid{id}}.
Consequently, \ensuremath{{\Varid{f}}\kr{\Varid{id}}} is always \emph{injective}, by \eqref{eq:180120a}.
\item	
Two functions \ensuremath{\Varid{f}} and \ensuremath{\Varid{g}} are said to be \emph{complementary} whenever
\ensuremath{\Varid{id}\leq {\Varid{f}}\kr{\Varid{g}}}.\footnote{Cf. \cite{MHNHT07}. Other terminologies are
\emph{monic pair} \cite{FS90} or \emph{jointly monic} \cite{BM97}.\\~
}
\end{itemize}
For instance, the \emph{projections}
\ensuremath{\Varid{fst}\;(\Varid{a},\Varid{b})\mathrel{=}\Varid{a}} and
\ensuremath{\Varid{snd}\;(\Varid{a},\Varid{b})\mathrel{=}\Varid{b}}
are complementary since \ensuremath{{\Varid{fst}}\kr{\Varid{snd}}\mathrel{=}\Varid{id}}.

\section{Minimal complements} \label{sec:191026ce}
Given some \ensuremath{\Varid{f}}, suppose that:
(a) \ensuremath{\Varid{id}\leq {\Varid{f}}\kr{\Varid{g}}} for some \ensuremath{\Varid{g}};
(b) if \ensuremath{\Varid{id}\leq {\Varid{f}}\kr{\Varid{h}}} for some \ensuremath{\Varid{h}} such that \ensuremath{\Varid{h}\leq \Varid{g}}, then \ensuremath{\Varid{g}\leq \Varid{h}} holds.
Then \ensuremath{\Varid{g}} is said to be a \emph{minimal complement} of \ensuremath{\Varid{f}}
\cite{BS81}. Minimal complements (not unique in general) capture \emph{``what
is missing"} from the original function for \emph{injectivity} to hold.

Calculating a minimal complement \ensuremath{\Varid{g}} for a given function \ensuremath{\Varid{f}} can be regarded
as a \emph{correct-by-construction} strategy for implementing \ensuremath{\Varid{f}} in a
reversible way, in the sense that: (a) \ensuremath{{\Varid{f}}\kr{\Varid{g}}} is injective even if \ensuremath{\Varid{f}} is not,
and (b) \ensuremath{\Varid{f}} is implemented by \ensuremath{{\Varid{f}}\kr{\Varid{g}}}, since \ensuremath{\Varid{f}\mathrel{=}\Varid{fst} \comp ({\Varid{f}}\kr{\Varid{g}})}.

For Boolean functions, minimal complements are easy to calculate using matrices.
In the following example we wish to calculate a minimal complement for the
(non-injective) exclusive-or Boolean operator:
\begin{eqnarray}
\ensuremath{\larrow{\B\times\B}{\xsor }{\B}}=
\begin{bmatrix}
\cell{1}&\cell{0}&\cell{0}&\cell{1}\\
\cell{0}&\cell{1}&\cell{1}&\cell{0}
\end{bmatrix}
	\label{eq:191026c}
\end{eqnarray}
We start from the kernel of \ensuremath{\xsor }: 
\begin{eqnarray*}
	\ensuremath{\ker{\xsor }} = \ker{ \begin{bmatrix}
\cell{1}&\cell{0}&\cell{0}&\cell{1}\\
\cell{0}&\cell{1}&\cell{1}&\cell{0}
\end{bmatrix}} =
\begin{bmatrix}
\cell{1}&\cell{0}&\cell{0}&\cell{1}\\
\cell{0}&\cell{1}&\cell{1}&\cell{0}\\
\cell{0}&\cell{1}&\cell{1}&\cell{0}\\
\cell{1}&\cell{0}&\cell{0}&\cell{1}
\end{bmatrix}
\end{eqnarray*}
By \eqref{eq:030218a}, the kernel of a complement \ensuremath{\Varid{g}} has to cancel all \ensuremath{\mathrm{1}}s
in \ensuremath{\ker{\xsor }} that fall outside the diagonal \ensuremath{\Varid{id}}. The identity function
itself would do this,
\begin{eqnarray*}
	\ensuremath{\ker{\Varid{id}}} =
\begin{bmatrix}
\cell{1}&\cell{0}&\cell{0}&\cell{0}\\
\cell{0}&\cell{1}&\cell{0}&\cell{0}\\
\cell{0}&\cell{0}&\cell{1}&\cell{0}\\
\cell{0}&\cell{0}&\cell{0}&\cell{1}
\end{bmatrix}
\end{eqnarray*}
but this would be an overkill -- \ensuremath{\Varid{id}} complements any function! Moreover, it
is not minimal in this case. To reduce injectivity we need to start adding
\ensuremath{\mathrm{1}}s to \ensuremath{\ker{\Varid{id}}} where \ensuremath{\ker{\xsor }} has \ensuremath{\mathrm{0}}s, e.g.
\begin{eqnarray}
\begin{bmatrix}
\cell{1}&\cell{1}&\cell{1}&\cell{0}\\
\cell{1}&\cell{1}&\cell{0}&\cell{1}\\
\cell{1}&\cell{0}&\cell{1}&\cell{1}\\
\cell{0}&\cell{1}&\cell{1}&\cell{1}
\end{bmatrix}
	\label{eq:190223b}
\end{eqnarray}
However, this isn't a kernel anymore. Why? Because it is not an equivalence
relation: it is \emph{reflexive} (cf.\ diagonal) and \emph{symmetric}, but not \emph{transitive}.

To handle transitivity we resort to a basic result in relation algebra:
\emph{a symmetric and reflexive relation is an \emph{equivalence} iff it is
a difunctional relation}, where
\begin{quote}
a relation \ensuremath{\Conid{R}} is \emph{difunctional} iff \ensuremath{\Conid{R} \comp \conv{\Conid{R}} \comp \Conid{R}\; \subseteq \;\Conid{R}} \cite{Ol17}.
\end{quote}
One can construct finite \emph{difunctional} relations easily, by inspection:
just make sure that
columns either do not intersect or are the same. Clearly, \eqref{eq:190223b} is not
difunctional.

To make it difunctional, we have to surgically cancel zeros symmetrically,
outside the diagonal:
\begin{eqnarray*}
\begin{bmatrix}
\cell{1}&\cell{1}&\cell{\red 1}&\cell{0}\\
\cell{1}&\cell{1}&\cell{0}&\cell{1}\\
\cell{\red 1}&\cell{0}&\cell{1}&\cell{1}\\
\cell{0}&\cell{1}&\cell{1}&\cell{1}
\end{bmatrix}
	\ensuremath{\to }
\begin{bmatrix}
\cell{1}&\cell{1}&\cell{0}&\cell{0}\\
\cell{1}&\cell{1}&\cell{0}&\cell{\red1}\\
\cell{0}&\cell{0}&\cell{1}&\cell{1}\\
\cell{0}&\cell{\red1}&\cell{1}&\cell{1}
\end{bmatrix}
	\ensuremath{\to }
\begin{bmatrix}
\cell{1}&\cell{1}&\cell{0}&\cell{0}\\
\cell{1}&\cell{1}&\cell{0}&\cell{0}\\
\cell{0}&\cell{0}&\cell{1}&\cell{1}\\
\cell{0}&\cell{0}&\cell{1}&\cell{1}
\end{bmatrix}
\end{eqnarray*}
What we obtain is \ensuremath{\ker{\Varid{fst}}}, the kernel of the first projection
$\ensuremath{\larrow{\B\times\B}{\Varid{fst}}{\B}}=
\begin{bmatrix}
\cell{1}&\cell{1}&\cell{0}&\cell{0}\\
\cell{0}&\cell{0}&\cell{1}&\cell{1}
\end{bmatrix}$.
So, function \ensuremath{\Varid{fst}\;(\Varid{a},\Varid{b})\mathrel{=}\Varid{a}} is a minimal complement of \ensuremath{\xsor }.

We said that minimal complements are not unique in general and this is
one such case. Indeed, we might have decided to perform alternative cancellations, e.g.
\begin{eqnarray*}
\begin{bmatrix}
\cell{1}&\cell{\red1}&\cell{1}&\cell{0}\\
\cell{\red1}&\cell{1}&\cell{0}&\cell{1}\\
\cell{1}&\cell{0}&\cell{1}&\cell{1}\\
\cell{0}&\cell{1}&\cell{1}&\cell{1}
\end{bmatrix}
	\ensuremath{\to }
\begin{bmatrix}
\cell{1}&\cell{0}&\cell{1}&\cell{0}\\
\cell{0}&\cell{1}&\cell{0}&\cell{1}\\
\cell{1}&\cell{0}&\cell{1}&\cell{\red1}\\
\cell{0}&\cell{1}&\cell{\red1}&\cell{1}
\end{bmatrix}
	\ensuremath{\to }
\begin{bmatrix}
\cell{1}&\cell{0}&\cell{1}&\cell{0}\\
\cell{0}&\cell{1}&\cell{0}&\cell{1}\\
\cell{1}&\cell{0}&\cell{1}&\cell{0}\\
\cell{0}&\cell{1}&\cell{0}&\cell{1}
\end{bmatrix}
\end{eqnarray*}
ending up this time with \ensuremath{\ker{\Varid{snd}}}, the kernel of the other projection \ensuremath{\Varid{snd}\;(\Varid{a},\Varid{b})\mathrel{=}\Varid{b}}.
So, both \ensuremath{\Varid{fst}} and \ensuremath{\Varid{snd}} are \emph{minimal complements} of \ensuremath{\xsor }.

Let us see what comes out of the \ensuremath{\Varid{fst}}-complementation of exclusive-or:
\begin{eqnarray}
	\ensuremath{\larrow{\B\times\B}{{\Varid{fst}}\kr{\xsor }}{\B\times\B}} =
\begin{bmatrix}
1	&0	&0	&0\\
0	&1	&0	&0\\
0	&0	&0	&1\\
0	&0	&1	&0
\end{bmatrix}
	\label{eq:191029d}
\end{eqnarray}
This is a well-known \emph{bijection},
in fact a familiar gate known as \textsc{cx}, or
\textsc{cnot} (for \emph{"controlled not"}), usually depicted as follows:
\begin{center}
	\cnot
\end{center}
Why does it bear this name? We calculate:
\begin{eqnarray}
\start
	\ensuremath{\Varid{cnot}\mathrel{=}{\Varid{fst}}\kr{\xsor }}
	\nonumber
\just\equiv{ go pointwise }
	\ensuremath{\Varid{cnot}\;(\Varid{a},\Varid{b})\mathrel{=}(\Varid{a},\Varid{a}\mathbin{\xsor} \Varid{b}))}
	\nonumber
\just\equiv{ since \ensuremath{\mathrm{0}\mathbin{\xsor} \Varid{b}\mathrel{=}\Varid{b}} and \ensuremath{\mathrm{1}\mathbin{\xsor} \Varid{b}\mathrel{=}\neg \;\Varid{b}} }
	\ensuremath{\begin{lcbr}\Varid{cnot}\;(\mathrm{0},\Varid{b})\mathrel{=}(\mathrm{0},\Varid{b})\\\Varid{cnot}\;(\mathrm{1},\Varid{b})\mathrel{=}(\mathrm{1},\neg \;\Varid{b})\end{lcbr}}
	\label{eq:191029c}
\end{eqnarray}
Informally: \emph{controlled} bit \ensuremath{\Varid{b}} is negated \emph{iff} the \emph{control} bit \ensuremath{\Varid{a}} is set; otherwise, the gate does nothing.

Thus we have designed this gate following a \emph{constructive} approach
-- we built it by \emph{minimal} complementation.
Note the role of the \ensuremath{\Varid{fst}} complement in copying the control bit to the output.

\paragraph{Other \ensuremath{\Varid{fst}}-complementations}
As a second example, we take the classical circuit
\begin{center}
\resizebox{.3\textwidth}{!}{\usebox{\toffolicore}}
\end{center}
Can this be made into a bijection in the same way?
The function implemented is \footnotesize
\begin{eqnarray*}
	\ensuremath{\longrarrow{{\B}^{\mathrm{2}}\times\B}{\Varid{f}\mathrel{=}\xsor  \comp ((\mathrel{\wedge})\times\Varid{id})}{\B}} &=&
\begin{bmatrix}
1	& 0	& 1	& 0	& 1	& 0	& 0	& 1 \\
0	& 1	& 0	& 1	& 0	& 1	& 1	& 0
\end{bmatrix}
\end{eqnarray*} \normalsize
where
\begin{eqnarray}
	\ensuremath{(\Varid{f}\times\Varid{g})\;(\Varid{a},\Varid{b})\mathrel{=}(\Varid{f}\;\Varid{a},\Varid{f}\;\Varid{b})}
	\label{eq:960923c-pw}
\end{eqnarray}
is the ``tensor" product of two functions.
Let us {complement} \ensuremath{\Varid{f}} with \ensuremath{\Varid{fst}} again, which in this context has type
\ensuremath{\rarrow{{\B}^{\mathrm{2}}\times\B}{\Varid{fst}}{{\B}^{\mathrm{2}}}}.
The outcome is another \emph{bijection}, known as the \textsc{ccnot},
or Toffoli, gate \footnotesize
\begin{eqnarray*}
\ensuremath{\Varid{ccnot}\mathrel{=}{\Varid{fst}}\kr{(\xsor  \comp ((\mathrel{\wedge})\times\Varid{id}))}} =
\begin{bmatrix}
1	&0	&0	&0	&0	&0	&0	&0\\
0	&1	&0	&0	&0	&0	&0	&0\\
0	&0	&1	&0	&0	&0	&0	&0\\
0	&0	&0	&1	&0	&0	&0	&0\\
0	&0	&0	&0	&1	&0	&0	&0\\
0	&0	&0	&0	&0	&1	&0	&0\\
0	&0	&0	&0	&0	&0	&0	&1\\
0	&0	&0	&0	&0	&0	&1	&0
\end{bmatrix}
\end{eqnarray*} \normalsize
usually depicted as follows:
\begin{center}
	\toffoli
\end{center}
As for \ensuremath{\Varid{cnot}}, a similar calculation will lead to the pointwise version:
\begin{quote}
\ensuremath{\Varid{ccnot}\mathbin{:}{\B}^{\mathrm{2}}\times\B\to {\B}^{\mathrm{2}}\times\B}
\\
\ensuremath{\Varid{ccnot}\;((\mathrm{1},\mathrm{1}),\Varid{c})\mathrel{=}((\mathrm{1},\mathrm{1}),\neg \;\Varid{c})}
\\
\ensuremath{\Varid{ccnot}\;((\Varid{a},\Varid{b}),\Varid{c})\mathrel{=}((\Varid{a},\Varid{b}),\Varid{c})}
\end{quote}

As a last example of \ensuremath{\Varid{fst}}-complementation, let us see a famous device in
quantum programming arising from the following generic evolution of the
\textsc{cnot} gate, parametric on \ensuremath{\rarrow{\Conid{A}}{\Varid{f}}{\Conid{B}}} and such that
\ensuremath{(\Conid{B};\theta,\mathrm{0})} is a monoid satisfying \ensuremath{\Varid{x}\;\theta\;\Varid{x}\mathrel{=}\mathrm{0}} for all \ensuremath{\Varid{x}\;{\in}\;\Conid{B}}:%
\footnote{Our convention is that \ensuremath{\Conid{M} \comp \Conid{N}} takes precedence
over all other binary combinators, so \ensuremath{{\Varid{fst}}\kr{\theta \comp (\Varid{f}\times\Varid{id})}} means
\ensuremath{{\Varid{fst}}\kr{(\theta \comp (\Varid{f}\times\Varid{id}))}}. For economy of notation, we overload the \ensuremath{\theta} symbol to  denote both the uncurried and curried versions of the operator.}
\begin{quote}
	\ensuremath{U{}\;\Varid{f}\mathbin{:}(\Conid{A}\to \Conid{B})\to (\Conid{A}\times\Conid{B})\to (\Conid{A}\times\Conid{B})}
\\
	\ensuremath{U{}\;\Varid{f}\mathrel{=}{\Varid{fst}}\kr{\theta \comp (\Varid{f}\times\Varid{id})}} ,
\end{quote}
that is,
\begin{eqnarray*}
	\mbox\gufgate
\end{eqnarray*}
Clearly, for \ensuremath{\theta\mathrel{=}\mathbin{\xsor} }:
\begin{quote}
\ensuremath{\Varid{cnot}\mathrel{=}U{\Varid{id}}}\\
\ensuremath{\Varid{ccnot}\mathrel{=}U{(\mathrel{\wedge})}}
\end{quote}
It is easy to see that, for every \ensuremath{\Varid{f}}, \ensuremath{U{\Varid{f}}} is a \emph{bijection} because
it is its own inverse:
\begin{eqnarray*}
\start
	\ensuremath{U{\Varid{f}} \comp U{\Varid{f}}\mathrel{=}\Varid{id}}
\just\equiv{\ensuremath{U{\Varid{f}}\;(\Varid{x},\Varid{y})\mathrel{=}(\Varid{x},(\Varid{f}\;\Varid{x})\;\theta\;\Varid{y})}}
	\ensuremath{U{\Varid{f}}\;(\Varid{x},(\Varid{f}\;\Varid{x})\;\theta\;\Varid{y})\mathrel{=}(\Varid{x},\Varid{y})}
\just\equiv{ again \ensuremath{U{\Varid{f}}\;(\Varid{x},\Varid{y})\mathrel{=}(\Varid{x},(\Varid{f}\;\Varid{x})\;\theta\;\Varid{y})}}
	\ensuremath{(\Varid{x},(\Varid{f}\;\Varid{x})\;\theta\;((\Varid{f}\;\Varid{x})\;\theta\;\Varid{y}))\mathrel{=}(\Varid{x},\Varid{y})}
\just\equiv{ \ensuremath{\theta} is associative and \ensuremath{\Varid{x}\;\theta\;\Varid{x}\mathrel{=}\mathrm{0}}}
	\ensuremath{(\Varid{x},\mathrm{0}\;\theta\;\Varid{y})\mathrel{=}(\Varid{x},\Varid{y})}
\just\equiv{\ensuremath{\mathrm{0}\;\theta\;\Varid{x}\mathrel{=}\Varid{x}}}
	\ensuremath{(\Varid{x},\Varid{y})\mathrel{=}(\Varid{x},\Varid{y})}
\qed
\end{eqnarray*}
\ensuremath{U{\Varid{f}}} is therefore a \emph{reversible refinement} of an arbitrary \ensuremath{\Varid{f}\mathbin{:}\Conid{A}\to \Conid{B}} (for a monoid \ensuremath{\Conid{B}} as above) in the
sense that\footnote{\ensuremath{\kons{\mathrm{0}}} is the everywhere \ensuremath{\mathrm{0}}-constant function. In
general, \ensuremath{\kons{\Varid{k}}\;\Varid{x}\mathrel{=}\Varid{k}} for all suitably typed \ensuremath{\Varid{x}} and \ensuremath{\Varid{k}}.}
\begin{quote}
\ensuremath{\Varid{snd} \comp U{\Varid{f}} \comp (\Varid{id}\times\kons{\mathrm{0}})\mathrel{=}\Varid{f}} ,
\end{quote}
or in pointwise notation,
\begin{quote}
\ensuremath{\Varid{f}\;\Varid{x}\mathrel{=}\Varid{b}\;\mathbf{where}\;(\anonymous ,\Varid{b})\mathrel{=}U{\Varid{f}}\;(\Varid{x},\mathrm{0})}.
\end{quote}

\section{The dual view}
Before moving on and generalising \ensuremath{\Varid{fst}}-complementation to more
interesting programming constructs, we present a non-standard
perspective of Boolean gates which is based on coproducts
(\ensuremath{\Conid{A}\mathbin{+}\Conid{B}}) rather than products (\ensuremath{\Conid{A}\times\Conid{B}}). By \ensuremath{\Conid{A}\mathbin{+}\Conid{B}} we mean
the \emph{disjoint union} of \ensuremath{\Conid{A}} and \ensuremath{\Conid{B}}:
\begin{eqnarray}
	\ensuremath{\Conid{A}\mathbin{+}\Conid{B}\mathrel{=}{\{\mskip1.5mu i_1\;\Varid{x}\mid \Varid{x}\;{\in}\;\Conid{A}\mskip1.5mu\}}\cup{\{\mskip1.5mu i_2\;\Varid{y}\mid \Varid{y}\;{\in}\;\Conid{B}\mskip1.5mu\}}}
	\label{eq:191026a}
\end{eqnarray}
where \ensuremath{i_1} and \ensuremath{i_2} are injective.
(Disjointness relies on assuming \ensuremath{\conv{i_1} \comp i_2\mathrel{=}\bot }, that is, for all \ensuremath{\Varid{x}}
and \ensuremath{\Varid{y}}, \ensuremath{i_1\;\Varid{x}\not=i_2\;\Varid{y}}.)

Given any two relations \ensuremath{\rarrow{\Conid{A}}{\Conid{R}}{\Conid{C}}} and \ensuremath{\rarrow{\Conid{B}}{\Conid{S}}{\Conid{C}}}, there exists a
unique relation \ensuremath{\rarrow{\Conid{A}\mathbin{+}\Conid{B}}{\Conid{X}}{\Conid{C}}} such that \ensuremath{\Conid{X} \comp i_1\mathrel{=}\Conid{R}} and \ensuremath{\Conid{X} \comp i_2\mathrel{=}\Conid{S}}.
We denote that relation by \ensuremath{\alt{\Conid{R}}{\Conid{S}}}:
\begin{eqnarray}
X = \ensuremath{\alt{\Conid{R}}{\Conid{S}}} & \ensuremath{\Leftrightarrow} &
        \left\{
                \begin{array}{rcl}
                        X \comp i_1 = R
                \\
                        X \comp i_2 = S
                \end{array}
        \right.
	\label{eq:081008c}
\end{eqnarray}
The \emph{direct sum} of two relations arises immediately from:
\begin{eqnarray}
	R + S = \ensuremath{\alt{i_1 \comp \Conid{R}}{i_2 \comp \Conid{S}}}
	\label{eq:040201d}
\end{eqnarray}
The isomorphism
\begin{eqnarray}
	\ensuremath{\larrow{\Conid{A}\mathbin{+}\Conid{A}}{\gamma}{\B\times\Conid{A}}\mathrel{=}\alt{{\Varid{false}}\kr{\Varid{id}}}{{\Varid{true}}\kr{\Varid{id}}})}
	\label{eq:191026b}
\end{eqnarray}
holds (where \ensuremath{\Varid{false}} and \ensuremath{\Varid{true}} are the obvious constant functions)
and can be re-written into
\begin{eqnarray}
	\ensuremath{\gamma\mathrel{=}{\alt{\Varid{false}}{\Varid{true}}}\kr{\alt{\Varid{id}}{\Varid{id}}}}
	\label{eq:191026d}
\end{eqnarray}
thanks to the so-called \emph{exchange law}:
\begin{eqnarray}
        \ensuremath{\alt{{\Conid{R}}\kr{\Conid{S}}}{{\Conid{T}}\kr{\Conid{V}}}} &=&
        \ensuremath{{\alt{\Conid{R}}{\Conid{T}}}\kr{\alt{\Conid{S}}{\Conid{V}}}}.
	\label{eq:701d-rel}
\end{eqnarray}
So, we have that \ensuremath{\B\times\B} is isomorphic to \ensuremath{\B\mathbin{+}\B} through \ensuremath{\gamma}
\eqref{eq:191026b}. This provides us with an alternative (dual)
view of logic gates, for instance:
conjunction
\begin{eqnarray}
\start	\ensuremath{\rarrow{\B\mathbin{+}\B}{(\mathrel{\wedge}) \comp \gamma}{\B}} = \ensuremath{\alt{\Varid{false}}{\Varid{id}}},
\end{eqnarray}
disjunction
\begin{eqnarray}
\start	\ensuremath{\rarrow{\B\mathbin{+}\B}{(\mathrel{\vee}) \comp \gamma}{\B}} = \ensuremath{\alt{\Varid{true}}{\neg }},
\end{eqnarray}
exclusive-or
\begin{eqnarray}
\start	\ensuremath{\rarrow{\B\mathbin{+}\B}{\xsor  \comp \gamma}{\B}\mathrel{=}\alt{\Varid{id}}{\neg }},
	\label{eq:161128a}
\end{eqnarray}
and so on.

Note how \ensuremath{\B\mathbin{+}\B} captures the second bit of a Boolean gate
once the first is set to false (on the left of the sum) or to true
(on the right of the sum). So \eqref{eq:161128a} immediately tells
that exclusive-or behaves as the identity in the first case
and as negation in the second. In matrix notation -- cf.\ \eqref{eq:191026c}:
\begin{eqnarray*}
	\ensuremath{\xsor \mathrel{=}\meither{\matrix{\mathrm{1}}{\mathrm{0}}{\mathrm{0}}{\mathrm{1}}}{\matrix{\mathrm{0}}{\mathrm{1}}{\mathrm{1}}{\mathrm{0}}}}.
\end{eqnarray*}

Applying the same transformation to \ensuremath{\Varid{fst}}-comple\-mented operations
yields similarly expressive denotations of Boolean gates.
For instance, \ensuremath{\Varid{cnot}\mathrel{=}{\Varid{fst}}\kr{\mathbin{\xsor} }} is transformed into
\ensuremath{\Varid{id}\mathbin{+}\neg } through \ensuremath{\gamma}:
\begin{eqnarray*}
\start
	\ensuremath{\Varid{cnot} \comp \gamma}
\just={ \ensuremath{\Varid{cnot}\mathrel{=}{\Varid{fst}}\kr{\mathbin{\xsor} }}; pairing; \eqref{eq:161128a} }
	\ensuremath{{(\Varid{fst} \comp \gamma)}\kr{\alt{\Varid{id}}{\neg }}}
\just={ \eqref{eq:191026d}; pairing; exchange law }
	\ensuremath{\alt{{\Varid{false}}\kr{\Varid{id}}}{{\Varid{true}}\kr{\neg }}}
\just={ \ensuremath{\Varid{true} \comp \Varid{f}\mathrel{=}\Varid{f}}; \eqref{eq:191026b}; pairing laws }
	\ensuremath{\gamma \comp (\Varid{id}\mathbin{+}\neg )}
\end{eqnarray*}
That is, \ensuremath{\gamma} has the relational type\footnote{
In general, \ensuremath{\Varid{f}} is said to have relational type \ensuremath{\Conid{S}\leftarrow \Conid{R}}
whenever \ensuremath{\Varid{f} \comp \Conid{R}\; \subseteq \;\Conid{S} \comp \Varid{f}} holds.}
\begin{eqnarray*}
	\ensuremath{\larrow{\Varid{id}\mathbin{+}\neg }{\gamma}{{\Varid{fst}}\kr{\mathbin{\xsor} }}}.
\end{eqnarray*}
Written as \ensuremath{\Varid{id}\mathbin{+}\neg }, \ensuremath{\Varid{cnot}} is immediately seen to be an isomorphism,
because \ensuremath{\Varid{id}} and \ensuremath{\neg } are so. In the same setting, the Toffoli gate
\ensuremath{\Varid{ccnot}} will be expressed as
\begin{quote}
\ensuremath{\Varid{id}\mathbin{+}(\Varid{id}\mathbin{+}\neg )},
\end{quote}
again an isomorphism by construction.

Isomorphism \ensuremath{\gamma} will play an important role in implementing a form of
conditional quantum control in section \ref{sec:191029b}. The coproduct
construct is also inherently present in the strategy 
that underlies section \ref{sec:quanta}.

\section{Generalising \ensuremath{\Varid{fst}}-complementation} \label{sec:200203a}
As seen in section \ref{sec:191026ce}, the projection \ensuremath{\rarrow{\Conid{A}\times\Conid{B}}{\Varid{fst}}{\Conid{A}}} plays a role in injectivity refinements, working as minimal
complement in several situations.  In general, \ensuremath{\Varid{fst}}-complementation
\begin{quote}
\ensuremath{\Varid{id}\leq {\Varid{fst}}\kr{\Varid{f}}}
\end{quote}
works whenever
\begin{quote}
\ensuremath{\Varid{f}\;(\Varid{a},\Varid{b})\mathrel{=}\Varid{f}\;(\Varid{a},\Varid{b'})\Rightarrow \Varid{b}\mathrel{=}\Varid{b'}}
\end{quote}
holds, that is\footnote{We abbreviate \ensuremath{\Varid{curry}\;\Varid{f}} by \ensuremath{\overline{\Varid{f}}}, that is: \ensuremath{\overline{\Varid{f}}\;\Varid{a}\;\Varid{b}\mathrel{=}\Varid{curry}\;\Varid{f}\;\Varid{a}\;\Varid{b}\mathrel{=}\Varid{f}\;(\Varid{a},\Varid{b})}.}:
\begin{quote}
\ensuremath{\overline{\Varid{f}}\;\Varid{a}\;\Varid{b}\mathrel{=}\overline{\Varid{f}}\;\Varid{a}\;\Varid{b'}\Rightarrow \Varid{b}\mathrel{=}\Varid{b'}}.
\end{quote}
In other words, \ensuremath{\Varid{f}} is left-cancellative: it is \emph{injective} on the \emph{second} argument once the \emph{first} is fixed.

Wherever \ensuremath{\rarrow{\Conid{A}\times\Conid{B}}{\Varid{fst}}{\Conid{A}}} complements a function of type
\ensuremath{\rarrow{\Conid{A}\times\Conid{B}}{}{\Conid{B}}}, it makes room (type-wise) for a \emph{bijection}
of type \ensuremath{\rarrow{\Conid{A}\times\Conid{B}}{}{\Conid{A}\times\Conid{B}}}. Can \ensuremath{{\Varid{fst}}\kr{(\anonymous )}} be extended to
more elaborate computations, e.g.\ \emph{recursively}?
Note that such \ensuremath{\Conid{A}\times\Conid{B}\to \Conid{A}\times\Conid{B}} computations, of shape
\begin{eqnarray*}
\vcenter{\ensuremath{\fstcbox{\Conid{A}}{\Conid{B}}{}{\Conid{A}}{\Conid{B}}}}
\end{eqnarray*}
can be chained together. Take, for instance, \ensuremath{\Varid{n}} copies of
\begin{eqnarray*}
\vcenter{\ensuremath{\fstcbox{\Varid{x}}{y_0 }{\Varid{f}}{\Varid{x}}{y_1 }}}
\end{eqnarray*}
and draw each of them in a different way,
\begin{eqnarray*}
\vcenter{\xymatrix@R=1.5em@C=1.5em{
&
	\ensuremath{\Varid{x}}
\\
	\ensuremath{y_0 }
		\ar[r]
&
	*+[F]{~\ensuremath{\Varid{f}}~}
		\ar[u]
		\ar[r]
&
	\ensuremath{y_1 }
\\
&
	\ensuremath{\Varid{x}}
		\ar[u]
}},
\end{eqnarray*}
so that they can be chained as depicted below:
\begin{eqnarray*}
\vcenter{\xymatrix@R=1.5em@C=1.5em{
&
	\ensuremath{x_0 }
&
	\ensuremath{x_1 }
&
	\ensuremath{x_2 }
\\
	\ensuremath{y_0 }
		\ar[r]
&
	*+[F]{~\ensuremath{\Varid{f}}~}
		\ar[u]
		\ar[r]
&
	*+[F]{~\ensuremath{\Varid{f}}~}
		\ar[u]
		\ar[r]
&
	*+[F]{~\ensuremath{\Varid{f}}~}
		\ar[u]
		\ar[r]
&
	\ensuremath{y_3 }
\\
&
	\ensuremath{x_0 }
		\ar[u]
&
	\ensuremath{x_1 }
		\ar[u]
&
	\ensuremath{x_2 }
		\ar[u]
}} \cdots
\end{eqnarray*}
Clearly, \ensuremath{[\mskip1.5mu x_0 ,x_1 ,x_2 ,\mathbin{...}\mskip1.5mu]} can be regarded as a control-sequence, which
is passed along to the output. Meanwhile, the input \ensuremath{y_0 } is subject to an
accumulation of transformations performed by \ensuremath{\Varid{f}}, one for each \ensuremath{x_i }.

\begin{figure}
\footnotesize\centering
	\includegraphics[width=0.45\textwidth]{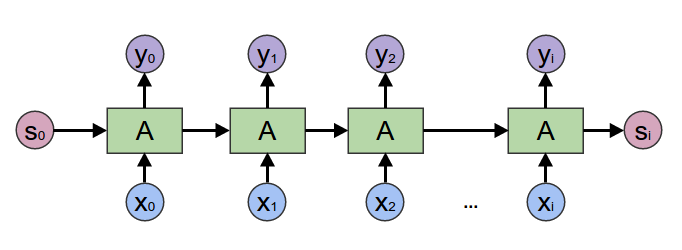}
\caption{Recurrent neural network (RNN) depicted in \cite{Co15} as an instance of an accumulating map.\label{fig:190226a}}
\end{figure}

Interestingly, this chain can be regarded as an instance of a functional
programming pattern known as the \emph{accumulating map}.\footnote{Cf.\ e.g.\ the
function \ensuremath{\Varid{mapAccumR}} in the Haskell language.}
This pattern turns up in various contexts
(namely in neural networks, see Figure \ref{fig:190226a}).
In the sequel, it will be shown to be an instance of a construct
that we shall introduce shortly and term \emph{quantamorphism},
as it will generalise to quantum computing later.

\section{Towards (constructive) recursive complementation} \label{sec:200203b}
Suppose one wants to offer an \emph{arbitrary} function \ensuremath{\Varid{k}\mathbin{:}\Conid{A}\to \Conid{B}}
wrapped in a \emph{bijective} ``envelope", as happened above in the
derivation of \ensuremath{\Varid{cnot}} and other gates.
The ``smallest" (\emph{generic}) type for such an enveloped function is \ensuremath{\Conid{A}\times\Conid{B}\to \Conid{A}\times\Conid{B}}.

Now suppose that \ensuremath{\Varid{k}} is a \emph{recursive} function over finite lists, for instance
\ensuremath{\Varid{k}\mathrel{=}\mathsf{foldr}\ \overline{\Varid{f}}\;\Varid{b}} for \ensuremath{\Varid{f}\mathbin{:}\Conid{A}\times\Conid{B}\to \Conid{B}}, that is,
\begin{quote}
\ensuremath{\Varid{k}\mathbin{:}{\Conid{A}}^{*}\to \Conid{B}}
\\
\ensuremath{\Varid{k}\;[\mskip1.5mu \mskip1.5mu]\mathrel{=}\Varid{b}}
\\
\ensuremath{\Varid{k}\;(\Varid{a}\mathbin{:}\Varid{x})\mathrel{=}\Varid{f}\;(\Varid{a},\Varid{k}\;\Varid{x})} .
\end{quote}
How do we ``constructively'' build the corresponding (\emph{recursive},
\emph{bijective}) envelope of type \ensuremath{{\Conid{A}}^{*}\times\Conid{B}\to {\Conid{A}}^{*}\times\Conid{B}}? Let us define
\ensuremath{\env{\Varid{f}}} such that
\begin{eqnarray}
	\ensuremath{\env{\Varid{f}}\;(\Varid{x},\Varid{b})\mathrel{=}\mathsf{foldr}\ \overline{\Varid{f}}\;\Varid{b}\;\Varid{x}},
	\label{eq:190304c}
\end{eqnarray}
that is,
\begin{eqnarray*}
\start	\ensuremath{\env{\Varid{f}}\;([\mskip1.5mu \mskip1.5mu],\Varid{b})\mathrel{=}\Varid{b}}
\more	\ensuremath{\env{\Varid{f}}\;(\Varid{a}\mathbin{:}\Varid{x},\Varid{b})\mathrel{=}\Varid{f}\;(\Varid{a},\env{\Varid{f}}\;(\Varid{x},\Varid{b}))} .
\end{eqnarray*}
We can depict \ensuremath{\env{\Varid{f}}} in the form of a commutative diagram:
\begin{eqnarray*}
\myxym{
	\ensuremath{{\Conid{A}}^{*}\times\Conid{B}}
		\ar[d]_{\ensuremath{\env{\Varid{f}}}}
&
	\ensuremath{\Conid{B}\mathbin{+}\Conid{A}\times({\Conid{A}}^{*}\times\Conid{B})}
		\ar[l]_-{\ensuremath{\alpha }}
		\ar[d]^{\ensuremath{\Varid{id}\mathbin{+}\Varid{id}\times\env{\Varid{f}}}}
\\
	\ensuremath{\Conid{B}}
&
	\ensuremath{\Conid{B}\mathbin{+}\Conid{A}\times\Conid{B}}
		\ar[l]^{\ensuremath{\alt{\Varid{id}}{\Varid{f}}}}
}
\end{eqnarray*}
In the diagram, the isomorphism
\begin{eqnarray*}
\myxym{
	\ensuremath{{\Conid{A}}^{*}\times\Conid{B}}
&
	\ensuremath{\Conid{B}\mathbin{+}\Conid{A}\times({\Conid{A}}^{*}\times\Conid{B})}
		\ar[l]_-{\ensuremath{\alpha }}
}
\end{eqnarray*}
is defined by
\begin{eqnarray}
	\ensuremath{\alpha \mathrel{=}\alt{{\Varid{nil}}\kr{\Varid{id}}}{(\Varid{cons}\times\Varid{id}) \comp \mathsf{a}}}
	\label{eq:180318a}
\end{eqnarray}
where
\begin{quote}
\ensuremath{\Varid{nil}\;\anonymous \mathrel{=}[\mskip1.5mu \mskip1.5mu]} \\ \ensuremath{\Varid{cons}\;(\Varid{a},\Varid{x})\mathrel{=}\Varid{a}\mathbin{:}\Varid{x}}
\end{quote}
are the components of the initial algebra \ensuremath{\mathsf{in}\mathrel{=}\alt{\Varid{nil}}{\Varid{cons}}} of
finite lists \cite{BM97}, and the isomorphism
\begin{eqnarray}
\arrayin{
\start
	\ensuremath{\mathsf{a}\mathbin{:}\Conid{A}\times(\Conid{B}\times\Conid{C})\to (\Conid{A}\times\Conid{B})\times\Conid{C}}
\more 
	\ensuremath{\mathsf{a}\mathrel{=}{(\Varid{id}\times\Varid{fst})}\kr{(\Varid{snd} \comp \Varid{snd})}}
}
	\label{eq:180318b}
\end{eqnarray}
in \eqref{eq:180318a} is the associator.

Below, we will need something more general, namely:
\begin{eqnarray}
\myxym{
	\ensuremath{{\Conid{A}}^{*}\times\Conid{B}}
		\ar[d]_{\ensuremath{\env{\Varid{h}}}}
&
	\ensuremath{\Conid{B}\mathbin{+}\Conid{A}\times({\Conid{A}}^{*}\times\Conid{B})}
		\ar[l]_-{\ensuremath{\alpha }}
		\ar[d]^{\ensuremath{\Varid{id}\mathbin{+}\Varid{id}\times\env{\Varid{h}}}}
\\
	\ensuremath{\Conid{C}}
&
	\ensuremath{\Conid{B}\mathbin{+}\Conid{A}\times\Conid{C}}
		\ar[l]^{\ensuremath{\Varid{h}}}
}
	\label{eq:190416a}
\end{eqnarray}
This is specified by the \emph{universal} property
\begin{eqnarray}
	\ensuremath{\Varid{k}\mathrel{=}\env{\Varid{h}}} & \ensuremath{\Leftrightarrow} & \ensuremath{\Varid{k} \comp \alpha \mathrel{=}\Varid{h} \comp \fun F \;\Varid{k}}
	\label{eq:180317a}
\end{eqnarray}
where \ensuremath{\fun F \;\Varid{f}\mathrel{=}\Varid{id}\mathbin{+}\Varid{id}\times\Varid{f}} captures the list-recursion pattern.
This \ensuremath{\env{\Varid{h}}} can be regarded as an extension of the well-known \emph{catamorphism}
combinator.\footnote{See e.g.\ \cite{Hi13} for a thorough study of such kind of extensions to the standard theory \cite{BM97}.}
All the standard laws apply, including reflexion \ensuremath{\env{\alpha }\mathrel{=}\Varid{id}} and the loop-intercombination law
\begin{eqnarray}
	\ensuremath{{\env{\Varid{f}}}\kr{\env{\Varid{g}}}\mathrel{=}\env{(\Varid{f}\times\Varid{g}) \comp ({\fun F \;\Varid{fst}}\kr{\fun F \;\Varid{snd}})}},
	\label{eq:190304a}
\end{eqnarray}
often referred to as the \emph{``banana-split"} law \cite{BM97}.
From \eqref{eq:180317a} and \eqref{eq:180318b}, one also infers
\begin{eqnarray}
	\ensuremath{\larrow{{\Conid{A}}^{*}\times\Conid{B}}{\Varid{fst}}{{\Conid{A}}^{*}}\mathrel{=}\env{\mathsf{in}}}
	\label{eq:180318c}
\end{eqnarray}
by easy calculation:
\begin{eqnarray*}
\start
	\ensuremath{\Varid{fst}\mathrel{=}\env{\mathsf{in}}}
\just\equiv{ \eqref{eq:180317a} }
	\ensuremath{\Varid{fst} \comp \alpha \mathrel{=}\mathsf{in} \comp (\Varid{id}\mathbin{+}\Varid{id}\times\Varid{fst})}
\just\equiv{ \ensuremath{\mathsf{in}\mathrel{=}\alt{\Varid{nil}}{\Varid{cons}}}; coproducts }
	\ensuremath{\Varid{fst} \comp \alpha \mathrel{=}\alt{\Varid{nil}}{\Varid{cons} \comp (\Varid{id}\times\Varid{fst})}}
\just\equiv{ definition of \ensuremath{\alpha } (\ref{eq:180318a}) and \ensuremath{\mathsf{a}} (\ref{eq:180318b}) }
	\ensuremath{\Varid{true}}
\qed
\end{eqnarray*}

\paragraph{Promoting \ensuremath{\Varid{fst}}-complementation}
Suppose that a given \emph{non-injective} \ensuremath{\Varid{f}\mathbin{:}\Conid{A}\times\Conid{B}\to \Conid{B}} is complemented by
\ensuremath{\Varid{fst}\mathbin{:}\Conid{A}\times\Conid{B}\to \Conid{A}}, i.e. that \ensuremath{{\Varid{fst}}\kr{\Varid{f}}} is injective.
We can place it in \eqref{eq:190416a}
and ask: will \ensuremath{\env{\alt{\Varid{id}}{\Varid{f}}}} be \ensuremath{\Varid{fst}}-complemented too?
We start by unfolding the term \ensuremath{{\Varid{fst}}\kr{\env{\alt{\Varid{id}}{\Varid{f}}}}}:
\begin{eqnarray}
\start
	\ensuremath{{\Varid{fst}}\kr{\env{\alt{\Varid{id}}{\Varid{f}}}}}
	\nonumber
\just={ \eqref{eq:180318c} followed by  banana-split \eqref{eq:190304a} }
	\ensuremath{\env{(\mathsf{in}\times\alt{\Varid{id}}{\Varid{f}}) \comp ({\fun F \;\Varid{fst}}\kr{\fun F \;\Varid{snd}})}}
	\nonumber
\just={ \ensuremath{\mathsf{in}\mathrel{=}\alt{\Varid{nil}}{\Varid{cons}}\mskip1.5mu]}; pairing laws (products) }
	\ensuremath{\env{{\alt{\Varid{nil}}{\Varid{cons} \comp (\Varid{id}\times\Varid{fst})}}\kr{\alt{\Varid{id}}{\Varid{f} \comp (\Varid{id}\times\Varid{snd})}}}}
	\nonumber
\just={ exchange law (\ref{eq:701d-rel}) }
	\ensuremath{\env{\alt{{\Varid{nil}}\kr{\Varid{id}}}{{(\Varid{cons} \comp (\Varid{id}\times\Varid{fst}))}\kr{(\Varid{f} \comp (\Varid{id}\times\Varid{snd}))}}}}
	\nonumber
\just={ products ; \ensuremath{\mathsf{a} \comp \conv{\mathsf{a}}\mathrel{=}\Varid{id}}}
	\ensuremath{\env{\alpha  \comp (\Varid{id}\mathbin{+}\conv{\mathsf{a}} \comp ({(\Varid{id}\times\Varid{fst})}\kr{\Varid{f} \comp (\Varid{id}\times\Varid{snd})}))}}
	\label{eq:180318d}
\end{eqnarray}
Thus,
\begin{eqnarray*}
	\ensuremath{{\Varid{fst}}\kr{\env{\alt{\Varid{id}}{\Varid{f}}}}} & = & \ensuremath{\env{\Psi\ ({\Varid{fst}}\kr{\Varid{f}})}},
\end{eqnarray*}
by introducing
\begin{eqnarray*}
	\ensuremath{\Psi\ \Varid{x}\mathrel{=}\alpha  \comp (\Varid{id}\mathbin{+}\conv{\mathsf{a}} \comp ({(\Varid{id}\times\Varid{fst})}\kr{\Varid{snd} \comp \Varid{x} \comp (\Varid{id}\times\Varid{snd})}))},
\end{eqnarray*}
which shrinks to
\begin{eqnarray}
	\ensuremath{\Psi\ \Varid{x}\mathrel{=}\alpha  \comp (\Varid{id}\mathbin{+}\mathsf{xl} \comp (\Varid{id}\times\Varid{x}) \comp \mathsf{xl})}
	\label{eq:190418a}
\end{eqnarray}
using the isomorphism \ensuremath{\larrow{\Conid{A}\times(\Conid{B}\times\Conid{C})}{\mathsf{xl}}{\Conid{B}\times(\Conid{A}\times\Conid{C})}} instead of \ensuremath{\mathsf{a}}.\footnote{Calculations can be found in the appendix.}
Putting everything into a diagram, we obtain:
\begin{eqnarray*}
\xymatrix@C=1.2em@R=1.1em{
	\ensuremath{{\Conid{A}}^{*}\times\Conid{B}}
		\ar[d]_{\ensuremath{\env{\Varid{h}}}}
&&
	\ensuremath{\Conid{B}\mathbin{+}\Conid{A}\times({\Conid{A}}^{*}\times\Conid{B})}
		\ar[ll]_-{\ensuremath{\alpha }}
		\ar[d]^{\ensuremath{\Varid{id}\mathbin{+}\Varid{id}\times\env{\Varid{h}}}}
\\
	\ensuremath{{\Conid{A}}^{*}\times\Conid{B}}
&&
	\ensuremath{\Conid{B}\mathbin{+}\Conid{A}\times({\Conid{A}}^{*}\times\Conid{B})}
		\ar[dl]^-{\rule{4em}{0pt}\ensuremath{\Varid{id}\mathbin{+}\mathsf{xl} \comp (\Varid{id}\times({\Varid{fst}}\kr{\Varid{f}})) \comp \mathsf{xl}}}
		\ar[ll]^-{\ensuremath{\Varid{h}\mathrel{=}\Psi\ ({\Varid{fst}}\kr{\Varid{f}})}}
\\
&
	\ensuremath{\Conid{B}\mathbin{+}\Conid{A}\times({\Conid{A}}^{*}\times\Conid{B})}
		\ar[ul]^-{\ensuremath{\alpha }} 
}
\end{eqnarray*}
Clearly, \ensuremath{\Psi\ } preserves injectivity, as does \ensuremath{\env{\anonymous }} -- see the appendix
for details. Therefore, \ensuremath{{\Varid{fst}}\kr{\Varid{f}}} being {injective} ensures \ensuremath{{\Varid{fst}}\kr{\env{\alt{\Varid{id}}{\Varid{f}}}}} is also injective. In words:
\begin{quote}\em
The \ensuremath{\Varid{fst}}-complementation of \ensuremath{\Varid{f}} in \ensuremath{\mathsf{foldr}\ \overline{\Varid{f}}\;\Varid{b}}
is promoted to the \ensuremath{\Varid{fst}}-complementation of the fold itself.
\end{quote}
That is to say, \ensuremath{\Varid{fst}}-complementation is \emph{propagated} inductively across
lists and we get the construction of a \emph{reversible fold}, defined by
\ensuremath{\mathsf{rfold}\ \Varid{f}\mathrel{=}\env{\Psi\ ({\Varid{fst}}\kr{\Varid{f}})}}. Unfolding the definition and adding
variables, we get, in standard Haskell notation:
\begin{hscode}\SaveRestoreHook
\column{B}{@{}>{\hspre}l<{\hspost}@{}}%
\column{7}{@{}>{\hspre}l<{\hspost}@{}}%
\column{E}{@{}>{\hspre}l<{\hspost}@{}}%
\>[B]{}\mathsf{rfold}\ \mathbin{::}((\Varid{a},\Varid{b})\to \Varid{b})\to ([\mskip1.5mu \Varid{a}\mskip1.5mu],\Varid{b})\to ([\mskip1.5mu \Varid{a}\mskip1.5mu],\Varid{b}){}\<[E]%
\\
\>[B]{}\mathsf{rfold}\ \Varid{f}\;([\mskip1.5mu \mskip1.5mu],\Varid{b})\mathrel{=}([\mskip1.5mu \mskip1.5mu],\Varid{b}){}\<[E]%
\\
\>[B]{}\mathsf{rfold}\ \Varid{f}\;(\Varid{a}\mathbin{:}\Varid{x},\Varid{b})\mathrel{=}(\Varid{a}\mathbin{:}\Varid{y},\Varid{f}\;(\Varid{a},\Varid{b'})){}\<[E]%
\\
\>[B]{}\hsindent{7}{}\<[7]%
\>[7]{}\mathbf{where}\;(\Varid{y},\Varid{b'})\mathrel{=}\mathsf{rfold}\ \Varid{f}\;(\Varid{x},\Varid{b}){}\<[E]%
\ColumnHook
\end{hscode}\resethooks
We can therefore rely on the reversibility of \ensuremath{\mathsf{rfold}\ \Varid{f}}
\begin{quote}
	\rfoldgate
\end{quote}
provided \ensuremath{\Varid{f}} is complemented by \ensuremath{\Varid{fst}}.

This result can be generalised by defining, given some \ensuremath{\Varid{f}\mathbin{:}\Conid{A}\times\Conid{B}\to \Conid{C}\times\Conid{B}},
\ensuremath{\quanta{\Varid{f}}\mathrel{=}\env{\Psi\ \Varid{f}}} as pictured in the following diagram:
\begin{eqnarray}
\xymatrix@C=1.0em@R=1.1em{
	\ensuremath{{\Conid{A}}^{*}\times\Conid{B}}
		\ar[d]_{\ensuremath{\quanta{\Varid{f}}}}
&&
	\ensuremath{\Conid{B}\mathbin{+}\Conid{A}\times({\Conid{A}}^{*}\times\Conid{B})}
		\ar[ll]_-{\ensuremath{\alpha }}
		\ar[d]^{\ensuremath{\Varid{id}\mathbin{+}\Varid{id}\times\quanta{\Varid{f}}}}
\\
	\ensuremath{{\Conid{C}}^{*}\times\Conid{B}}
&&
	\ensuremath{\Conid{B}\mathbin{+}\Conid{A}\times({\Conid{C}}^{*}\times\Conid{B})}
		\ar[ll]^-{\ensuremath{\Psi\ \Varid{f}}}
}
	\label{eq:190418c}
\end{eqnarray}
Clearly,
\begin{eqnarray}
	\ensuremath{\quanta{\Varid{id}}\mathrel{=}\Varid{id}}
\end{eqnarray}
since \ensuremath{\Psi\ \Varid{id}\mathrel{=}\alpha }. If \ensuremath{\Varid{f}} is reversible, then \ensuremath{\quanta{\Varid{f}}} will also be reversible, as we have
seen. For instance, \ensuremath{\quanta{\Varid{cnot}}\mathbin{:}{\B}^{*}\times\B\to {\B}^{*}\times\B}
will be reversible, because so is \ensuremath{\Varid{cnot}\mathbin{:}\B\times\B\to \B\times\B}.

By \emph{free theorem} calculation \cite{Wa89}, we get the following properties
among others:\footnote{\ensuremath{{\Varid{k}}^{*}} denotes the usual \ensuremath{\Varid{map}\;\Varid{k}} operation on sequences.}
\begin{eqnarray}
	\ensuremath{\quanta{\Varid{f}} \comp ({\Varid{k}}^{*}\times\Varid{id})\mathrel{=}\quanta{\Varid{f} \comp (\Varid{k}\times\Varid{id})}}
\\
	\ensuremath{({\Varid{k}}^{*}\times\Varid{id}) \comp \quanta{\Varid{f}}\mathrel{=}\quanta{(\Varid{k}\times\Varid{id}) \comp \Varid{f}}}
\end{eqnarray}

Converting the construction \ensuremath{\quanta{\cdot }} to Haskell notation yields
\begin{hscode}\SaveRestoreHook
\column{B}{@{}>{\hspre}l<{\hspost}@{}}%
\column{6}{@{}>{\hspre}l<{\hspost}@{}}%
\column{E}{@{}>{\hspre}l<{\hspost}@{}}%
\>[B]{}\quanta{\cdot }\mathbin{::}((\Varid{a},\Varid{b})\to (\Varid{c},\Varid{b}))\to ([\mskip1.5mu \Varid{a}\mskip1.5mu],\Varid{b})\to ([\mskip1.5mu \Varid{c}\mskip1.5mu],\Varid{b}){}\<[E]%
\\
\>[B]{}\quanta{\Varid{f}}\;([\mskip1.5mu \mskip1.5mu],\Varid{b})\mathrel{=}([\mskip1.5mu \mskip1.5mu],\Varid{b}){}\<[E]%
\\
\>[B]{}\quanta{\Varid{f}}\;(\Varid{a}\mathbin{:}\Varid{x},\Varid{b})\mathrel{=}(\Varid{c}\mathbin{:}\Varid{y},\Varid{b''})\;\mathbf{where}{}\<[E]%
\\
\>[B]{}\hsindent{6}{}\<[6]%
\>[6]{}(\Varid{y},\Varid{b'})\mathrel{=}\quanta{\Varid{f}}\;(\Varid{x},\Varid{b}){}\<[E]%
\\
\>[B]{}\hsindent{6}{}\<[6]%
\>[6]{}(\Varid{c},\Varid{b''})\mathrel{=}\Varid{f}\;(\Varid{a},\Varid{b'}){}\<[E]%
\ColumnHook
\end{hscode}\resethooks
which corresponds to the standard \ensuremath{\Varid{mapAccumR}} function.

In the sequel, we shall refer to \ensuremath{\quanta{\Varid{f}}} as being a \emph{quantamorphism}.
The reason for this is that this generalises nicely to
quantum programming (for \ensuremath{\Varid{f}} reversible) as we shall see next.

\section{Going quantum} \label{sec:200203c}
Recall that \emph{functions} can be represented by matrices, e.g.\ the \emph{controlled-not}
\begin{eqnarray*}
	\ensuremath{\begin{lcbr}\Varid{cnot}\;(\mathrm{0},\Varid{b})\mathrel{=}(\mathrm{0},\Varid{b})\\\Varid{cnot}\;(\mathrm{1},\Varid{b})\mathrel{=}(\mathrm{1},\neg \;\Varid{b})\end{lcbr}}
\end{eqnarray*}
is described by the matrix: {\small%
\begin{eqnarray*}
\begin{array}{r|cccc}&\rotatebox{90}{\ensuremath{(\textsf{0},\textsf{0})}}&\rotatebox{90}{\ensuremath{(\textsf{0},\textsf{1})}}&\rotatebox{90}{\ensuremath{(\textsf{1},\textsf{0})}}&\rotatebox{90}{\ensuremath{(\textsf{1},\textsf{1})}}\\\hline \hbox{\ensuremath{(\textsf{0},\textsf{0})}}&1&0&0&0\\\hbox{\ensuremath{(\textsf{0},\textsf{1})}}&0&1&0&0\\\hbox{\ensuremath{(\textsf{1},\textsf{0})}}&0&0&0&1\\\hbox{\ensuremath{(\textsf{1},\textsf{1})}}&0&0&1&0
\end{array}
\end{eqnarray*}}%
Now think of a \emph{probabilistic} ``evolution" of \ensuremath{\Varid{cnot}}: {\small%
\begin{eqnarray*}
\begin{array}{r|cccc}&\rotatebox{90}{\ensuremath{(\textsf{0},\textsf{0})}}&\rotatebox{90}{\ensuremath{(\textsf{0},\textsf{1})}}&\rotatebox{90}{\ensuremath{(\textsf{1},\textsf{0})}}&\rotatebox{90}{\ensuremath{(\textsf{1},\textsf{1})}}\\\hline \hbox{\ensuremath{(\textsf{0},\textsf{0})}}&1&0&0&0\\\hbox{\ensuremath{(\textsf{0},\textsf{1})}}&0&\ensuremath{\frac{\mathrm{1}}{\mathrm{2}}}&0&0\\\hbox{\ensuremath{(\textsf{1},\textsf{0})}}&0&\ensuremath{\frac{\mathrm{1}}{\mathrm{2}}}&0&1\\\hbox{\ensuremath{(\textsf{1},\textsf{1})}}&0&0&1&0
\end{array}
\end{eqnarray*}}%
In this evolution, function \ensuremath{\Varid{cnot}} becomes \emph{probabilistic}:
\ensuremath{\Varid{cnot}\;(\textsf{0},\textsf{1})} will evaluate to either \ensuremath{(\textsf{0},\textsf{1})} or \ensuremath{(\textsf{1},\textsf{0})}
with equal probability (\ensuremath{\mathrm{50}\mathbin{\%}}).

\begin{figure}
\footnotesize\centering
	\includegraphics[width=0.45\textwidth]{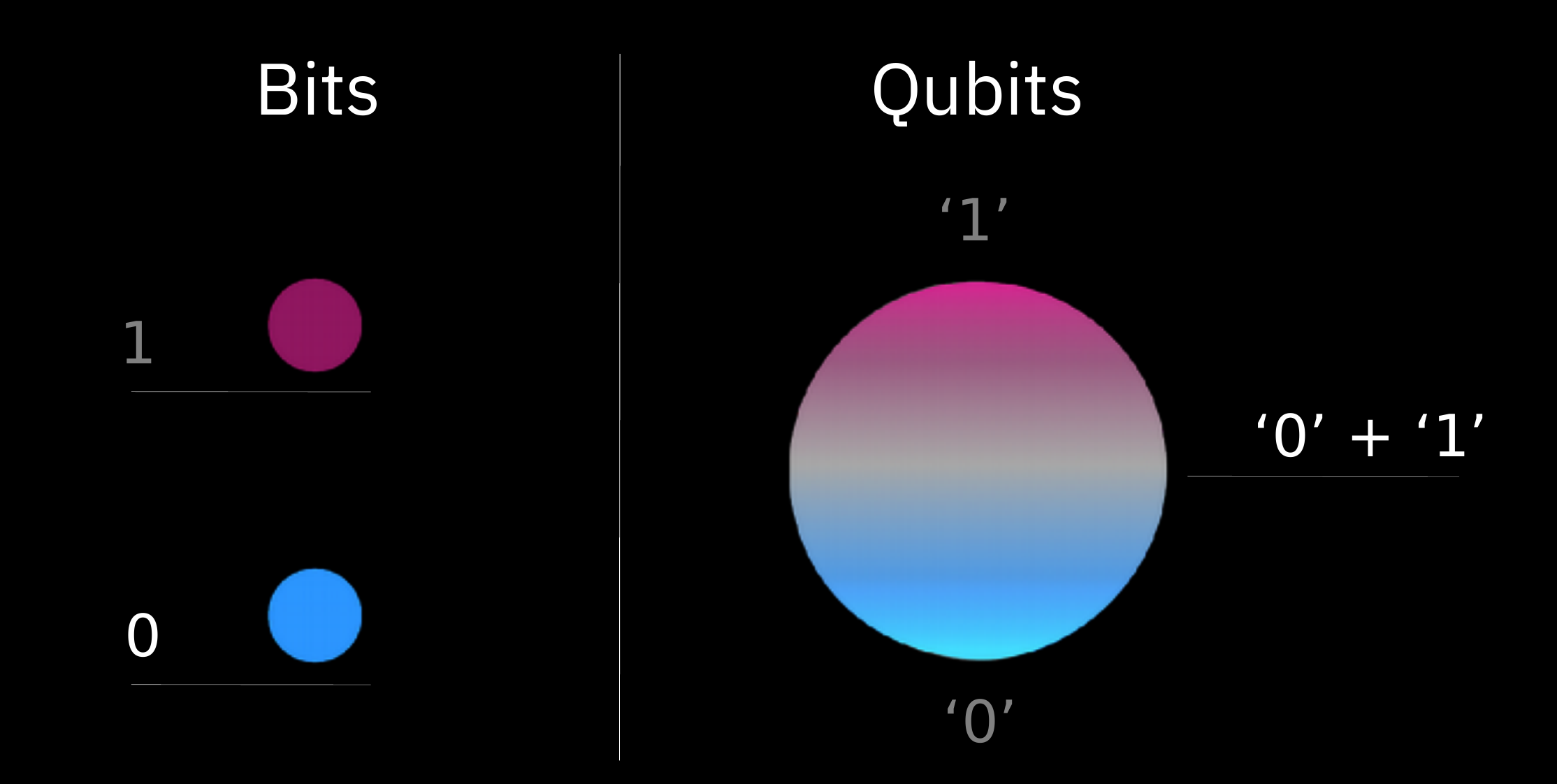}
\caption{Suggestive depiction of a \emph{qubit} as a \emph{superposition} of classical bits. Credits: \cite{Mo18}.\label{fig:190416b}}
\end{figure}

Moving further to \emph{quantum computing} corresponds to generalising
probabilities to \emph{amplitudes}, for instance: {\small
\begin{eqnarray}
B = \begin{array}{r|cccc}&\rotatebox{90}{\ensuremath{(\textsf{0},\textsf{0})}}&\rotatebox{90}{\ensuremath{(\textsf{0},\textsf{1})}}&\rotatebox{90}{\ensuremath{(\textsf{1},\textsf{0})}}&\rotatebox{90}{\ensuremath{(\textsf{1},\textsf{1})}}\\\hline \hbox{\ensuremath{(\textsf{0},\textsf{0})}}&\ensuremath{\frac{\mathrm{1}}{\sqrt{\mathrm{2}}}}&0&\ensuremath{\frac{\mathrm{1}}{\sqrt{\mathrm{2}}}}&0\\\hbox{\ensuremath{(\textsf{0},\textsf{1})}}&0&\ensuremath{\frac{\mathrm{1}}{\sqrt{\mathrm{2}}}}&0&\ensuremath{\frac{\mathrm{1}}{\sqrt{\mathrm{2}}}}\\\hbox{\ensuremath{(\textsf{1},\textsf{0})}}&0&\ensuremath{\frac{\mathrm{1}}{\sqrt{\mathrm{2}}}}&0&\ensuremath{\mathbin{-}\frac{\mathrm{1}}{\sqrt{\mathrm{2}}}}\\\hbox{\ensuremath{(\textsf{1},\textsf{1})}}&\ensuremath{\frac{\mathrm{1}}{\sqrt{\mathrm{2}}}}&0&\ensuremath{\mathbin{-}\frac{\mathrm{1}}{\sqrt{\mathrm{2}}}}&0
\end{array}
	\label{eq:190426a}
\end{eqnarray}}%
Amplitudes are \emph{complex} numbers indicating the \emph{superposition}
of information at quantum information level (Figure \ref{fig:190416b}).

Quantum programs (\emph{QP}) are made of elementary units called quantum
\emph{gates}, for instance the \ensuremath{\Conid{T}}-gate,\footnote{
Where \ensuremath{{\Varid{e}}^{\Varid{i}\;\Varid{x}}\mathrel{=}\Varid{cos}\;\Varid{x}\mathbin{+}\Varid{i}\;(\Varid{sin}\;\Varid{x})} -- recall Euler's formula.}
\begin{eqnarray*}
    \ensuremath{\Conid{T}} =
    \begin{array}{r|cc}&\rotatebox{0}{\ensuremath{\textsf{0}}}&\rotatebox{0}{\ensuremath{\textsf{1}}}\\\hline
	\hbox{\ensuremath{\textsf{0}}}&\ensuremath{\mathrm{1}}&\ensuremath{\mathrm{0}}\\\hbox{\ensuremath{\textsf{1}}}&\ensuremath{\mathrm{0}}&\ensuremath{{\Varid{e}}^{\frac{\Varid{i}\;\pi }{\mathrm{4}}}}
    \end{array},
\end{eqnarray*}
and the \emph{Hadamard} gate,
\begin{eqnarray}
	\ensuremath{\Conid{H}}=\begin{array}{r|cc}&\rotatebox{0}{\ensuremath{\textsf{0}}}&\rotatebox{0}{\ensuremath{\textsf{1}}}\\\hline \hbox{\ensuremath{\textsf{0}}}&\ensuremath{\frac{\mathrm{1}}{\sqrt{\mathrm{2}}}}&\ensuremath{\frac{\mathrm{1}}{\sqrt{\mathrm{2}}}}\\\hbox{\ensuremath{\textsf{1}}}&\ensuremath{\frac{\mathrm{1}}{\sqrt{\mathrm{2}}}}&\ensuremath{\mathbin{-}\frac{\mathrm{1}}{\sqrt{\mathrm{2}}}}
\end{array}.
\end{eqnarray}
which is a ``component" of (\ref{eq:190426a}) in a sense to be
made precise soon.

The question is: how do we combine the functional approach of the
previous sections with such a dramatic quantum extension? Reviewing some background material
is required at this point.

All \emph{functions} in the taxonomy of Fig.~\ref{fig:190223a}
form a \emph{category} \cite{CP11} under functional composition,
\ensuremath{(\Varid{f} \comp \Varid{g})\;\Varid{a}\mathrel{=}\Varid{f}\;(\Varid{g}\;\Varid{a})}, with identity \ensuremath{\Varid{id}\;\Varid{a}\mathrel{=}\Varid{a}}. Moreover, all relations in the
same taxonomy form another category, under relational composition,
\ensuremath{\Varid{b}\;(\Conid{R} \comp \Conid{S})\;\Varid{a}\Leftrightarrow\rcb{\exists }{\Varid{c}}{}{\Varid{b}\;\Conid{R}\;\Varid{c}\mathrel{\wedge}\Varid{c}\;\Conid{S}\;\Varid{a}}}.
This includes functions as special case, as already seen.

Both functions and relations can be regarded as \ensuremath{\{\mskip1.5mu \mathrm{0},\mathrm{1}\mskip1.5mu\}}-matrices, provided
\ensuremath{\mathrm{0}} and \ensuremath{\mathrm{1}} are regarded as Boolean values. Interestingly, such matrices can be
extended to arbitrary (typed) matrices where \ensuremath{\mathrm{0}} and \ensuremath{\mathrm{1}} are, respectively,
the unit of addition and of multiplication of a semiring structure.
One such semiring is the field of complex numbers,
allowing us to include matrices such as \ensuremath{\Conid{H}} and \ensuremath{\Conid{B}} above.

This leads us to a \emph{linear algebra of programming} \cite{Ol12},
in which matrices are typed and written in the same way as functions or
relations, i.e.\ as arrows (morphisms) \ensuremath{\larrow{\Conid{A}}{\Conid{M}}{\Conid{B}}}. This denotes a matrix
\ensuremath{\Conid{M}} whose columns are indexed by \ensuremath{\Conid{A}} and rows by \ensuremath{\Conid{B}}. Under matrix
multiplication,
such matrices will form a category, too \cite{MO13c}.

Although the category of (entire, total) functions and categories of matrices
don't exactly have the same properties, they share a common ground of
useful constructs. Coproducts correspond to direct sums of matrices,
denoted by \ensuremath{\Conid{M}\mathbin{\oplus}\Conid{N}}, and there is a \emph{tensor} (\ref{eq:960923c-pw}) given by
the Kronecker product, written \ensuremath{{\Conid{M}}\hskip 1pt\otimes\hskip 1pt{\Conid{N}}}. For instance,
\begin{eqnarray*}
\ensuremath{{\Varid{id}}\hskip 1pt\otimes\hskip 1pt{\Conid{H}}} =\begin{array}{r|cccc}&\rotatebox{90}{\ensuremath{(\textsf{0},\textsf{0})}}&\rotatebox{90}{\ensuremath{(\textsf{0},\textsf{1})}}&\rotatebox{90}{\ensuremath{(\textsf{1},\textsf{0})}}&\rotatebox{90}{\ensuremath{(\textsf{1},\textsf{1})}}\\\hline \hbox{\ensuremath{(\textsf{0},\textsf{0})}}&\ensuremath{\frac{\mathrm{1}}{\sqrt{\mathrm{2}}}}&\ensuremath{\frac{\mathrm{1}}{\sqrt{\mathrm{2}}}}&0&0\\\hbox{\ensuremath{(\textsf{0},\textsf{1})}}&\ensuremath{\frac{\mathrm{1}}{\sqrt{\mathrm{2}}}}&-\ensuremath{\frac{\mathrm{1}}{\sqrt{\mathrm{2}}}}&0&0\\\hbox{\ensuremath{(\textsf{1},\textsf{0})}}&0&0&\ensuremath{\frac{\mathrm{1}}{\sqrt{\mathrm{2}}}}&\ensuremath{\frac{\mathrm{1}}{\sqrt{\mathrm{2}}}}\\\hbox{\ensuremath{(\textsf{1},\textsf{1})}}&0&0&\ensuremath{\frac{\mathrm{1}}{\sqrt{\mathrm{2}}}}&-\ensuremath{\frac{\mathrm{1}}{\sqrt{\mathrm{2}}}}
\end{array}
\end{eqnarray*}

Finally, one can interpret diagrams such as \eqref{eq:190418c} in a category
of matrices, meaning that such categories have \emph{catamorphisms}.
Let us consider two examples of \ensuremath{\quanta{\Varid{x}}}, for two instances of
\ensuremath{\rarrow{\B\times\B}{\Varid{x}}{\B\times\B}}. In the first case, \ensuremath{\Varid{x}\mathrel{=}\Varid{cnot}} and one gets
the typed matrix \ensuremath{\quanta{\Varid{cnot}}} pictured in Fig.~\ref{fig:190418d}.
This matrix clearly shows a (fragment of a) reversible function.
For \ensuremath{\Varid{x}\mathrel{=}\Conid{B}} from \eqref{eq:190426a}, the matrix \ensuremath{\quanta{\Conid{B}}} is depicted in Fig.~\ref{fig:180522b}.

However, what does \ensuremath{\quanta{\Conid{B}}} mean, giving that \ensuremath{\Conid{B}} is a matrix and not a
function?
This leads us to truly quantum \emph{quantamorphisms}, the main topic of the sections that follow.

\begin{figure}\centering\tiny
\(\begin{array}{r|cccccccccccccc}&\rotatebox{90}{\ensuremath{([\mskip1.5mu \mskip1.5mu],\textsf{0})}}&\rotatebox{90}{\ensuremath{([\mskip1.5mu \mskip1.5mu],\textsf{1})}}&\rotatebox{90}{\ensuremath{([\mskip1.5mu \textsf{0}\mskip1.5mu],\textsf{0})}}&\rotatebox{90}{\ensuremath{([\mskip1.5mu \textsf{0}\mskip1.5mu],\textsf{1})}}&\rotatebox{90}{\ensuremath{([\mskip1.5mu \textsf{0},\textsf{0}\mskip1.5mu],\textsf{0})}}&\rotatebox{90}{\ensuremath{([\mskip1.5mu \textsf{0},\textsf{0}\mskip1.5mu],\textsf{1})}}&\rotatebox{90}{\ensuremath{([\mskip1.5mu \textsf{1},\textsf{0}\mskip1.5mu],\textsf{0})}}&\rotatebox{90}{\ensuremath{([\mskip1.5mu \textsf{1},\textsf{0}\mskip1.5mu],\textsf{1})}}&\rotatebox{90}{\ensuremath{([\mskip1.5mu \textsf{1}\mskip1.5mu],\textsf{0})}}&\rotatebox{90}{\ensuremath{([\mskip1.5mu \textsf{1}\mskip1.5mu],\textsf{1})}}&\rotatebox{90}{\ensuremath{([\mskip1.5mu \textsf{0},\textsf{1}\mskip1.5mu],\textsf{0})}}&\rotatebox{90}{\ensuremath{([\mskip1.5mu \textsf{0},\textsf{1}\mskip1.5mu],\textsf{1})}}&\rotatebox{90}{\ensuremath{([\mskip1.5mu \textsf{1},\textsf{1}\mskip1.5mu],\textsf{0})}}&\rotatebox{90}{\ensuremath{([\mskip1.5mu \textsf{1},\textsf{1}\mskip1.5mu],\textsf{1})}}\\\hline \hbox{\ensuremath{([\mskip1.5mu \mskip1.5mu],\textsf{0})}}&1&0&0&0&0&0&0&0&0&0&0&0&0&0\\\hbox{\ensuremath{([\mskip1.5mu \mskip1.5mu],\textsf{1})}}&0&1&0&0&0&0&0&0&0&0&0&0&0&0\\\hbox{\ensuremath{([\mskip1.5mu \textsf{0}\mskip1.5mu],\textsf{0})}}&0&0&1&0&0&0&0&0&0&0&0&0&0&0\\\hbox{\ensuremath{([\mskip1.5mu \textsf{0}\mskip1.5mu],\textsf{1})}}&0&0&0&1&0&0&0&0&0&0&0&0&0&0\\\hbox{\ensuremath{([\mskip1.5mu \textsf{0},\textsf{0}\mskip1.5mu],\textsf{0})}}&0&0&0&0&1&0&0&0&0&0&0&0&0&0\\\hbox{\ensuremath{([\mskip1.5mu \textsf{0},\textsf{0}\mskip1.5mu],\textsf{1})}}&0&0&0&0&0&1&0&0&0&0&0&0&0&0\\\hbox{\ensuremath{([\mskip1.5mu \textsf{1},\textsf{0}\mskip1.5mu],\textsf{0})}}&0&0&0&0&0&0&0&1&0&0&0&0&0&0\\\hbox{\ensuremath{([\mskip1.5mu \textsf{1},\textsf{0}\mskip1.5mu],\textsf{1})}}&0&0&0&0&0&0&1&0&0&0&0&0&0&0\\\hbox{\ensuremath{([\mskip1.5mu \textsf{1}\mskip1.5mu],\textsf{0})}}&0&0&0&0&0&0&0&0&0&1&0&0&0&0\\\hbox{\ensuremath{([\mskip1.5mu \textsf{1}\mskip1.5mu],\textsf{1})}}&0&0&0&0&0&0&0&0&1&0&0&0&0&0\\\hbox{\ensuremath{([\mskip1.5mu \textsf{0},\textsf{1}\mskip1.5mu],\textsf{0})}}&0&0&0&0&0&0&0&0&0&0&0&1&0&0\\\hbox{\ensuremath{([\mskip1.5mu \textsf{0},\textsf{1}\mskip1.5mu],\textsf{1})}}&0&0&0&0&0&0&0&0&0&0&1&0&0&0\\\hbox{\ensuremath{([\mskip1.5mu \textsf{1},\textsf{1}\mskip1.5mu],\textsf{0})}}&0&0&0&0&0&0&0&0&0&0&0&0&1&0\\\hbox{\ensuremath{([\mskip1.5mu \textsf{1},\textsf{1}\mskip1.5mu],\textsf{1})}}&0&0&0&0&0&0&0&0&0&0&0&0&0&1
\end{array}
\)
	\caption{\footnotesize\label{fig:190418d}Matrix for \ensuremath{\quanta{\Varid{cnot}}} truncated to input (output) lists of maximum length \ensuremath{\mathrm{2}} for visualisation purposes.}
\end{figure}

\begin{figure*}\centering\footnotesize
\(
\begin{array}{r|rrrrrrrrrrrrrr}&\rotatebox{90}{\ensuremath{([\mskip1.5mu \mskip1.5mu],\textsf{0})}}&\rotatebox{90}{\ensuremath{([\mskip1.5mu \mskip1.5mu],\textsf{1})}}&\rotatebox{90}{\ensuremath{([\mskip1.5mu \textsf{0}\mskip1.5mu],\textsf{0})}}&\rotatebox{90}{\ensuremath{([\mskip1.5mu \textsf{0}\mskip1.5mu],\textsf{1})}}&\rotatebox{90}{\ensuremath{([\mskip1.5mu \textsf{0},\textsf{0}\mskip1.5mu],\textsf{0})}}&\rotatebox{90}{\ensuremath{([\mskip1.5mu \textsf{0},\textsf{0}\mskip1.5mu],\textsf{1})}}&\rotatebox{90}{\ensuremath{([\mskip1.5mu \textsf{1},\textsf{0}\mskip1.5mu],\textsf{0})}}&\rotatebox{90}{\ensuremath{([\mskip1.5mu \textsf{1},\textsf{0}\mskip1.5mu],\textsf{1})}}&\rotatebox{90}{\ensuremath{([\mskip1.5mu \textsf{1}\mskip1.5mu],\textsf{0})}}&\rotatebox{90}{\ensuremath{([\mskip1.5mu \textsf{1}\mskip1.5mu],\textsf{1})}}&\rotatebox{90}{\ensuremath{([\mskip1.5mu \textsf{0},\textsf{1}\mskip1.5mu],\textsf{0})}}&\rotatebox{90}{\ensuremath{([\mskip1.5mu \textsf{0},\textsf{1}\mskip1.5mu],\textsf{1})}}&\rotatebox{90}{\ensuremath{([\mskip1.5mu \textsf{1},\textsf{1}\mskip1.5mu],\textsf{0})}}&\rotatebox{90}{\ensuremath{([\mskip1.5mu \textsf{1},\textsf{1}\mskip1.5mu],\textsf{1})}}\\\hline \hbox{\ensuremath{([\mskip1.5mu \mskip1.5mu],\textsf{0})}}&1&0&0&0&0&0&0&0&0&0&0&0&0&0\\\hbox{\ensuremath{([\mskip1.5mu \mskip1.5mu],\textsf{1})}}&0&1&0&0&0&0&0&0&0&0&0&0&0&0\\\hbox{\ensuremath{([\mskip1.5mu \textsf{0}\mskip1.5mu],\textsf{0})}}&0&0&\ensuremath{\frac{\mathrm{1}}{\sqrt{\mathrm{2}}}}&0&0&0&0&0&\ensuremath{\frac{\mathrm{1}}{\sqrt{\mathrm{2}}}}&0&0&0&0&0\\\hbox{\ensuremath{([\mskip1.5mu \textsf{0}\mskip1.5mu],\textsf{1})}}&0&0&0&\ensuremath{\frac{\mathrm{1}}{\sqrt{\mathrm{2}}}}&0&0&0&0&0&\ensuremath{\frac{\mathrm{1}}{\sqrt{\mathrm{2}}}}&0&0&0&0\\\hbox{\ensuremath{([\mskip1.5mu \textsf{0},\textsf{0}\mskip1.5mu],\textsf{0})}}&0&0&0&0&\ensuremath{\frac{\mathrm{1}}{\mathrm{2}}}&0&\ensuremath{\frac{\mathrm{1}}{\mathrm{2}}}&0&0&0&\ensuremath{\frac{\mathrm{1}}{\mathrm{2}}}&0&\ensuremath{\frac{\mathrm{1}}{\mathrm{2}}}&0\\\hbox{\ensuremath{([\mskip1.5mu \textsf{0},\textsf{0}\mskip1.5mu],\textsf{1})}}&0&0&0&0&0&\ensuremath{\frac{\mathrm{1}}{\mathrm{2}}}&0&\ensuremath{\frac{\mathrm{1}}{\mathrm{2}}}&0&0&0&\ensuremath{\frac{\mathrm{1}}{\mathrm{2}}}&0&\ensuremath{\frac{\mathrm{1}}{\mathrm{2}}}\\\hbox{\ensuremath{([\mskip1.5mu \textsf{1},\textsf{0}\mskip1.5mu],\textsf{0})}}&0&0&0&0&0&\ensuremath{\frac{\mathrm{1}}{\mathrm{2}}}&0&-\ensuremath{\frac{\mathrm{1}}{\mathrm{2}}}&0&0&0&\ensuremath{\frac{\mathrm{1}}{\mathrm{2}}}&0&-\ensuremath{\frac{\mathrm{1}}{\mathrm{2}}}\\\hbox{\ensuremath{([\mskip1.5mu \textsf{1},\textsf{0}\mskip1.5mu],\textsf{1})}}&0&0&0&0&\ensuremath{\frac{\mathrm{1}}{\mathrm{2}}}&0&-\ensuremath{\frac{\mathrm{1}}{\mathrm{2}}}&0&0&0&\ensuremath{\frac{\mathrm{1}}{\mathrm{2}}}&0&-\ensuremath{\frac{\mathrm{1}}{\mathrm{2}}}&0\\\hbox{\ensuremath{([\mskip1.5mu \textsf{1}\mskip1.5mu],\textsf{0})}}&0&0&0&\ensuremath{\frac{\mathrm{1}}{\sqrt{\mathrm{2}}}}&0&0&0&0&0&\ensuremath{\mathbin{-}\frac{\mathrm{1}}{\sqrt{\mathrm{2}}}}&0&0&0&0\\\hbox{\ensuremath{([\mskip1.5mu \textsf{1}\mskip1.5mu],\textsf{1})}}&0&0&\ensuremath{\frac{\mathrm{1}}{\sqrt{\mathrm{2}}}}&0&0&0&0&0&\ensuremath{\mathbin{-}\frac{\mathrm{1}}{\sqrt{\mathrm{2}}}}&0&0&0&0&0\\\hbox{\ensuremath{([\mskip1.5mu \textsf{0},\textsf{1}\mskip1.5mu],\textsf{0})}}&0&0&0&0&0&\ensuremath{\frac{\mathrm{1}}{\mathrm{2}}}&0&\ensuremath{\frac{\mathrm{1}}{\mathrm{2}}}&0&0&0&-\ensuremath{\frac{\mathrm{1}}{\mathrm{2}}}&0&-\ensuremath{\frac{\mathrm{1}}{\mathrm{2}}}\\\hbox{\ensuremath{([\mskip1.5mu \textsf{0},\textsf{1}\mskip1.5mu],\textsf{1})}}&0&0&0&0&\ensuremath{\frac{\mathrm{1}}{\mathrm{2}}}&0&\ensuremath{\frac{\mathrm{1}}{\mathrm{2}}}&0&0&0&-\ensuremath{\frac{\mathrm{1}}{\mathrm{2}}}&0&-\ensuremath{\frac{\mathrm{1}}{\mathrm{2}}}&0\\\hbox{\ensuremath{([\mskip1.5mu \textsf{1},\textsf{1}\mskip1.5mu],\textsf{0})}}&0&0&0&0&\ensuremath{\frac{\mathrm{1}}{\mathrm{2}}}&0&-\ensuremath{\frac{\mathrm{1}}{\mathrm{2}}}&0&0&0&-\ensuremath{\frac{\mathrm{1}}{\mathrm{2}}}&0&\ensuremath{\frac{\mathrm{1}}{\mathrm{2}}}&0\\\hbox{\ensuremath{([\mskip1.5mu \textsf{1},\textsf{1}\mskip1.5mu],\textsf{1})}}&0&0&0&0&0&\ensuremath{\frac{\mathrm{1}}{\mathrm{2}}}&0&-\ensuremath{\frac{\mathrm{1}}{\mathrm{2}}}&0&0&0&-\ensuremath{\frac{\mathrm{1}}{\mathrm{2}}}&0&\ensuremath{\frac{\mathrm{1}}{\mathrm{2}}}
\end{array}
\)
	\caption{\footnotesize\label{fig:180522b}Matrix for \ensuremath{\quanta{\Conid{B}}} in the same range as Fig.~\ref{fig:190418d}.}
\end{figure*}

\section{Quantum abstraction}
It is well known that every relation \ensuremath{\Conid{R}\mathbin{:}\Conid{A}\to \Conid{B}} can be represented
faithfully as a set-valued function \ensuremath{\Lambda{\Conid{R}}\mathbin{:}\Conid{A}\to \fun P \;\Conid{B}}, where \ensuremath{\fun P \;\Conid{B}} denotes
the powerset of \ensuremath{\Conid{B}}, under the correspondence: \ensuremath{\Varid{b}\;\Conid{R}\;\Varid{a}\Leftrightarrow\Varid{b}\;{\in}\;\Lambda{\Conid{R}}\;\Varid{a}}.
Under this correspondence, relation composition of \ensuremath{\Conid{R}\mathbin{:}\Conid{A}\to \Conid{B}} and \ensuremath{\Conid{S}\mathbin{:}\Conid{B}\to \Conid{C}} is given by
\begin{eqnarray*}
\ensuremath{\Varid{c}\;(\Conid{S} \comp \Conid{R})\;\Varid{a}\Leftrightarrow\Varid{c}\;{\in}\;{\bigcup}\;\{\mskip1.5mu \Lambda{\Conid{S}}\;\Varid{b}\mid \Varid{b}\;{\in}\;\Lambda{\Conid{R}}\;\Varid{a}\mskip1.5mu\}} .
\end{eqnarray*}
This can be written in a more generic way.
Let \ensuremath{\Varid{f}} and \ensuremath{\Varid{g}} abbreviate \ensuremath{\Lambda{\Conid{R}}} and \ensuremath{\Lambda{\Conid{S}}}, respectively, with types
\ensuremath{\Varid{f}\mathbin{:}\Conid{A}\to \fun P \;\Conid{B}} and \ensuremath{\Varid{g}\mathbin{:}\Conid{B}\to \fun P \;\Conid{C}}.
Then composition \ensuremath{\Conid{S} \comp \Conid{R}} is represented by \ensuremath{\Varid{g}\kcomp\Varid{f}}, of type \ensuremath{\Conid{A}\to \fun P \;\Conid{C}},
defined \emph{monadically} by
\begin{eqnarray*}
	\ensuremath{(\Varid{g}\kcomp\Varid{f})\;\Varid{a}\mathrel{=}\mathbf{do}\;\{\mskip1.5mu \Varid{b}\leftarrow \Varid{f}\;\Varid{a};\Varid{g}\;\Varid{b}\mskip1.5mu\}}
\end{eqnarray*}
Where does this terminology and notation come from?

It turns out that \ensuremath{\fun P \;\Conid{X}} is a \emph{monad} \cite{GH11} in the category of sets,
and monads induce particular categories known as \emph{Kleisli} categories.
The category of sets and relations ``is" the Kleisli category induced by the
monad \ensuremath{\fun P } in the original category of sets and functions, where it is
represented by \ensuremath{\fun P }-valued functions composed as \ensuremath{\Varid{g}\kcomp\Varid{f}} above. Comparing
such a definition of composition with that of arbitrary functions in the
original category, namely
\begin{eqnarray*}
	\ensuremath{(\Varid{g} \comp \Varid{f})\;\Varid{a}\mathrel{=}\mathbf{let}\;\Varid{b}\mathrel{=}\Varid{f}\;\Varid{a}\;\mathbf{in}\;\Varid{g}\;\Varid{b}},
\end{eqnarray*}
one immediately sees how the monadic \ensuremath{\mathbf{do}} notation generalises the \ensuremath{\mathbf{let}}
notation used in ordinary mathematics and programming languages.

On the other hand, any function \ensuremath{\Varid{f}\mathbin{:}\Conid{A}\to \Conid{B}} can be represented in the Kleisli
category by \ensuremath{\Varid{f'}\mathrel{=}\mathbf{ret}  \comp \Varid{f}}, where \ensuremath{\mathbf{ret} \mathbin{:}\Conid{A}\to \fun P \;\Conid{A}} is the function
that yields the smallest set that contains its argument, \ensuremath{\mathbf{ret} \;\Varid{a}\mathrel{=}\{\mskip1.5mu \Varid{a}\mskip1.5mu\}}.\footnote{For economy of space in the mathematical layout, \ensuremath{\mathbf{ret} } abbreviates the more usual \textsf{return} keyword.}

\paragraph{Matrices and the vector space monad}
In the same way that a relation can be faithfully represented by a set-valued
function, any matrix can be represented by a vector-valued function.
Each such vector corresponds to a column of the original matrix.
For instance, the \emph{Hadamard} gate
\begin{eqnarray}
H =\begin{array}{r|cc}&\rotatebox{0}{\ensuremath{\textsf{0}}}&\rotatebox{0}{\ensuremath{\textsf{1}}}\\\hline \hbox{\ensuremath{\textsf{0}}}&\ensuremath{\frac{\mathrm{1}}{\sqrt{\mathrm{2}}}}&\ensuremath{\frac{\mathrm{1}}{\sqrt{\mathrm{2}}}}\\\hbox{\ensuremath{\textsf{1}}}&\ensuremath{\frac{\mathrm{1}}{\sqrt{\mathrm{2}}}}&\ensuremath{\mathbin{-}\frac{\mathrm{1}}{\sqrt{\mathrm{2}}}}
\end{array}
	\label{eq:H}
\end{eqnarray}
is represented by the function
\begin{eqnarray*}
\start \ensuremath{\Varid{had}\mathbin{::}\B\to \Conid{Vec}\;\B}
\more \ensuremath{\Varid{had}\;\textsf{0}\mathrel{=}\begin{bmatrix}\frac{\mathrm{1}}{\sqrt{\mathrm{2}}}\\\frac{\mathrm{1}}{\sqrt{\mathrm{2}}}\end{bmatrix}}
\more \ensuremath{\Varid{had}\;\textsf{1}\mathrel{=}\begin{bmatrix}\frac{\mathrm{1}}{\sqrt{\mathrm{2}}}\\\mathbin{-}\frac{\mathrm{1}}{\sqrt{\mathrm{2}}}\end{bmatrix}}
\end{eqnarray*}
That is to say, the\footnote{The use of definite article ``the" arises from
a simplification in the current paper: all matrices are regarded as complex number valued.
In general, the abstract notion of a vector space (or, more generally, of a semimodule) is doubly
parametric: both its basis and its underlying semiring can change \cite{laop-0.1.0.7}.

Another important assumption is that every vector \ensuremath{\Varid{v}\;{\in}\;\Conid{Vec}\;\Conid{X}} is supposed to have \emph{finite} support, that is, the number of nonzero entries in \ensuremath{\Varid{v}} is finite.}
category of matrices is the Kleisli category induced
by the monad \ensuremath{\Conid{Vec}}, where \ensuremath{\Conid{Vec}\;\Conid{X}} denotes the type of all vectors with basis
\ensuremath{\Conid{X}}. This means that for each \ensuremath{\Varid{x}\;{\in}\;\Conid{X}} there is a column vector \ensuremath{\mathbf{ret} \;\Varid{x}}
such that \ensuremath{\mathbf{ret} \mathbin{:}\Conid{X}\to \Conid{Vec}\;\Conid{X}} represents the identity matrix \ensuremath{\Varid{id}\mathbin{:}\Conid{X}\to \Conid{X}}.

In the quantum field, the ``Dirac notation" \ensuremath{{|\Varid{x}\rangle}} usually replaces
\ensuremath{\mathbf{ret} \;\Varid{x}}, as in
\begin{eqnarray*}
	\ensuremath{{|\textsf{0}\rangle}\mathrel{=}\begin{bmatrix}\mathrm{1}\\\mathrm{0}\end{bmatrix}} ~~~~~
	\ensuremath{{|\textsf{1}\rangle}\mathrel{=}\begin{bmatrix}\mathrm{0}\\\mathrm{1}\end{bmatrix}}
\end{eqnarray*}
for \ensuremath{\Conid{X}\mathrel{=}\B}. Using this notation, the Hadamard gate can be redefined
as follows:
\begin{eqnarray}
\arrayin{
\start \ensuremath{\Varid{had}\mathbin{::}\B\to \Conid{Vec}\;\B}
\more \ensuremath{\Varid{had}\;\textsf{0}\mathrel{=}\frac{{{|\textsf{0}\rangle}} + {{|\textsf{1}\rangle}}}{\sqrt{\mathrm{2}}}}
\more \ensuremath{\Varid{had}\;\textsf{1}\mathrel{=}\frac{{{|\textsf{0}\rangle}} - {{|\textsf{1}\rangle}}}{\sqrt{\mathrm{2}}}}
}
	\label{eq:had}
\end{eqnarray}
The inhabitants of type \ensuremath{\Conid{Vec}\;\B} are usually known as \emph{qubits},
generalising classical \ensuremath{\Varid{bits}} (Fig.~\ref{fig:190416b}). Bits are therefore special cases of qubits:
\ensuremath{{|\textsf{0}\rangle}} and \ensuremath{{|\textsf{1}\rangle}} are classical, while e.g.
	\ensuremath{\frac{{{|\textsf{0}\rangle}} + {{|\textsf{1}\rangle}}}{\sqrt{\mathrm{2}}}} and
	\ensuremath{\frac{{{|\textsf{0}\rangle}} - {{|\textsf{1}\rangle}}}{\sqrt{\mathrm{2}}}}
are \emph{superpositions} of \ensuremath{{|\textsf{0}\rangle}} and \ensuremath{{|\textsf{1}\rangle}}.

By Kleisli correspondence, all matrix-categorical operations can be
encoded monadically, as for instance in the following definition of the
Kronecker (tensor) product
\begin{eqnarray}
\hskip -3em\mbox{
\begin{minipage}{0.4\textwidth}
\begin{hscode}\SaveRestoreHook
\column{B}{@{}>{\hspre}l<{\hspost}@{}}%
\column{6}{@{}>{\hspre}l<{\hspost}@{}}%
\column{E}{@{}>{\hspre}l<{\hspost}@{}}%
\>[B]{}({\Varid{f}}\hskip 1pt\otimes\hskip 1pt{\Varid{g}})\;(\Varid{a},\Varid{b})\mathrel{=}\mathbf{do}\;\{\mskip1.5mu {}\<[E]%
\\
\>[B]{}\hsindent{6}{}\<[6]%
\>[6]{}\Varid{x}\leftarrow \Varid{f}\;\Varid{a};{}\<[E]%
\\
\>[B]{}\hsindent{6}{}\<[6]%
\>[6]{}\Varid{y}\leftarrow \Varid{g}\;\Varid{b};{}\<[E]%
\\
\>[B]{}\hsindent{6}{}\<[6]%
\>[6]{}\mathbf{ret} \;(\Varid{x},\Varid{y})\mskip1.5mu\}{}\<[E]%
\ColumnHook
\end{hscode}\resethooks
\end{minipage}
} \label{eq:kronecker} 
\end{eqnarray}
where, for \ensuremath{\Varid{f}\mathbin{:}\Conid{A}\to \Conid{Vec}\;\Conid{X}} and \ensuremath{\Varid{g}\mathbin{:}\Conid{B}\to \Conid{Vec}\;\Conid{Y}}, the function
\ensuremath{{\Varid{f}}\hskip 1pt\otimes\hskip 1pt{\Varid{g}}} is of type \ensuremath{(\Conid{A}\times\Conid{B})\to \Conid{Vec}\;(\Conid{X}\times\Conid{Y})}.

Let us see this representation at work by looking at the
\emph{structure} of a famous example in quantum computing --
the ``Alice" part of the teleportation protocol \cite{NC11}:

\begin{center}
\def\x{\(\xymatrix{
c \ar@{-}[rrr] & & & *-{\bullet} \ar@{-}[d] \ar@{-}[r] & *+[F]{H} \ar@{-}[r] &  c'
\\
a \ar@{-}[r] & *+[F]{H} \ar@{-}[rrrr] & *-{\bullet} \ar@{-}[d] & *{\oplus}  &  & a'
\\
b \ar@{-}[rrrrr] & & *{\oplus} & & & b'
}\)}
\mbox{
\begin{picture}(190.00,110.00)
\put(00.00,90.00){\x}
\put(100.00,45.00){\dashbox{2.00}(60.00,60.00)[cc]{}}
\put(25.00,05.00){\dashbox{2.00}(60.00,60.00)[cc]{}}
\end{picture}
}
\end{center}
The first block, marked by the dashed square covering inputs \ensuremath{\Varid{a}} and \ensuremath{\Varid{b}},
is the matrix \ensuremath{\Conid{B}\mathrel{=}\Varid{cnot} \comp ({\Conid{H}}\hskip 1pt\otimes\hskip 1pt{\Varid{id}})} -- recall \eqref{eq:190426a} --
which creates a so-called Bell state. Let \ensuremath{\Varid{bell}\mathrel{=}\Lambda{\Conid{B}}}.
As \ensuremath{\Varid{cnot}} is classical, we have to use \ensuremath{\mathbf{ret}  \comp \Varid{cnot}} in the monadic encoding
of \ensuremath{\Conid{B}}:
\begin{eqnarray}
	\ensuremath{\Varid{bell}\;(\Varid{a},\Varid{b})\mathrel{=}\mathbf{do}\;\{\mskip1.5mu \Varid{x}\leftarrow \Varid{had}\;\Varid{a};\mathbf{ret} \;(\Varid{cnot}\;(\Varid{x},\Varid{b}))\mskip1.5mu\}}
	\label{eq:190425a}
\end{eqnarray}
(Details in the appendix.) Then the second block, marked by the other dashed square, is \ensuremath{{\Conid{B}}^{\dagger}\mathrel{=}({\Conid{H}}\hskip 1pt\otimes\hskip 1pt{\Varid{id}}) \comp \Varid{cnot}}, where \ensuremath{{\Conid{X}}^{\dagger}} is the \emph{conjugate} transpose of \ensuremath{\Conid{X}}.\footnote{Recall that \ensuremath{{\Conid{X}}^{\dagger}} coincides with \ensuremath{\conv{\Conid{X}}} wherever \ensuremath{\Conid{X}} does not involve imaginary parts.}
This, using the same
encoding rules, is represented by:
\begin{hscode}\SaveRestoreHook
\column{B}{@{}>{\hspre}l<{\hspost}@{}}%
\column{5}{@{}>{\hspre}l<{\hspost}@{}}%
\column{E}{@{}>{\hspre}l<{\hspost}@{}}%
\>[B]{}\Varid{unbell}\;(\Varid{c},\Varid{a})\mathrel{=}\mathbf{let}\;(\anonymous ,\Varid{a'})\mathrel{=}\Varid{cnot}\;(\Varid{c},\Varid{a}){}\<[E]%
\\
\>[B]{}\hsindent{5}{}\<[5]%
\>[5]{}\mathbf{in}\;\mathbf{do}\;\{\mskip1.5mu \Varid{b}\leftarrow \Varid{h}\;\Varid{c};\mathbf{ret} \;(\Varid{b},\Varid{a'})\mskip1.5mu\}{}\<[E]%
\ColumnHook
\end{hscode}\resethooks
Then the two blocks are put together via the associator isomorphism \ensuremath{\mathsf{a}},
recall \eqref{eq:180318b}:
\begin{center}
	\ensuremath{\Conid{A}\mathrel{=}({\Varid{unbell}}\hskip 1pt\otimes\hskip 1pt{\Varid{id}}) \comp \mathsf{a} \comp ({\Varid{id}}\hskip 1pt\otimes\hskip 1pt{\Varid{bell}})}
\end{center}
Finally, \ensuremath{\Varid{alice}\mathrel{=}\Lambda{\Conid{A}}} becomes the monadic function:
\begin{hscode}\SaveRestoreHook
\column{B}{@{}>{\hspre}l<{\hspost}@{}}%
\column{5}{@{}>{\hspre}l<{\hspost}@{}}%
\column{8}{@{}>{\hspre}c<{\hspost}@{}}%
\column{8E}{@{}l@{}}%
\column{9}{@{}>{\hspre}l<{\hspost}@{}}%
\column{18}{@{}>{\hspre}l<{\hspost}@{}}%
\column{E}{@{}>{\hspre}l<{\hspost}@{}}%
\>[B]{}\Varid{alice}\;(\Varid{c},(\Varid{a},\Varid{b}))\mathrel{=}{}\<[E]%
\\
\>[B]{}\hsindent{5}{}\<[5]%
\>[5]{}\mathbf{do}\;\{\mskip1.5mu {}\<[E]%
\\
\>[5]{}\hsindent{4}{}\<[9]%
\>[9]{}(\Varid{a'},\Varid{b'}){}\<[18]%
\>[18]{}\leftarrow \Varid{bell}\;(\Varid{a},\Varid{b});{}\<[E]%
\\
\>[5]{}\hsindent{4}{}\<[9]%
\>[9]{}(\Varid{c'},\Varid{a''})\leftarrow \Varid{unbell}\;(\Varid{c},\Varid{a'});{}\<[E]%
\\
\>[5]{}\hsindent{4}{}\<[9]%
\>[9]{}\mathbf{ret} \;(\Varid{c'},(\Varid{a''},\Varid{b'})){}\<[E]%
\\
\>[5]{}\hsindent{3}{}\<[8]%
\>[8]{}\mskip1.5mu\}{}\<[8E]%
\ColumnHook
\end{hscode}\resethooks

\section{Conditional control} \label{sec:191029b}
\begin{figure}
\begin{center}
	{\includegraphics[width=0.3\textwidth]{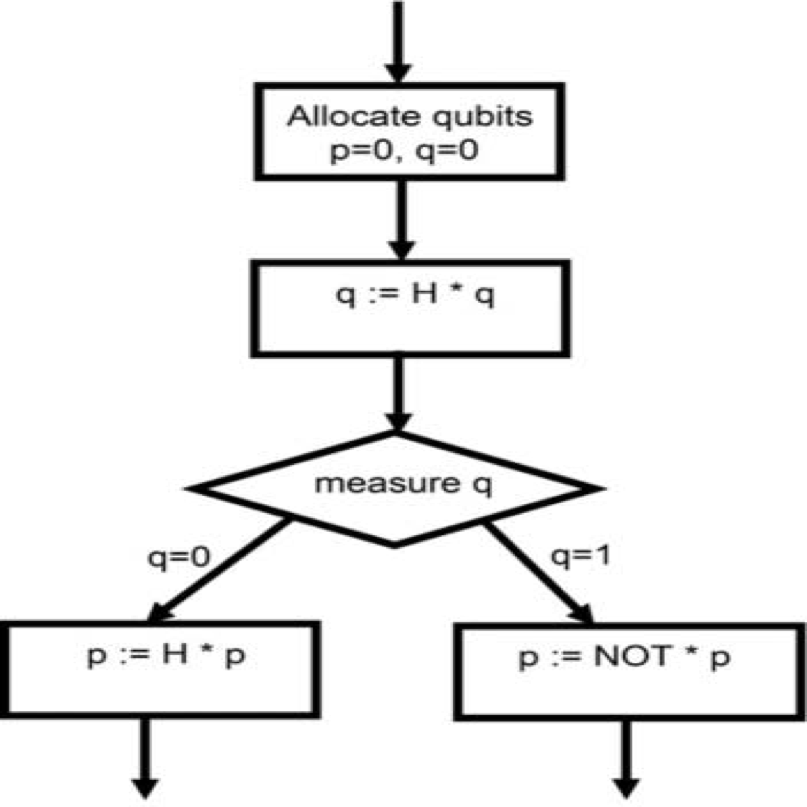}}
\end{center}
	\caption{\footnotesize Conditional flowchart in QFC, taken from Fig.7.4 of \cite{YM08}, page 236. H is the Hadamard gate.} \label{fig:191029a}
\end{figure}

Figure \ref{fig:191029a} quotes a conditional flowchart expressed in the
functional quantum programming language QFC \cite{Se04}. Note how the conditional
control involves a measurement, and thus happens at the \emph{classical}
level, with the subsequent branch being chosen depending on the (classical)
outcome of said measurement.

It turns out that there is a different kind
of conditional quantum control which does not require measuring the control qubit, and
which provides a useful construct for quantum programming.
The quantum programming language QML \cite{AG05} was the first to support this kind of control.
For instance,
the following monadic program
\begin{eqnarray}
\hskip -3em\mbox{
\begin{minipage}{0.4\textwidth}
\begin{hscode}\SaveRestoreHook
\column{B}{@{}>{\hspre}l<{\hspost}@{}}%
\column{14}{@{}>{\hspre}l<{\hspost}@{}}%
\column{52}{@{}>{\hspre}l<{\hspost}@{}}%
\column{E}{@{}>{\hspre}l<{\hspost}@{}}%
\>[B]{}\Varid{cond}\;(\Varid{q},\Varid{p})\mathrel{=}\mathbf{do}\;\{\mskip1.5mu {}\<[E]%
\\
\>[B]{}\hsindent{14}{}\<[14]%
\>[14]{}\Varid{q'}\leftarrow \Varid{had}\;\Varid{q};{}\<[E]%
\\
\>[B]{}\hsindent{14}{}\<[14]%
\>[14]{}\Varid{p'}\leftarrow \mathbf{if}\;\Varid{q'}\;\mathbf{then}\;\mathbf{ret} \;(\neg \;\Varid{p})\;\mathbf{else}\;{}\<[52]%
\>[52]{}\Varid{had}\;\Varid{p};{}\<[E]%
\\
\>[B]{}\hsindent{14}{}\<[14]%
\>[14]{}\mathbf{ret} \;(\Varid{q'},\Varid{p'})\mskip1.5mu\}{}\<[E]%
\ColumnHook
\end{hscode}\resethooks
\end{minipage}
} \label{eq:cond} 
\end{eqnarray}
-- for \ensuremath{\Varid{had}} the Hadamard gate, recall (\ref{eq:had}) -- encodes in Haskell an analogue
of Figure \ref{fig:191029a} but using such a form of quantum conditional control.
This piece of code implements the following unitary matrix:
\begin{eqnarray*}
\begin{array}{r|rrrr}&\rotatebox{90}{\ensuremath{(\textsf{0},\textsf{0})}}&\rotatebox{90}{\ensuremath{(\textsf{0},\textsf{1})}}&\rotatebox{90}{\ensuremath{(\textsf{1},\textsf{0})}}&\rotatebox{90}{\ensuremath{(\textsf{1},\textsf{1})}}\\\hline \hbox{\ensuremath{(\textsf{0},\textsf{0})}}&\ensuremath{\frac{\mathrm{1}}{\mathrm{2}}}&\ensuremath{\frac{\mathrm{1}}{\mathrm{2}}}&\ensuremath{\frac{\mathrm{1}}{\mathrm{2}}}&\ensuremath{\frac{\mathrm{1}}{\mathrm{2}}}\\\hbox{\ensuremath{(\textsf{0},\textsf{1})}}&0&0&\ensuremath{\frac{\sqrt{\mathrm{2}}}{\mathrm{2}}}&\ensuremath{\mathbin{-}\frac{\sqrt{\mathrm{2}}}{\mathrm{2}}}\\\hbox{\ensuremath{(\textsf{1},\textsf{0})}}&\ensuremath{\frac{\mathrm{1}}{\mathrm{2}}}&\ensuremath{\frac{\mathrm{1}}{\mathrm{2}}}&-\ensuremath{\frac{\mathrm{1}}{\mathrm{2}}}&-\ensuremath{\frac{\mathrm{1}}{\mathrm{2}}}\\\hbox{\ensuremath{(\textsf{1},\textsf{1})}}&\ensuremath{\frac{\sqrt{\mathrm{2}}}{\mathrm{2}}}&\ensuremath{\mathbin{-}\frac{\sqrt{\mathrm{2}}}{\mathrm{2}}}&0&0
\end{array}
\end{eqnarray*}

This suggests the following quantum conditional combinator which,
rather than measuring the control bit, implements a superposition of
conditionals:
\begin{hscode}\SaveRestoreHook
\column{B}{@{}>{\hspre}l<{\hspost}@{}}%
\column{9}{@{}>{\hspre}l<{\hspost}@{}}%
\column{E}{@{}>{\hspre}l<{\hspost}@{}}%
\>[B]{}({\Varid{f}}\mathbin\diamond{\Varid{g}})\;(\Varid{x},\Varid{y})\mathrel{=}\mathbf{do}\;\{\mskip1.5mu {}\<[E]%
\\
\>[B]{}\hsindent{9}{}\<[9]%
\>[9]{}\Varid{a}\leftarrow \Varid{x};{}\<[E]%
\\
\>[B]{}\hsindent{9}{}\<[9]%
\>[9]{}\Varid{b}\leftarrow \Varid{y};{}\<[E]%
\\
\>[B]{}\hsindent{9}{}\<[9]%
\>[9]{}\Varid{c}\leftarrow \mathbf{if}\;\Varid{a}\;\mathbf{then}\;\Varid{f}\;(\Varid{a},\Varid{b})\;\mathbf{else}\;\Varid{g}\;(\Varid{a},\Varid{b});{}\<[E]%
\\
\>[B]{}\hsindent{9}{}\<[9]%
\>[9]{}\mathbf{ret} \;(\Varid{a},\Varid{c})\mskip1.5mu\}{}\<[E]%
\ColumnHook
\end{hscode}\resethooks
The corresponding linear algebra expression is
\begin{eqnarray}
	\ensuremath{{\Varid{f}}\mathbin\diamond{\Varid{g}}\mathrel{=}({\Varid{id}}\hskip 1pt\otimes\hskip 1pt{\alt{\Varid{f}}{\Varid{g}}}) \comp ({\Varid{fst}}\kr{\conv{\gamma}})},
	\label{eq:191029e}
\end{eqnarray}
where the isomorphism \ensuremath{\gamma} from \eqref{eq:191026b} plays a central role.
In the form of a gate, this combinator looks like:

\begin{picture}(110.00,57.00)(-10,15)
\put(10.00,18.00){\framebox(40.00,47.00)[cc]{}}
\put(62.00,45.00){\makebox(23.00,16.00)[lc]{\ensuremath{\Varid{a}}}}
\put(62.00,25.00){\makebox(23.00,16.00)[lc]{\ensuremath{\mathbf{if}\;\Varid{a}\;\mathbf{then}\;\Varid{f}\;(\Varid{a},\Varid{b})\;\mathbf{else}\;\Varid{g}\;(\Varid{a},\Varid{b})}}}
\put(-15.00,44.00){\makebox(20.00,16.00)[cc]{\ensuremath{\Varid{a}}}}
\put(22.00,39.00){\ensuremath{{\Varid{f}}\mathbin\diamond{\Varid{g}}}}
\put(-15.00,24.00){\makebox(20.00,16.00)[cc]{\ensuremath{\Varid{b}}}}
\put(0.00,52.00){\line(1,0){10.00}}
\put(50.00,52.00){\line(1,0){10.00}}
\put(0.00,32.00){\line(1,0){10.00}}
\put(50.00,32.00){\line(1,0){10.00}}
\end{picture}

As an example of this quantum ``choice" operator, note that
\begin{eqnarray}
	\ensuremath{\Varid{cnot}\mathrel{=}{\Varid{id}}\mathbin\diamond{\xsor }}
	\label{eq:191029f}
\end{eqnarray}
is the expected conditional version of case-based definition
\eqref{eq:191029c}. The calculation of \eqref{eq:191029f} is almost
immediate:
\begin{eqnarray*}
\start
	\ensuremath{\Varid{cnot}\mathrel{=}{\Varid{id}}\mathbin\diamond{\xsor }}
\just\equiv{ \eqref{eq:191029d} and \eqref{eq:191029e} ; pairing absortion }
	\ensuremath{{\Varid{fst}}\kr{\xsor }\mathrel{=}{\Varid{fst}}\kr{(\alt{\Varid{id}}{\neg } \comp \conv{\gamma})}}
\just\implied{ Leibniz }
	\ensuremath{\xsor \mathrel{=}\alt{\Varid{id}}{\neg } \comp \conv{\gamma}}
\just\equiv{ isomorphism \ensuremath{\gamma} }
	\ensuremath{\xsor  \comp \gamma\mathrel{=}\alt{\Varid{id}}{\neg }}
\just\equiv{ \eqref{eq:161128a} }
	\ensuremath{\Varid{true}}
\qed
\end{eqnarray*}

Quantum ``choice" leads to a quantum extension of the (classical) McCarthy
conditional functional combinator similar to the probabilistic one given in \cite{Ol12}:
\begin{eqnarray*}
	\ensuremath{\Varid{p}\to \Varid{f},\Varid{g}\mathrel{=}({\Varid{f}}\mathbin\diamond{\Varid{g}}) \comp ({\Varid{p}}\hskip 1pt\otimes\hskip 1pt{\Varid{id}})}
\end{eqnarray*}
The diagram below spells out the whole pipeline:
\begin{eqnarray*}
\xymatrix{
	\ensuremath{\B\times\Conid{A}}
		\ar[r]^{\ensuremath{{\Varid{p}}\hskip 1pt\otimes\hskip 1pt{\Varid{id}}}}
        	\ar@/_1pc/[rrd]_-{\ensuremath{\Varid{p}\to \Varid{f},\Varid{g}}}
&
	\ensuremath{\B\times\Conid{A}}
		\ar[r]^-{\ensuremath{{\Varid{fst}}\kr{\conv{\gamma}}}}
&
	\ensuremath{\B\times(\Conid{A}\mathbin{+}\Conid{A})}
		\ar[d]_-{\ensuremath{{\Varid{id}}\hskip 1pt\otimes\hskip 1pt{\alt{\Varid{f}}{\Varid{g}}}}}
\\
&&
	\ensuremath{\B\times\Conid{A}}
}
\end{eqnarray*}
Back to our motivating example, the following quantum McCarthy conditional
\begin{eqnarray*}
      \ensuremath{\Conid{H}\to \Conid{X},\Conid{H}}
\end{eqnarray*}
expresses in rather compact matrix notation the function \ensuremath{\Varid{cond}} given above, recall
(\ref{eq:cond},\ref{eq:not},\ref{eq:H}).

\section{Quantamorphisms}\label{sec:quanta}
We are now in a position to interpret diagram \eqref{eq:190418c} in the category
of matrices. This makes sense because initial algebras in the category of
sets and functions lift to Kleisli categories over it \cite{HJS07}. We obtain
the following (recursive) definition of quantamorphisms as matrices:
\begin{eqnarray}
	\ensuremath{\quanta{\Conid{Q}}\mathrel{=}\Psi\ \Conid{Q} \comp (\Varid{id}\mathbin\oplus {\Varid{id}}\hskip 1pt\otimes\hskip 1pt{\quanta{\Conid{Q}}}) \comp \conv{\alpha }},
\end{eqnarray}
cf.
\begin{eqnarray*}
\xymatrix@C=1.0em@R=1.1em{
	\ensuremath{{\Conid{A}}^{*}\times\Conid{B}}
		\ar[d]_{\ensuremath{\quanta{\Conid{Q}}}}
	\ar[rr]^-{\ensuremath{\conv{\alpha }}}
&&
	\ensuremath{\Conid{B}\mathbin{+}\Conid{A}\times({\Conid{A}}^{*}\times\Conid{B})}
		\ar[d]^{\ensuremath{\Varid{id}\mathbin\oplus {\Varid{id}}\hskip 1pt\otimes\hskip 1pt{\quanta{\Conid{Q}}}}}
\\
	\ensuremath{{\Conid{C}}^{*}\times\Conid{B}}
&&
	\ensuremath{\Conid{B}\mathbin{+}\Conid{A}\times({\Conid{C}}^{*}\times\Conid{B})}
		\ar[ll]^-{\ensuremath{\Psi\ \Conid{Q}}}
}
\end{eqnarray*}
where the parameter matrix \ensuremath{\Conid{Q}} is of type \ensuremath{\Conid{A}\times\Conid{B}\to \Conid{A}\times\Conid{B}} and
\begin{eqnarray}
	\ensuremath{\Psi\ \Conid{Q}\mathrel{=}\alpha  \comp (\Varid{id}\mathbin\oplus \mathsf{xl} \comp ({\Varid{id}}\hskip 1pt\otimes\hskip 1pt{\Conid{Q}}) \comp \mathsf{xl})}.
\end{eqnarray}

For this to be a quantum program there is a restriction, however:
\ensuremath{\Conid{Q}} must be a \emph{unitary transformation}. A \ensuremath{\mathbb{C}}-valued
matrix \ensuremath{\Conid{M}} is unitary iff
\begin{quote}
	\ensuremath{\Conid{M} \comp {\Conid{M}}^{\dagger}\mathrel{=}{\Conid{M}}^{\dagger} \comp \Conid{M}\mathrel{=}\Varid{id}}
\end{quote}
holds.  Comparing this with
\begin{eqnarray*}
	\ensuremath{\Varid{f} \comp \conv{\Varid{f}}\mathrel{=}\conv{\Varid{f}} \comp \Varid{f}\mathrel{=}\Varid{id}}
\end{eqnarray*}
we realise that \emph{isomorphisms} are exactly the \emph{classical} unitary
transformations.

Recall that \ensuremath{\Conid{Vec}\;\Conid{A}} is the data type of all \ensuremath{\mathbb{C}}-valued vectors with base
\ensuremath{\Conid{A}} and that \ensuremath{\Conid{A}\to \Conid{Vec}\;\Conid{B}} is a function representing a \emph{matrix} of type
\ensuremath{\Conid{A}\to \Conid{B}}. So all linear algebra expressions can be encoded as \ensuremath{\Conid{Vec}}-valued
functions and the \emph{quantamorphism} diagram above becomes the following
\ensuremath{\Conid{Vec}}-monadic program when rendered in the concrete syntax of Haskell:
\begin{hscode}\SaveRestoreHook
\column{B}{@{}>{\hspre}l<{\hspost}@{}}%
\column{6}{@{}>{\hspre}l<{\hspost}@{}}%
\column{E}{@{}>{\hspre}l<{\hspost}@{}}%
\>[B]{}\quanta{\cdot }\mathbin{::}((\Varid{a},\Varid{b})\to \Conid{Vec}\;(\Varid{c},\Varid{b}))\to ([\mskip1.5mu \Varid{a}\mskip1.5mu],\Varid{b})\to \Conid{Vec}\;([\mskip1.5mu \Varid{c}\mskip1.5mu],\Varid{b}){}\<[E]%
\\
\>[B]{}\quanta{\Varid{f}}\;([\mskip1.5mu \mskip1.5mu],\Varid{b})\mathrel{=}\mathbf{ret} \;([\mskip1.5mu \mskip1.5mu],\Varid{b}){}\<[E]%
\\
\>[B]{}\quanta{\Varid{f}}\;(\Varid{h}\mathbin{:}\Varid{t},\Varid{b})\mathrel{=}\mathbf{do}\;\{\mskip1.5mu {}\<[E]%
\\
\>[B]{}\hsindent{6}{}\<[6]%
\>[6]{}(\Varid{t'},\Varid{b'})\leftarrow \quanta{\Varid{f}}\;(\Varid{t},\Varid{b});{}\<[E]%
\\
\>[B]{}\hsindent{6}{}\<[6]%
\>[6]{}(\Varid{h''},\Varid{b''})\leftarrow \Varid{f}\;(\Varid{h},\Varid{b'});{}\<[E]%
\\
\>[B]{}\hsindent{6}{}\<[6]%
\>[6]{}\mathbf{ret} \;(\Varid{h''}\mathbin{:}\Varid{t'},\Varid{b''}){}\<[E]%
\\
\>[B]{}\hsindent{6}{}\<[6]%
\>[6]{}\mskip1.5mu\}{}\<[E]%
\ColumnHook
\end{hscode}\resethooks
The parameter \ensuremath{\Varid{f}\mathrel{=}\Lambda{\Conid{Q}}} must represent some unitary \ensuremath{\Conid{Q}}. Then \ensuremath{\quanta{\Varid{f}}}
controls \emph{qubit} \ensuremath{\Varid{b}} according to the \emph{list} of bits
passed as first parameter and the \emph{quantum} operator \ensuremath{\Conid{Q}}. The outcome
is unitary.

We can use the above monadic program to simulate quantum folds. For instance,
suppose we use \ensuremath{\Varid{bell}} from (\ref{eq:190425a}) to control the input qubit.
We can check what comes
out using \emph{GHCi} (here we are only showing the non-zero amplitudes): {
\begin{eqnarray}\small
\ensuremath{\Varid{x}\mathrel{=}\quanta{\Varid{bell}}\;([\mskip1.5mu \textsf{0},\textsf{1},\textsf{1},\textsf{1}\mskip1.5mu],\textsf{0})} =
\begin{array}{c|c}\ensuremath{{\B}^{*}\times\B}&\rotatebox{0}{}\\\hline
\hbox{\ensuremath{([\mskip1.5mu \textsf{0},\textsf{0},\textsf{0}\mskip1.5mu],\textsf{1})}}&\ensuremath{\frac{\mathrm{1}}{\mathrm{2}\;\sqrt{\mathrm{2}}}}\\
\hbox{\ensuremath{([\mskip1.5mu \textsf{1},\textsf{0},\textsf{0}\mskip1.5mu],\textsf{0})}}&\ensuremath{\mathbin{-}\frac{\mathrm{1}}{\mathrm{2}\;\sqrt{\mathrm{2}}}}\\
\hbox{\ensuremath{([\mskip1.5mu \textsf{0},\textsf{1},\textsf{0}\mskip1.5mu],\textsf{0})}}&\ensuremath{\frac{\mathrm{1}}{\mathrm{2}\;\sqrt{\mathrm{2}}}}\\
\hbox{\ensuremath{([\mskip1.5mu \textsf{1},\textsf{1},\textsf{0}\mskip1.5mu],\textsf{1})}}&\ensuremath{\mathbin{-}\frac{\mathrm{1}}{\mathrm{2}\;\sqrt{\mathrm{2}}}}\\
\hbox{\ensuremath{([\mskip1.5mu \textsf{0},\textsf{0},\textsf{1}\mskip1.5mu],\textsf{0})}}&\ensuremath{\frac{\mathrm{1}}{\mathrm{2}\;\sqrt{\mathrm{2}}}}\\
\hbox{\ensuremath{([\mskip1.5mu \textsf{1},\textsf{0},\textsf{1}\mskip1.5mu],\textsf{1})}}&\ensuremath{\mathbin{-}\frac{\mathrm{1}}{\mathrm{2}\;\sqrt{\mathrm{2}}}}\\
\hbox{\ensuremath{([\mskip1.5mu \textsf{0},\textsf{1},\textsf{1}\mskip1.5mu],\textsf{1})}}&\ensuremath{\frac{\mathrm{1}}{\mathrm{2}\;\sqrt{\mathrm{2}}}}\\
\hbox{\ensuremath{([\mskip1.5mu \textsf{1},\textsf{1},\textsf{1}\mskip1.5mu],\textsf{0})}}&\ensuremath{\mathbin{-}\frac{\mathrm{1}}{\mathrm{2}\;\sqrt{\mathrm{2}}}}
\end{array}
	\label{eq:190522a}
\end{eqnarray}}%
We can also superpose a quantamorphism with itself by passing quantum
information to the control part itself, scaling up what happens at elementary
gate level:
\begin{eqnarray*}
	\ensuremath{\Varid{y}\mathrel{=}\mathbf{do}\;\{\mskip1.5mu \Varid{i}\leftarrow \Varid{x};\quanta{\Varid{bell}}\;\Varid{i}\mskip1.5mu\}}
\end{eqnarray*}
Here \ensuremath{\Varid{x}} is the highly superposed stated calculated above \eqref{eq:190522a}.
The outcome will be:
\begin{eqnarray*}
\ensuremath{\Varid{y}} = \begin{array}{c|c}\ensuremath{{\B}^{*}\times\B}&\rotatebox{0}{}\\\hline
\hbox{\ensuremath{([\mskip1.5mu \textsf{1},\textsf{0},\textsf{0}\mskip1.5mu],\textsf{0})}}&\ensuremath{\frac{\mathrm{1}}{\mathrm{2}}}\\
\hbox{\ensuremath{([\mskip1.5mu \textsf{1},\textsf{0},\textsf{0}\mskip1.5mu],\textsf{1})}}&\ensuremath{\frac{\mathrm{1}}{\mathrm{2}}}\\
\hbox{\ensuremath{([\mskip1.5mu \textsf{0},\textsf{1},\textsf{1}\mskip1.5mu],\textsf{0})}}&\ensuremath{\mathbin{-}\frac{\mathrm{1}}{\mathrm{2}}}\\
\hbox{\ensuremath{([\mskip1.5mu \textsf{0},\textsf{1},\textsf{1}\mskip1.5mu],\textsf{1})}}&\ensuremath{\frac{\mathrm{1}}{\mathrm{2}}}
\end{array}
\end{eqnarray*}

The next step is -- instead of simulating -- to ``compile" quantamorphisms
such as \ensuremath{\quanta{\Varid{bell}}} so that they can run on a real quantum device.
The process
of compiling and running such quantum programs is described below.

\section{Implementation} \label{sec:200203d}

Recall that our main goal is to generate real (non-trivial)
quantum programs and to run them on quantum hardware, namely on
IBM Q Experience devices.

The current strategy consists in using the tool-chain depicted in Figure
\ref{fig:workflow}, which has five main steps:

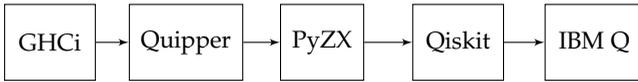
\begin{figure}
\centering\small
   \begin{tikzpicture}[node distance = 1.8cm]
       \node [block, minimum width = 1.2cm, minimum height = 1 cm] (a) {GHCi};
       \node [block, minimum width = 1.5cm, minimum height = 1 cm, right of=a] (b) {Quipper};
       \node [block, minimum width = 1.1cm, minimum height = 1 cm, right of=b] (c) {PyZX};
       \node [block, minimum width = 1.2cm, minimum height = 1 cm, right of=c] (d) {Qiskit};
       \node [block, minimum width = 1.3cm, minimum height = 1 cm, right of=d] (e) {IBM Q};
       \path [arrow] (a) -> (b);
       \path [arrow] (b) -> (c);
       \path [arrow] (c) -> (d);
       \path [arrow] (d) -> (e);
    \end{tikzpicture}
\caption{\footnotesize Tool-chain describing the quantamorphism compilation work-flow.}
\label{fig:workflow}
\end{figure}

\begin{itemize}
\item	\emph{GHCi} -- depending on the resources (i.e.\ the number of qubits
available), the monadic quantamorphisms are used to generate the finite,
unitary matrices that describe the intended (recursive) quantum computations;
\item	\emph{Quipper} \cite{GLRSV13} -- this tool generates the \emph{quantum
circuit} from the unitary matrix;
\item	\emph{PyZX} \cite{kissinger2019pyzx, DBLP:conf/icalp/CoeckeD08} -- this tool (based on
the theory of the ZX-calculus) is used to optimise the quantum circuit issued by Quipper;
\item	\emph{Qiskit \cite{Qiskit}} -- the quantum circuit generated by the previous steps is passed to this Python interface, which optimises circuits for the restrictions
of a particular physical quantum processor and manages the executions of
experiments on remote-access backends;
	\item	\emph{IBM Q} -- this is the actual hardware where \emph{Qiskit} runs the final code.
\end{itemize}

\begin{figure}
	\begin{center}
    \includegraphics[trim=2.7cm 0cm 2.1cm 0cm, clip, scale=1.5]{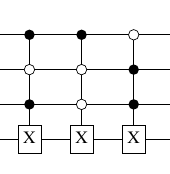}
    \end{center}
    \caption{ \label{fig:220120b} \footnotesize Circuit for \ensuremath{\quanta{\Varid{cnot}}} restricted to three control qubits.}
\end{figure}

\paragraph{GHCi and Quipper}

The practical implementation of a quantamorphism starts with the generation
of the corresponding unitary matrix. The case study presented in this section
is the quantamorphism over the control-not quantum gate accepting lists of bits
restricted by the number of control qubits (3 qubits).
The corresponding 16x16 matrix given in Fig.~\ref{fig:220120a}
is the outcome of running the quantamorphism of section~\ref{sec:quanta} in
GHCi under such size restrictions.

Then another functional programming language,  Quipper, is used to generate
the corresponding quantum circuit, shown in Fig.~\ref{fig:220120b}, using
Quipper's \text{\ttfamily exat\char95{}synthesis}~ functionality to synthesise the circuit
from the matrix.

Although this circuit looks small and feasible, IBM Q Experience devices
are unprepared to handle this kind of multi-qubit gates. As Quipper is not
bound to any particular hardware, it allows the production of circuits like
this, which require a decomposition stage, as explained next.

\begin{figure}
	\begin{center}
	\includegraphics[trim=2.9cm 0cm 53cm 0cm, clip, width=0.49\textwidth]{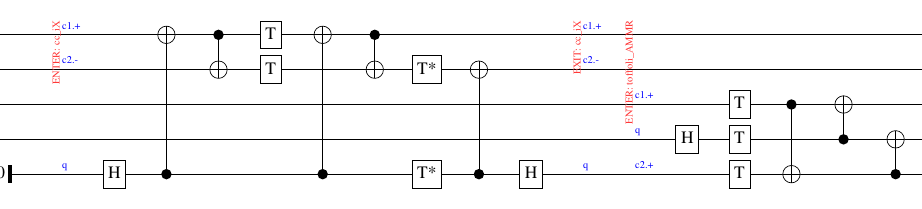}
    \end{center}
    \caption{ \label{fig:220120d} \footnotesize First part of decomposed circuit for \ensuremath{\quanta{\Varid{cnot}}} (Fig.~\ref{fig:220120b}).}
\end{figure}

The manual decomposition of this circuit is easy, see Fig.~\ref{fig:220120c}.
However, the outcome demands two ancillary qubits and Toffoli gates, which
entail further decomposition. This is circumvented by using another
Quipper function, \text{\ttfamily decompose\char95{}generic}, which generates a suitable
(but longer) decomposition -- see Fig.~\ref{fig:220120d}.

\begin{figure}
	\[
	\Qcircuit @C=.1cm @R=.3cm {
        \lstick{q_{control}}&\ctrl{1}&\qw&\qw&\qw&\ctrl{1}&\ctrl{1}&\qw&\qw&\qw&\ctrl{1}&\ctrlo{1}&\qw&\qw&\qw&\ctrlo{1}&\qw\\
        \lstick{q_{control}}&\ctrlo{3}&\qw&\qw&\qw&\ctrlo{3}&\ctrlo{3}&\qw&\qw&\qw&\ctrlo{3}&\ctrl{3}&\qw&\qw&\qw&\ctrl{3}&\qw\\
        \lstick{q_{control}}&\qw&\ctrl{2}&\qw&\ctrl{2}&\qw&\qw&\ctrlo{2}&\qw&\ctrlo{2}&\qw&\qw&\ctrl{2}&\qw&\ctrl{2}&\qw&\qw\\
        \lstick{q_{target}}&\qw&\qw&\targ&\qw&\qw&\qw&\qw&\targ&\qw&\qw&\qw&\qw&\targ&\qw&\qw&\qw\\
        \lstick{q_{aux}}&\targ&\ctrl{1}&\qw&\ctrl{1}&\targ&\targ&\ctrl{1}&\qw&\ctrl{1}&\targ&\targ&\ctrl{1}&\qw&\ctrl{1}&\targ&\qw\\
        \lstick{q_{aux}}&\qw&\targ&\ctrl{-2}&\targ&\qw&\qw&\targ&\ctrl{-2}&\targ&\qw&\qw&\targ&\ctrl{-2}&\targ&\qw&\qw
        }
	\]
	\caption{\label{fig:220120c} \footnotesize Manual decomposition of the \ensuremath{\quanta{\Varid{cnot}}} circuit of Fig.~\ref{fig:220120b} \cite{NC11}.}
\end{figure}

\begin{figure*}\tiny
\[
\begin{array}{r|rrrrrrrrrrrrrrrr}&\rotatebox{90}{\ensuremath{([\mskip1.5mu \mskip1.5mu],\textsf{0})}}&\rotatebox{90}{\ensuremath{([\mskip1.5mu \mskip1.5mu],\textsf{1})}}&\rotatebox{90}{\ensuremath{([\mskip1.5mu \textsf{0}\mskip1.5mu],\textsf{0})}}&\rotatebox{90}{\ensuremath{([\mskip1.5mu \textsf{0}\mskip1.5mu],\textsf{1})}}&\rotatebox{90}{\ensuremath{([\mskip1.5mu \textsf{0},\textsf{0}\mskip1.5mu],\textsf{0})}}&\rotatebox{90}{\ensuremath{([\mskip1.5mu \textsf{0},\textsf{0}\mskip1.5mu],\textsf{1})}}&\rotatebox{90}{\ensuremath{([\mskip1.5mu \textsf{1},\textsf{0}\mskip1.5mu],\textsf{0})}}&\rotatebox{90}{\ensuremath{([\mskip1.5mu \textsf{1},\textsf{0}\mskip1.5mu],\textsf{1})}}&\rotatebox{90}{\ensuremath{([\mskip1.5mu \textsf{1}\mskip1.5mu],\textsf{0})}}&\rotatebox{90}{\ensuremath{([\mskip1.5mu \textsf{1}\mskip1.5mu],\textsf{1})}}&\rotatebox{90}{\ensuremath{([\mskip1.5mu \textsf{0},\textsf{1}\mskip1.5mu],\textsf{0})}}&\rotatebox{90}{\ensuremath{([\mskip1.5mu \textsf{0},\textsf{1}\mskip1.5mu],\textsf{1})}}&\rotatebox{90}{\ensuremath{([\mskip1.5mu \textsf{1},\textsf{1}\mskip1.5mu],\textsf{0})}}&\rotatebox{90}{\ensuremath{([\mskip1.5mu \textsf{1},\textsf{1}\mskip1.5mu],\textsf{1})}}&\rotatebox{90}{\ensuremath{([\mskip1.5mu \textsf{0},\textsf{0},\textsf{0}\mskip1.5mu],\textsf{0})}}&\rotatebox{90}{\ensuremath{([\mskip1.5mu \textsf{0},\textsf{0},\textsf{0}\mskip1.5mu],\textsf{1})}}\\\hline
\hbox{\ensuremath{([\mskip1.5mu \mskip1.5mu],\textsf{0})}}&1&0&0&0&0&0&0&0&0&0&0&0&0&0&0&0\\
\hbox{\ensuremath{([\mskip1.5mu \mskip1.5mu],\textsf{1})}} &0&1&0&0&0&0&0&0&0&0&0&0&0&0&0&0\\
\hbox{\ensuremath{([\mskip1.5mu \textsf{0}\mskip1.5mu],\textsf{0})}}&0&0&1&0&0&0&0&0&0&0&0&0&0&0&0&0\\
\hbox{\ensuremath{([\mskip1.5mu \textsf{0}\mskip1.5mu],\textsf{1})}} &0&0&0&1&0&0&0&0&0&0&0&0&0&0&0&0\\
\hbox{\ensuremath{([\mskip1.5mu \textsf{0},\textsf{0}\mskip1.5mu],\textsf{0})}}&0&0&0&0&1&0&0&0&0&0&0&0&0&0&0&0\\
\hbox{\ensuremath{([\mskip1.5mu \textsf{0},\textsf{0}\mskip1.5mu],\textsf{1})}} &0&0&0&0&0&1&0&0&0&0&0&0&0&0&0&0\\
\hbox{\ensuremath{([\mskip1.5mu \textsf{1},\textsf{0}\mskip1.5mu],\textsf{0})}} &0&0&0&0&0&0&0&1&0&0&0&0&0&0&0&0\\
\hbox{\ensuremath{([\mskip1.5mu \textsf{1},\textsf{0}\mskip1.5mu],\textsf{1})}}  &0&0&0&0&0&0&1&0&0&0&0&0&0&0&0&0\\
\hbox{\ensuremath{([\mskip1.5mu \textsf{1}\mskip1.5mu],\textsf{0})}}       &0&0&0&0&0&0&0&0&0&1&0&0&0&0&0&0\\
\hbox{\ensuremath{([\mskip1.5mu \textsf{1}\mskip1.5mu],\textsf{1})}}        &0&0&0&0&0&0&0&0&1&0&0&0&0&0&0&0\\
\hbox{\ensuremath{([\mskip1.5mu \textsf{0},\textsf{1}\mskip1.5mu],\textsf{0})}} &0&0&0&0&0&0&0&0&0&0&0&1&0&0&0&0\\
\hbox{\ensuremath{([\mskip1.5mu \textsf{0},\textsf{1}\mskip1.5mu],\textsf{1})}}  &0&0&0&0&0&0&0&0&0&0&1&0&0&0&0&0\\
\hbox{\ensuremath{([\mskip1.5mu \textsf{1},\textsf{1}\mskip1.5mu],\textsf{0})}}  &0&0&0&0&0&0&0&0&0&0&0&0&1&0&0&0\\
\hbox{\ensuremath{([\mskip1.5mu \textsf{1},\textsf{1}\mskip1.5mu],\textsf{1})}}   &0&0&0&0&0&0&0&0&0&0&0&0&0&1&0&0\\
\hbox{\ensuremath{([\mskip1.5mu \textsf{0},\textsf{0},\textsf{0}\mskip1.5mu],\textsf{0})}}&0&0&0&0&0&0&0&0&0&0&0&0&0&0&1&0\\
\hbox{\ensuremath{([\mskip1.5mu \textsf{0},\textsf{0},\textsf{0}\mskip1.5mu],\textsf{1})}} &0&0&0&0&0&0&0&0&0&0&0&0&0&0&0&1
\end{array}
\]
	\caption{\label{fig:220120a} \footnotesize Unitary matrix describing the semantics
of quantamorphism \ensuremath{\quanta{\Varid{cnot}}} restricted to the finite lists representable by
3 control qubits ($2^3=8$ lists, ranging from the empty list \ensuremath{[\mskip1.5mu \mskip1.5mu]} to \ensuremath{[\mskip1.5mu \mathrm{0},\mathrm{0},\mathrm{0}\mskip1.5mu]}).}
\end{figure*}

\def\omitir{
\begin{figure*}\tiny
\[
\begin{array}{r|cccccccccccccccccccccccccccccc}&\rotatebox{90}{\ensuremath{([\mskip1.5mu \mskip1.5mu],\textsf{0})}}&\rotatebox{90}{\ensuremath{([\mskip1.5mu \mskip1.5mu],\textsf{1})}}&\rotatebox{90}{\ensuremath{([\mskip1.5mu \textsf{0}\mskip1.5mu],\textsf{0})}}&\rotatebox{90}{\ensuremath{([\mskip1.5mu \textsf{0}\mskip1.5mu],\textsf{1})}}&\rotatebox{90}{\ensuremath{([\mskip1.5mu \textsf{0},\textsf{0}\mskip1.5mu],\textsf{0})}}&\rotatebox{90}{\ensuremath{([\mskip1.5mu \textsf{0},\textsf{0}\mskip1.5mu],\textsf{1})}}&\rotatebox{90}{\ensuremath{([\mskip1.5mu \textsf{0},\textsf{0},\textsf{0}\mskip1.5mu],\textsf{0})}}&\rotatebox{90}{\ensuremath{([\mskip1.5mu \textsf{0},\textsf{0},\textsf{0}\mskip1.5mu],\textsf{1})}}&\rotatebox{90}{\ensuremath{([\mskip1.5mu \textsf{0},\textsf{0},\textsf{1}\mskip1.5mu],\textsf{0})}}&\rotatebox{90}{\ensuremath{([\mskip1.5mu \textsf{0},\textsf{0},\textsf{1}\mskip1.5mu],\textsf{1})}}&\rotatebox{90}{\ensuremath{([\mskip1.5mu \textsf{0},\textsf{1}\mskip1.5mu],\textsf{0})}}&\rotatebox{90}{\ensuremath{([\mskip1.5mu \textsf{0},\textsf{1}\mskip1.5mu],\textsf{1})}}&\rotatebox{90}{\ensuremath{([\mskip1.5mu \textsf{0},\textsf{1},\textsf{0}\mskip1.5mu],\textsf{0})}}&\rotatebox{90}{\ensuremath{([\mskip1.5mu \textsf{0},\textsf{1},\textsf{0}\mskip1.5mu],\textsf{1})}}&\rotatebox{90}{\ensuremath{([\mskip1.5mu \textsf{0},\textsf{1},\textsf{1}\mskip1.5mu],\textsf{0})}}&\rotatebox{90}{\ensuremath{([\mskip1.5mu \textsf{0},\textsf{1},\textsf{1}\mskip1.5mu],\textsf{1})}}&\rotatebox{90}{\ensuremath{([\mskip1.5mu \textsf{1}\mskip1.5mu],\textsf{0})}}&\rotatebox{90}{\ensuremath{([\mskip1.5mu \textsf{1}\mskip1.5mu],\textsf{1})}}&\rotatebox{90}{\ensuremath{([\mskip1.5mu \textsf{1},\textsf{0}\mskip1.5mu],\textsf{0})}}&\rotatebox{90}{\ensuremath{([\mskip1.5mu \textsf{1},\textsf{0}\mskip1.5mu],\textsf{1})}}&\rotatebox{90}{\ensuremath{([\mskip1.5mu \textsf{1},\textsf{0},\textsf{0}\mskip1.5mu],\textsf{0})}}&\rotatebox{90}{\ensuremath{([\mskip1.5mu \textsf{1},\textsf{0},\textsf{0}\mskip1.5mu],\textsf{1})}}&\rotatebox{90}{\ensuremath{([\mskip1.5mu \textsf{1},\textsf{0},\textsf{1}\mskip1.5mu],\textsf{0})}}&\rotatebox{90}{\ensuremath{([\mskip1.5mu \textsf{1},\textsf{0},\textsf{1}\mskip1.5mu],\textsf{1})}}&\rotatebox{90}{\ensuremath{([\mskip1.5mu \textsf{1},\textsf{1}\mskip1.5mu],\textsf{0})}}&\rotatebox{90}{\ensuremath{([\mskip1.5mu \textsf{1},\textsf{1}\mskip1.5mu],\textsf{1})}}&\rotatebox{90}{\ensuremath{([\mskip1.5mu \textsf{1},\textsf{1},\textsf{0}\mskip1.5mu],\textsf{0})}}&\rotatebox{90}{\ensuremath{([\mskip1.5mu \textsf{1},\textsf{1},\textsf{0}\mskip1.5mu],\textsf{1})}}&\rotatebox{90}{\ensuremath{([\mskip1.5mu \textsf{1},\textsf{1},\textsf{1}\mskip1.5mu],\textsf{0})}}&\rotatebox{90}{\ensuremath{([\mskip1.5mu \textsf{1},\textsf{1},\textsf{1}\mskip1.5mu],\textsf{1})}}\\\hline \hbox{\ensuremath{([\mskip1.5mu \mskip1.5mu],\textsf{0})}}&1&0&0&0&0&0&0&0&0&0&0&0&0&0&0&0&0&0&0&0&0&0&0&0&0&0&0&0&0&0\\\hbox{\ensuremath{([\mskip1.5mu \mskip1.5mu],\textsf{1})}}&0&1&0&0&0&0&0&0&0&0&0&0&0&0&0&0&0&0&0&0&0&0&0&0&0&0&0&0&0&0\\\hbox{\ensuremath{([\mskip1.5mu \textsf{0}\mskip1.5mu],\textsf{0})}}&0&0&1&0&0&0&0&0&0&0&0&0&0&0&0&0&0&0&0&0&0&0&0&0&0&0&0&0&0&0\\\hbox{\ensuremath{([\mskip1.5mu \textsf{0}\mskip1.5mu],\textsf{1})}}&0&0&0&1&0&0&0&0&0&0&0&0&0&0&0&0&0&0&0&0&0&0&0&0&0&0&0&0&0&0\\\hbox{\ensuremath{([\mskip1.5mu \textsf{0},\textsf{0}\mskip1.5mu],\textsf{0})}}&0&0&0&0&1&0&0&0&0&0&0&0&0&0&0&0&0&0&0&0&0&0&0&0&0&0&0&0&0&0\\\hbox{\ensuremath{([\mskip1.5mu \textsf{0},\textsf{0}\mskip1.5mu],\textsf{1})}}&0&0&0&0&0&1&0&0&0&0&0&0&0&0&0&0&0&0&0&0&0&0&0&0&0&0&0&0&0&0\\\hbox{\ensuremath{([\mskip1.5mu \textsf{0},\textsf{0},\textsf{0}\mskip1.5mu],\textsf{0})}}&0&0&0&0&0&0&1&0&0&0&0&0&0&0&0&0&0&0&0&0&0&0&0&0&0&0&0&0&0&0\\\hbox{\ensuremath{([\mskip1.5mu \textsf{0},\textsf{0},\textsf{0}\mskip1.5mu],\textsf{1})}}&0&0&0&0&0&0&0&1&0&0&0&0&0&0&0&0&0&0&0&0&0&0&0&0&0&0&0&0&0&0\\\hbox{\ensuremath{([\mskip1.5mu \textsf{0},\textsf{0},\textsf{1}\mskip1.5mu],\textsf{0})}}&0&0&0&0&0&0&0&0&0&1&0&0&0&0&0&0&0&0&0&0&0&0&0&0&0&0&0&0&0&0\\\hbox{\ensuremath{([\mskip1.5mu \textsf{0},\textsf{0},\textsf{1}\mskip1.5mu],\textsf{1})}}&0&0&0&0&0&0&0&0&1&0&0&0&0&0&0&0&0&0&0&0&0&0&0&0&0&0&0&0&0&0\\\hbox{\ensuremath{([\mskip1.5mu \textsf{0},\textsf{1}\mskip1.5mu],\textsf{0})}}&0&0&0&0&0&0&0&0&0&0&0&1&0&0&0&0&0&0&0&0&0&0&0&0&0&0&0&0&0&0\\\hbox{\ensuremath{([\mskip1.5mu \textsf{0},\textsf{1}\mskip1.5mu],\textsf{1})}}&0&0&0&0&0&0&0&0&0&0&1&0&0&0&0&0&0&0&0&0&0&0&0&0&0&0&0&0&0&0\\\hbox{\ensuremath{([\mskip1.5mu \textsf{0},\textsf{1},\textsf{0}\mskip1.5mu],\textsf{0})}}&0&0&0&0&0&0&0&0&0&0&0&0&0&1&0&0&0&0&0&0&0&0&0&0&0&0&0&0&0&0\\\hbox{\ensuremath{([\mskip1.5mu \textsf{0},\textsf{1},\textsf{0}\mskip1.5mu],\textsf{1})}}&0&0&0&0&0&0&0&0&0&0&0&0&1&0&0&0&0&0&0&0&0&0&0&0&0&0&0&0&0&0\\\hbox{\ensuremath{([\mskip1.5mu \textsf{0},\textsf{1},\textsf{1}\mskip1.5mu],\textsf{0})}}&0&0&0&0&0&0&0&0&0&0&0&0&0&0&1&0&0&0&0&0&0&0&0&0&0&0&0&0&0&0\\\hbox{\ensuremath{([\mskip1.5mu \textsf{0},\textsf{1},\textsf{1}\mskip1.5mu],\textsf{1})}}&0&0&0&0&0&0&0&0&0&0&0&0&0&0&0&1&0&0&0&0&0&0&0&0&0&0&0&0&0&0\\\hbox{\ensuremath{([\mskip1.5mu \textsf{1}\mskip1.5mu],\textsf{0})}}&0&0&0&0&0&0&0&0&0&0&0&0&0&0&0&0&0&1&0&0&0&0&0&0&0&0&0&0&0&0\\\hbox{\ensuremath{([\mskip1.5mu \textsf{1}\mskip1.5mu],\textsf{1})}}&0&0&0&0&0&0&0&0&0&0&0&0&0&0&0&0&1&0&0&0&0&0&0&0&0&0&0&0&0&0\\\hbox{\ensuremath{([\mskip1.5mu \textsf{1},\textsf{0}\mskip1.5mu],\textsf{0})}}&0&0&0&0&0&0&0&0&0&0&0&0&0&0&0&0&0&0&0&1&0&0&0&0&0&0&0&0&0&0\\\hbox{\ensuremath{([\mskip1.5mu \textsf{1},\textsf{0}\mskip1.5mu],\textsf{1})}}&0&0&0&0&0&0&0&0&0&0&0&0&0&0&0&0&0&0&1&0&0&0&0&0&0&0&0&0&0&0\\\hbox{\ensuremath{([\mskip1.5mu \textsf{1},\textsf{0},\textsf{0}\mskip1.5mu],\textsf{0})}}&0&0&0&0&0&0&0&0&0&0&0&0&0&0&0&0&0&0&0&0&0&1&0&0&0&0&0&0&0&0\\\hbox{\ensuremath{([\mskip1.5mu \textsf{1},\textsf{0},\textsf{0}\mskip1.5mu],\textsf{1})}}&0&0&0&0&0&0&0&0&0&0&0&0&0&0&0&0&0&0&0&0&1&0&0&0&0&0&0&0&0&0\\\hbox{\ensuremath{([\mskip1.5mu \textsf{1},\textsf{0},\textsf{1}\mskip1.5mu],\textsf{0})}}&0&0&0&0&0&0&0&0&0&0&0&0&0&0&0&0&0&0&0&0&0&0&1&0&0&0&0&0&0&0\\\hbox{\ensuremath{([\mskip1.5mu \textsf{1},\textsf{0},\textsf{1}\mskip1.5mu],\textsf{1})}}&0&0&0&0&0&0&0&0&0&0&0&0&0&0&0&0&0&0&0&0&0&0&0&1&0&0&0&0&0&0\\\hbox{\ensuremath{([\mskip1.5mu \textsf{1},\textsf{1}\mskip1.5mu],\textsf{0})}}&0&0&0&0&0&0&0&0&0&0&0&0&0&0&0&0&0&0&0&0&0&0&0&0&1&0&0&0&0&0\\\hbox{\ensuremath{([\mskip1.5mu \textsf{1},\textsf{1}\mskip1.5mu],\textsf{1})}}&0&0&0&0&0&0&0&0&0&0&0&0&0&0&0&0&0&0&0&0&0&0&0&0&0&1&0&0&0&0\\\hbox{\ensuremath{([\mskip1.5mu \textsf{1},\textsf{1},\textsf{0}\mskip1.5mu],\textsf{0})}}&0&0&0&0&0&0&0&0&0&0&0&0&0&0&0&0&0&0&0&0&0&0&0&0&0&0&1&0&0&0\\\hbox{\ensuremath{([\mskip1.5mu \textsf{1},\textsf{1},\textsf{0}\mskip1.5mu],\textsf{1})}}&0&0&0&0&0&0&0&0&0&0&0&0&0&0&0&0&0&0&0&0&0&0&0&0&0&0&0&1&0&0\\\hbox{\ensuremath{([\mskip1.5mu \textsf{1},\textsf{1},\textsf{1}\mskip1.5mu],\textsf{0})}}&0&0&0&0&0&0&0&0&0&0&0&0&0&0&0&0&0&0&0&0&0&0&0&0&0&0&0&0&0&1\\\hbox{\ensuremath{([\mskip1.5mu \textsf{1},\textsf{1},\textsf{1}\mskip1.5mu],\textsf{1})}}&0&0&0&0&0&0&0&0&0&0&0&0&0&0&0&0&0&0&0&0&0&0&0&0&0&0&0&0&1&0
\end{array}
\]
	\caption{\footnotesize Unitary matrix describing the semantics
of quantamorphism \ensuremath{\quanta{\Varid{cnot}}} for input lists of length at most 3.}
	\label{fig:191027a}
\end{figure*}
}

\paragraph{Qiskit, PyZX and IBM Q Experience}

Qiskit is an open-source software for quantum computation. Using Qiskit is one of two ways to interact with the IBM Q Experience,
which in turn is a cloud platform providing interaction with real quantum devices.

Since the manual translation of Quipper circuits into Qiskit syntax is
error-prone, a tool -- QuippertoQiskit -- was developed to assist in this
phase of the pipeline.\footnote{This tool can translate every standard gate from Quipper to Qiskit syntax and is available in \cite{Quanta20}.}

The experiments were performed with version 0.14.1 of Qiskit and run in the
\textsf{ibmq\_boeblingen} device, version 1.0.6.
Although a considerable number of IBM Q devices are available to the public,
this specific system is exclusive to the IBM Q Network.

The selection of \textsf{ibmq\_boeblingen} derives from its relatively high
average decoherence times (77.888 $\mu s$/99.935$\mu s$) and a relatively
low average of CX errors (0.0118). It is important to use devices with high
decoherence times because this is the lifetime of the qubit state. After
such time there are little to no guarantees the state is the theoretically expected
one. On the other hand, CX gates are the gates that create entangled
states, where the probability of errors is higher.

Recall that the circuit of Fig.~\ref{fig:220120b} required decomposition
(Fig.\ref{fig:220120d}). Implementing this decomposition in Qiskit leads to
118 operations (gates) with depth 81. The size of the program is an issue
because the longer a quantum program takes to run, the greater  the chances are
of qubits losing their state (quantum decoherence). Moreover, 51 of the
gates are control gates, which cause an increase in error rate.

The size obstacle is inflated yet again when the circuit is compiled to
the actual quantum device. Such a compilation results in a circuit with
size 172 (with 125 control gates) and depth 132.

Luckily, there are some ways to optimise the transpiler process.
The following step saw the circuit go through the four types of transpiler
optimisation supplied by Qiskit and also be rewritten with PyZX.

The most straightforward optimisation of a circuit is its transformation
 locally with some known equations, e.g.\footnote{\ensuremath{\Conid{Z}} denotes the Pauli-Z gate and \ensuremath{\Conid{S}} the Swap gate \cite{NC11}.}
 $Z\cdot Z = id$, $X \cdot X = id$,
  $H \cdot Z \cdot H = X$, $S \cdot H \cdot S \cdot H \cdot S=H$, etc.
PyZX follows a different approach, which avoids having to deal with a large
number of equations. In the first step, it converts the circuits into smaller
sections, named \emph{spiders}. The spiders compose a ZX-diagram, which is internally
just a graph that can be optimised using the equations of the ZX calculus
\cite{DBLP:conf/icalp/CoeckeD08}.
The outcome of this optimisation is not a circuit, but PyZX generates a new one from it.
In other words, PyZX is a tool implementing the theory of ZX-calculus for automated rewriting
of large-scale quantum circuits \cite{kissinger2019pyzx}.

The circuit resulting from PyZX underwent Qiskit optimisations at levels 2
and 3 (in other words, optimisations that consider the errors of the selected
backend). A summary of the results can be found in the following table:
\emskip
{\footnotesize\hskip -2em
\(
\begin{array}{c|ccccccc}
& \bf init
& \bf backend
& \bf opt2
& \bf opt3
& \bf zxB
& \bf zxo2
& \bf zxo3 \\ \hline
\bf Size  & \it 118 & 172 & 174 & 208     & \it 46 & 86   & 103 \\
\bf CX    & \it  51 & 125 & 125 & \bf 109 & \it 17 & 60   & \bf 58 \\
\bf Depth & \it  81 & 132 & \bf 122 & 139 & \it 31 & \bf 55 & 64 \\
\end{array}
\)}
\emskip
\noindent The table shows the total number of gates, number of CX gates, and depth
of each quantum circuit. Column `init' refers to the initial circuit
implemented in Qiskit; `backend' is the circuit that actually runs on the backend
\textsf{ibmq\_boeblingen} without optimisations; `opt2' and `opt3' are the circuits
after the optimised transpiler levels 2 and 3, respectively; `zxB' corresponds
to the circuit that went through PyZX and the transpiler with no optimisations;
finally the last two (`zxo2' and `zxo3') went through PyZX and optimisation levels
2 and 3, respectively. The best results are highlighted in bold.

\section{Results on the IBM Q Boeblingen} \label{sec:200203e}
Qiskit comprises four modules: \emph{Terra}, \emph{Aer}, \emph{Aqua}, and
\emph{Ignis}. \emph{Terra} serves to create circuits, \emph{Aer} allows various
types of simulations, \emph{Aqua} handles the quantum algorithms, and finally, the
main function of \emph{Ignis} is to study and mitigate quantum errors.

Since quantum errors are a serious problem of large quantum circuits,
the \emph{Ignis} module is essential, making it possible to find the average measurement fidelity
of the qubits (0.796) and set a filter to mitigate errors in the results.

As expected, the results of the simulation agreed with the unitary matrix of
Fig.~\ref{fig:220120a}. In particular, simulations of the initial and the PyZX
circuits with all classical outcomes support the feasibility of compiling
quantum programs with this method -- see the table in Fig.~\ref{tab:270120b}.

\begin{figure}
\small\centering
\(
\begin{array}{c|c|c}
\multicolumn{2}{c|}{\hbox{\bf Initial and \bf PyZX circuit}} & \hbox{\bf Typed input}\\
\bf input & \bf output & \\ \hline \hline
0000 & 0000 & \hbox{\ensuremath{([\mskip1.5mu \mskip1.5mu],\textsf{0})}}\\
0001 & 0001 & \hbox{\ensuremath{([\mskip1.5mu \mskip1.5mu],\textsf{1})}}\\
0010 & 0010 & \hbox{\ensuremath{([\mskip1.5mu \textsf{0}\mskip1.5mu],\textsf{0})}}\\
0011 & 0011 & \hbox{\ensuremath{([\mskip1.5mu \textsf{0}\mskip1.5mu],\textsf{1})}}\\
0100 & 0100 & \hbox{\ensuremath{([\mskip1.5mu \textsf{0},\textsf{0}\mskip1.5mu],\textsf{0})}}\\
0101 & 0101 & \hbox{\ensuremath{([\mskip1.5mu \textsf{0},\textsf{0}\mskip1.5mu],\textsf{1})}}\\
0110 & \bf 0111 & \hbox{\ensuremath{([\mskip1.5mu \textsf{1},\textsf{0}\mskip1.5mu],\textsf{0})}}\\
0111 & \bf 0110 & \hbox{\ensuremath{([\mskip1.5mu \textsf{1},\textsf{0}\mskip1.5mu],\textsf{1})}}\\
1000 & \bf 1001 & \hbox{\ensuremath{([\mskip1.5mu \textsf{1}\mskip1.5mu],\textsf{0})}}\\
1001 & \bf 1000 & \hbox{\ensuremath{([\mskip1.5mu \textsf{1}\mskip1.5mu],\textsf{1})}}\\
1010 & \bf 1011 & \hbox{\ensuremath{([\mskip1.5mu \textsf{0},\textsf{1}\mskip1.5mu],\textsf{0})}}\\
1011 & \bf 1010 & \hbox{\ensuremath{([\mskip1.5mu \textsf{0},\textsf{1}\mskip1.5mu],\textsf{1})}}\\
1100 & 1100 & \hbox{\ensuremath{([\mskip1.5mu \textsf{1},\textsf{1}\mskip1.5mu],\textsf{0})}}\\
1101 & 1101 & \hbox{\ensuremath{([\mskip1.5mu \textsf{1},\textsf{1}\mskip1.5mu],\textsf{1})}}\\
1110 & 1110 & \hbox{\ensuremath{([\mskip1.5mu \textsf{0},\textsf{0},\textsf{0}\mskip1.5mu],\textsf{0})}}\\
1111 & 1111 & \hbox{\ensuremath{([\mskip1.5mu \textsf{0},\textsf{0},\textsf{0}\mskip1.5mu],\textsf{1})}}
\end{array}
\)
\caption{\small Comparison between input and outputs of the initial and the PyZX
circuits for simulations in Qiskit Aer. This aims to show that both circuits have
the same behavior.
\label{tab:270120b}}
\end{figure}

The extensive depth of the circuits could already lead the reader to the
conclusion that the results of execution in the real device are not as pleasing.

\begin{figure}
	\begin{center}
    \includegraphics{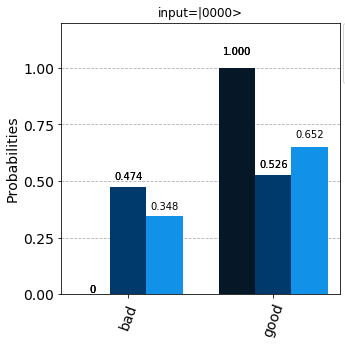}
    \end{center}
    \caption{ \label{fig:200202a} \footnotesize Qiskit plot of the probability
of bad/good results issued by \ensuremath{\quanta{\Varid{cnot}}} subject to input
\ensuremath{([\mskip1.5mu \mskip1.5mu],\mathrm{0})}, i.e.\ \ensuremath{{|\mathrm{0000}\rangle}} at bit level, running on \textsf{ibmq\_boeblingen}.
Good results are the ones where the measurement of the target qubit is zero,
and bad results are the measurements where the target qubit is one.
The darkest blue bars 
correspond to the simulation, i.e.\ the ideal \ensuremath{\mathrm{100}\mathbin{\%}}
of obtaining the right output (on the good side), \ensuremath{\mathrm{0}\mathbin{\%}} on the bad side.
The middle bars 
represent the results of running the circuit in the real device with no optimisations.
Finally, the light blue bars 
display the performance using the PyZX compiler, optimisation 3 and mitigation.}
\end{figure}

\begin{figure}
	\begin{center}
    \includegraphics[scale=1]{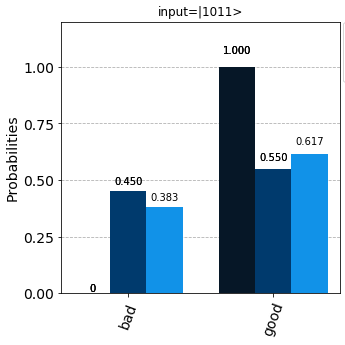}
    \end{center}
    \caption{ \label{fig:200202b} \footnotesize Plot similar to that of Fig.~\ref{fig:200202a}
for input \ensuremath{([\mskip1.5mu \mathrm{0},\mathrm{1}\mskip1.5mu],\mathrm{1})}, i.e.\ \ensuremath{{|\mathrm{1011}\rangle}} at bit level.}
\end{figure}

Figures~\ref{fig:200202a} and~\ref{fig:200202b} plot the outcome of
experimenting with inputs $\ket{0000}$ and $\ket{1011}$, respectively,
in the real device. In the first case, the expected outcome is $\ket{0000}$,
since the target should not change. In the second case, the expected result
is $\ket{1010}$.
We focus our attention on the result of measuring the target qubit.
Obtaining the expected results in the real device happens roughly in 50\% of
the measurements. However, manipulating the circuit to decrease its size,
optimising considering errors of the device, and filtering the outcomes reveals
a tendency towards the theoretical results.

\section{Conclusions}

Quantum programming is a new, promising paradigm for computing,
and as such one that is receiving much attention and investment.

QRAM\footnote{QRAM stands for Quantum Random Access Memory Model \cite{YM08}.}
is a widely accepted model for quantum computing systems, consisting of a classical
computer playing the role of the master and a quantum device accessed by
the master on request \cite{YM08}.
This model regards measurements as an intrinsic part of a quantum
computation, since this is the only way master and slave can communicate.
Moreover, measurements are indeed part of well-known
quantum strategies, for instance teleportation \cite{NC11}.

However, measurements add complexity to the mathematical
models of quantum computing, arising from the need to combine effects of
two kinds: \emph{quantum} effects explained by quantum physics and
the \emph{probabilistic} effect of reading quantum data.

It turns out that the classic master/quantum slave interplay is often thought
of too atomically, perhaps influenced by the imperative programming principle
that access to data always implies reading it from memory. While classical reading
does not harm the data, reading quantum data spoils the quantum
effect. From this perspective, measuring as little as possible is a good
idea. This paper investigates measurement-free quantum computations.

Another level of complexity arises from the aim to extend arbitrary classical
iteration and recursion to quantum programming. This has led to an extension
of classical fixpoint theory to the new setting which ends up with mixed feelings
about the possibility of ever truly realising quantum control \cite{BP15}.

The theory of classical recursion is well-structured around a taxonomy of
recursive patterns, termed \emph{morphisms} due to their inspiration in category
theory \cite{HWG15}. One member of this \emph{zoo}, called the \emph{catamorphism},
has the property of always terminating, while encompassing a wide class of
algorithms over inductive data structures such as lists, trees, and so on. Recursion
in the current paper is bound to the catamorphism pattern.

By restricting itself to structural recursion without measurements, this
paper achieves quantum control as in reference
\cite{DBLP:journals/corr/abs-1804-00952}:
\begin{quote} \em
The novel aspect of quantum control that we are able to capture here is a
notion of quantum loops. These loops were believed to be hard, if not impossible.
What makes it work in our approach is the fact that we are firmly within
a closed quantum system, without measurements.
(...) 
As we restrict fixpoints to structural recursion, valid isos are regular
enough to capture unitarity.
\end{quote}
However, we achieve this aim in a substantially different way, in various respects:
\begin{itemize}
\item	Reversibility: by generic calculation of reversible envelopes for
non-reversible operations, based on minimal complements (Section~\ref{sec:191026ce}).
Classical gates such as controlled-not and the Toffoli gate arise in this way.
\item	Recursion: by generalising complementation to a class of catamorphisms
called \emph{folds}, which are recursive functions over finite lists\footnote{Iterative
\emph{loops}, sometimes called \ensuremath{\Varid{for}}-loops, are also in this class, see e.g.\ \cite{Ol18ic}.} (Sections \ref{sec:200203a} and \ref{sec:200203b}).
\item
	Quantamorphisms: generalisation of reversible folds to unitary folds,
	enabling recursive quantum computing under structural quantum control (Sections \ref{sec:200203c} to \ref{sec:quanta}).
\item
	Implementation on IBM Q: proof-of-concept implementation of quantamorphisms
	on IBM Q Experience devices (Sections \ref{sec:200203d} to \ref{sec:200203e}).\footnote{For a detailed account of this experimentation see the webpage \cite{Quanta20} or the master's dissertation \cite{Ne18} for older versions.}
\end{itemize}

Concerning the last step, running the generated circuits in real devices
shows evidence of decoherence problems, albeit tending to the correct behaviour.

Real quantum devices are still in an initial stage and significant enhancements
to the systems took place while doing the research reported in this paper.
Some important functions in Qiskit were also altered, bugs were removed, and
the whole system had a significant update. A comprehensive account of all the tests carried
out on IBM Q devices can be found in reference \cite{Ne18}. Some of the circuits tested
in \cite{Ne18} have been re-tested showing reduced error rates.

Such fast advances in a relatively short time increase confidence with respect
to a follow-up to this work. Better results are expected by re-testing
the work already reported, albeit possibly encountering other unforeseen limitations
of quantum devices.

\section{Future Work}
Quantum control is still in its infancy. Tuned to the current paper,
this observation calls for an extension of quantamorphisms to inductive types other
than natural numbers and finite lists. Such a generalisation is a challenge
for future work, possibly inspired by so-called \emph{traversable} structures
\cite{Jaskelioff2012}.


On the experimental side, the tool-chain used in our lab setup uses GHCi and Quipper.
As both run Haskell programs, merging these two first blocks of the tool-chain
seems viable and interesting to explore. Achieving this will require a thorough
analyse of Quipper's recursive circuit implementation \cite{TheQuipp0:online}.

It is likely that implementing circuits with these methods will result in
an initial circuit larger than the circuit implemented through a matrix.
Similarly to classical reversible programs, quantum programs (reversible
by definition) tend to add a substantial amount of garbage.

A clear challenge in the implementation process is to handle quantum errors.
Compiling strategies able to curb this obstacle are under open debate.
While IBM Q Experience developed three different kinds of optimisation levels,
other academic researchers have developed tools like PyZX.
Despite progress, both strategies were not enough to achieve clean results.
Therefore, future work should explore further, more elaborate compilation strategies.
For example, the $t\ket{ket}$ compiler, which controls the routing problem aiming
for hardware compatibility with minimal additional gate overhead \cite{Cowtan2019Feb}, seems to
show impressive results when tested against Qiskit optimisation level 1 and PyZX \cite{Cowtan2019Jun}.

\section*{Acknowledgements}
This work is financed by the ERDF – European Regional Development Fund through the Operational Programme for Competitiveness and Internationalisation - COMPETE 2020 Programme and by National Funds through the Portuguese funding agency,
FCT -- Funda\c c\~ao para a Ci\^encia e a Tecnologia,
within project POCI-01-0145-FEDER-030947.

In this project, all experiments on IBM Q devices have been carried out under the IBM Q Hub at Minho license.

\bibliography{cqiqe19}

\newpage

\thispagestyle{empty}

\appendix \small

\section{Appendix}

\paragraph{Checking \ensuremath{\Varid{g}} \eqref{eq:180318d} }
Recall \ensuremath{\Varid{g}\;(\Varid{a},(\Varid{x},\Varid{b}))\mathrel{=}(\Varid{a},(\Varid{x},\Varid{f}\;(\Varid{a},\Varid{b})))} in:
\begin{eqnarray*}
\start
     \ensuremath{\conv{\mathsf{a}}\;({(\Varid{id}\times\Varid{fst})}\kr{(\Varid{f} \comp (\Varid{id}\times\Varid{snd}))}\;(\Varid{a},(\Varid{x},\Varid{b}))}
\just={ composition; \ensuremath{\Varid{fst}} and \ensuremath{\Varid{snd}} projections}
     \ensuremath{\conv{\mathsf{a}}\;((\Varid{a},\Varid{x}),\Varid{f}\;(\Varid{a},\Varid{b}))}
\just={ associate to the righ isomorphism \ensuremath{\conv{\mathsf{a}}}}
     \ensuremath{(\Varid{a},(\Varid{x},\Varid{f}\;(\Varid{a},\Varid{b})))}
\qed
\end{eqnarray*}
\paragraph{Calculation of \eqref{eq:190418a}}
Let \ensuremath{\beta\mathrel{=}{(\Varid{id}\times\Varid{fst})}\kr{(\Varid{id}\times\Varid{snd})}}. Then \ensuremath{\mathsf{xl}\mathrel{=}(\Varid{snd}\times\Varid{id}) \comp \beta} and
\begin{eqnarray}
     \ensuremath{\conv{\mathsf{a}} \comp (\Varid{id}\times\Varid{snd})\mathrel{=}\mathsf{xl} \comp (\Varid{snd}\times\Varid{id})}
     \label{eq:190418b}
\end{eqnarray}
holds.  Then:
\begin{eqnarray*}
\start
     \ensuremath{\Psi\ \Varid{x}}
\just={ starting definion of \ensuremath{\Psi\ \Varid{x}} }
     \ensuremath{\conv{\mathsf{a}} \comp ({(\Varid{id}\times\Varid{fst})}\kr{\Varid{snd} \comp \Varid{x} \comp (\Varid{id}\times\Varid{snd})})}
\just={ factor \ensuremath{\beta} to the right }
     \ensuremath{\conv{\mathsf{a}} \comp (\Varid{id}\times\Varid{snd} \comp \Varid{x}) \comp \beta}
\just={ \eqref{eq:190418b} ; product functor }
     \ensuremath{\mathsf{xl} \comp (\Varid{id}\times\Varid{x}) \comp (\Varid{snd}\times\Varid{id}) \comp \beta}
\just={ \ensuremath{\mathsf{xl}\mathrel{=}(\Varid{id}\times\Varid{x}) \comp \beta} }
     \ensuremath{\mathsf{xl} \comp (\Varid{id}\times\Varid{x}) \comp \mathsf{xl}}
\qed
\end{eqnarray*}

\paragraph{\ensuremath{\Psi\ } \eqref{eq:190418a} preserves injectivity}
For injective \ensuremath{\Varid{x}} the kernel of \ensuremath{\mathsf{xl} \comp (\Varid{id}\times\Varid{x}) \comp \mathsf{xl}} is
\ensuremath{\conv{\mathsf{xl}} \comp (\Varid{id}\times\ker{\Varid{x}}) \comp \mathsf{xl}\mathrel{=}\conv{\mathsf{xl}} \comp \mathsf{xl}\mathrel{=}\Varid{id}} since kernel distributes
by products. Then \ensuremath{\ker{(\Psi\ \Varid{x})}\mathrel{=}\Varid{id}} since kernels also distributes by coproducts.

\paragraph{\ensuremath{\env{\anonymous }} preserves injectivity}
Let \ensuremath{\Varid{k}\mathrel{=}\env{\Varid{f}}}. By the UP \eqref{eq:180317a}, \ensuremath{\Varid{k}\mathrel{=}\Varid{f} \comp (\fun F \;\Varid{k}) \comp \conv{\alpha }}.
We calculate \ensuremath{\Conid{K}\mathrel{=}\ker{\Varid{k}}} assuming \ensuremath{\ker{\Varid{f}}\mathrel{=}\Varid{id}}:
{\footnotesize
\begin{eqnarray*}
\start
     \ensuremath{\Conid{K}\mathrel{=}\conv{\Varid{k}}\comp{\Varid{k}}}
\just\equiv{ unfold \ensuremath{\Varid{f} \comp \fun F \;\Varid{k} \comp \conv{\alpha }} }
     \ensuremath{\Conid{K}\mathrel{=}\alpha  \comp \fun F \;\conv{\Varid{k}} \comp \conv{\Varid{f}} \comp \Varid{f} \comp \fun F \;\Varid{k} \comp \conv{\alpha }}
\just\equiv{ assumption: \ensuremath{\conv{\Varid{f}} \comp \Varid{f}\mathrel{=}\Varid{id}}}
     \ensuremath{\Conid{K}\mathrel{=}\alpha  \comp \fun F \;\conv{\Varid{k}} \comp \fun F \;\Varid{k} \comp \conv{\alpha }}
\just\equiv{ \ensuremath{\fun F \;(\Conid{R} \comp \Conid{S})\mathrel{=}(\fun F \;\Conid{R}) \comp (\fun F \;\Conid{S})} and \ensuremath{\fun F \;\conv{\Conid{R}}\mathrel{=}\conv{(\fun F \;\Conid{R})}} } 
     \ensuremath{\Conid{K}\mathrel{=}\alpha  \comp \fun F \;(\conv{\Varid{k}}\comp{\Varid{k}}) \comp \conv{\alpha }}
\just\equiv{\ensuremath{\Conid{K}\mathrel{=}\conv{\Varid{k}}\comp{\Varid{k}}}; UP (for relations)}
     \ensuremath{\Conid{K}\mathrel{=}\env{\alpha }}
\just\equiv{Reflexion: \ensuremath{\env{\alpha }\mathrel{=}\Varid{id}}}
     \ensuremath{\Conid{K}\mathrel{=}\Varid{id}}
\end{eqnarray*}
}

\paragraph{Do-notation calculus}
\begin{eqnarray}
\start
     \arrayin{
        \ensuremath{\mathbf{do}\;\{\mskip1.5mu \Varid{x}\leftarrow \mathbf{ret} \;\Varid{a};\Varid{f}\;\Varid{x}\mskip1.5mu\}} &=&
     \\ 
        \ensuremath{\mathbf{do}\;\{\mskip1.5mu \Varid{x}\leftarrow \Varid{f}\;\Varid{a};\mathbf{ret} \;\Varid{x}\mskip1.5mu\}} &=& \ensuremath{\Varid{f}\;\Varid{a}}
     } \label{eq:do-not-1}
\more
     \arrayin{
	\ensuremath{\mathbf{do}\;\{\mskip1.5mu \Varid{x}\leftarrow \Varid{f}\;\Varid{a};\mathbf{do}\;\{\mskip1.5mu \Varid{y}\leftarrow \Varid{g}\;\Varid{x};\Varid{h}\;\Varid{y}\mskip1.5mu\}\mskip1.5mu\}} &=&
     \\
        \ensuremath{\mathbf{do}\;\{\mskip1.5mu \Varid{x}\leftarrow \Varid{f}\;\Varid{a};\Varid{y}\leftarrow \Varid{g}\;\Varid{x};\Varid{h}\;\Varid{y}\mskip1.5mu\}} &&
     } \label{eq:do-not-2}
\more
     \arrayin{
        \ensuremath{\mathbf{do}\;\{\mskip1.5mu \Varid{y}\leftarrow \mathbf{do}\;\{\mskip1.5mu \Varid{x}\leftarrow \Varid{f}\;\Varid{a};\Varid{g}\;\Varid{x}\mskip1.5mu\};\Varid{h}\;\Varid{y}\mskip1.5mu\}} & = &
     \\
        \ensuremath{\mathbf{do}\;\{\mskip1.5mu \Varid{x}\leftarrow \Varid{f}\;\Varid{a};\Varid{y}\leftarrow \Varid{g}\;\Varid{x};\Varid{h}\;\Varid{y}\mskip1.5mu\}} &&
     } \label{eq:do-not-3}
\end{eqnarray}

\noindent (For a comprehensive account of the \ensuremath{\mathbf{do}}-notation calculus please see \cite{GH11}.)

\paragraph{Details of the calculation of \eqref{eq:190425a}}
First, the term \ensuremath{{\Conid{H}}\hskip 1pt\otimes\hskip 1pt{\Varid{id}}}:
\begin{eqnarray*}
\start
     \ensuremath{\Varid{f}\mathrel{=}\Lambda{({\Conid{H}}\hskip 1pt\otimes\hskip 1pt{\Varid{id}})}}
\just\equiv{ Kleisli correspondence }
     \ensuremath{\Varid{f}\mathrel{=}{\Varid{had}}\hskip 1pt\otimes\hskip 1pt{\mathbf{ret} }} 
\just\equiv{ (\ref{eq:kronecker}) }
     \ensuremath{\Varid{f}\;(\Varid{a},\Varid{b})\mathrel{=}\mathbf{do}\;\{\mskip1.5mu \Varid{x}\leftarrow \Varid{had}\;\Varid{a};\Varid{y}\leftarrow \mathbf{ret} \;\Varid{b};\mathbf{ret} \;(\Varid{x},\Varid{y})\mskip1.5mu\}}
\just\equiv{ (\ref{eq:do-not-1}) }
     \ensuremath{\Varid{f}\;(\Varid{a},\Varid{b})\mathrel{=}\mathbf{do}\;\{\mskip1.5mu \Varid{x}\leftarrow \Varid{had}\;\Varid{a};\mathbf{ret} \;(\Varid{x},\Varid{b})\mskip1.5mu\}}
\qed
\end{eqnarray*}
Then
\begin{quote}
	\ensuremath{\Varid{g}\mathrel{=}\Lambda{(\Varid{cnot} \comp ({\Conid{H}}\hskip 1pt\otimes\hskip 1pt{\Varid{id}}))\mathrel{=}(\mathbf{ret}  \comp \Varid{cnot})\kcomp\Varid{f}}} ,
\end{quote}
that is:
\begin{eqnarray*}
\start
     \ensuremath{\Varid{g}\;(\Varid{a},\Varid{b})\mathrel{=}\mathbf{do}\;\{\mskip1.5mu (\Varid{y},\Varid{z})\leftarrow \Varid{f}\;(\Varid{a},\Varid{b});(\mathbf{ret}  \comp \Varid{cnot})\;(\Varid{y},\Varid{z})\mskip1.5mu\}}
\just\equiv{ inline \ensuremath{\Varid{f}} calculated above }
\hskip-2.5em\mbox{
\begin{minipage}{0.4\textwidth}
\begin{hscode}\SaveRestoreHook
\column{B}{@{}>{\hspre}l<{\hspost}@{}}%
\column{5}{@{}>{\hspre}l<{\hspost}@{}}%
\column{E}{@{}>{\hspre}l<{\hspost}@{}}%
\>[B]{}\Varid{g}\;(\Varid{a},\Varid{b})\mathrel{=}\mathbf{do}\;\{\mskip1.5mu {}\<[E]%
\\
\>[B]{}\hsindent{5}{}\<[5]%
\>[5]{}(\Varid{y},\Varid{z})\leftarrow \mathbf{do}\;\{\mskip1.5mu \Varid{x}\leftarrow \Varid{had}\;\Varid{a};\mathbf{ret} \;(\Varid{x},\Varid{b})\mskip1.5mu\};{}\<[E]%
\\
\>[B]{}\hsindent{5}{}\<[5]%
\>[5]{}\mathbf{ret} \;(\Varid{cnot}\;(\Varid{y},\Varid{z})){}\<[E]%
\\
\>[B]{}\hsindent{5}{}\<[5]%
\>[5]{}\mskip1.5mu\}{}\<[E]%
\ColumnHook
\end{hscode}\resethooks
\end{minipage}
}
\just\equiv{ (\ref{eq:do-not-3}) }
\hskip-2.5em\mbox{
\begin{minipage}{0.4\textwidth}
\begin{hscode}\SaveRestoreHook
\column{B}{@{}>{\hspre}l<{\hspost}@{}}%
\column{4}{@{}>{\hspre}l<{\hspost}@{}}%
\column{E}{@{}>{\hspre}l<{\hspost}@{}}%
\>[B]{}\Varid{g}\;(\Varid{a},\Varid{b})\mathrel{=}\mathbf{do}\;\{\mskip1.5mu {}\<[E]%
\\
\>[B]{}\hsindent{4}{}\<[4]%
\>[4]{}\Varid{x}\leftarrow \Varid{had}\;\Varid{a};(\Varid{y},\Varid{z})\leftarrow \mathbf{ret} \;(\Varid{x},\Varid{b});{}\<[E]%
\\
\>[B]{}\hsindent{4}{}\<[4]%
\>[4]{}\mathbf{ret} \;(\Varid{cnot}\;(\Varid{y},\Varid{z}))\mskip1.5mu\}{}\<[E]%
\ColumnHook
\end{hscode}\resethooks
\end{minipage}
}
\just\equiv{ (\ref{eq:do-not-1},\ref{eq:do-not-2},\ref{eq:do-not-3}) }
     \ensuremath{\Varid{g}\;(\Varid{a},\Varid{b})\mathrel{=}\mathbf{do}\;\{\mskip1.5mu \Varid{x}\leftarrow \Varid{had}\;\Varid{a};\mathbf{ret} \;(\Varid{cnot}\;(\Varid{x},\Varid{b}))\mskip1.5mu\}}
\qed
\end{eqnarray*}

\end{document}